\documentclass[twocolumn,english]{IEEEtran}
\usepackage[T1]{fontenc}
\usepackage{babel}
\usepackage{array}
\usepackage{amsmath}
\usepackage{amsthm}
\usepackage{amssymb}
\usepackage{stackrel}
\usepackage{graphicx}
\usepackage[unicode=true,
 bookmarks=true,bookmarksnumbered=true,bookmarksopen=true,bookmarksopenlevel=1,
 breaklinks=false,pdfborder={0 0 0},pdfborderstyle={},backref=false,colorlinks=false]
 {hyperref}
\hypersetup{pdftitle={Your Title},
 pdfauthor={Your Name},
 pdfpagelayout=OneColumn, pdfnewwindow=true, pdfstartview=XYZ, plainpages=false}

\makeatletter

\providecommand{\tabularnewline}{\\}

\theoremstyle{plain}
\newtheorem{thm}{\protect\theoremname}
\theoremstyle{plain}
\newtheorem{prop}[thm]{\protect\propositionname}

\usepackage{balance}
\usepackage{color}
\usepackage[linesnumbered,lined,ruled,commentsnumbered,noend]{algorithm2e}
\usepackage{ragged2e}
\newcommand\justfy{\noindent\justifying}
\newtheorem{property}{Property}
\newtheorem{example}{Example}
\makeatother

\providecommand{\propositionname}{Proposition}
\providecommand{\theoremname}{Theorem}

\begin{document}
\title{On the Optimal Duration of Spectrum Leases in Exclusive License Markets
with Stochastic Demand}
\author{Gourav~Saha,~Alhussein~A.~Abouzeid,~Zaheer Khan,~and Marja~Matinmikko-Blue
\thanks{This material is based upon work supported by the National Science Foundation under grant numbers CNS-2007454 and CNS-1456887, FiDiPro Fellow award from Business Finland (MOSSAF), Academy of Finland 6Genesis Flagship (grant 318927), and partly funded by Infotech Oulu (2018-2021). A preliminary version of this work appeared in IEEE International Symposium on Dynamic Spectrum Access Networks (DySPAN) 2018 \cite{optimalleaseDySPAN}.\protect \\
G. Saha and A. A. Abouzeid are with the Department of Electrical, Computer and Systems Engineering, Rensselaer Polytechnic Institute, Troy, NY 12180, USA; Email: sahag@rpi.edu, abouzeid@ecse.rpi.edu.\protect \\
Zaheer Khan and M. Matinmikko-Blue are with Centre for Wireless Communications (CWC), University of Oulu, Finland; Email: zaheer.khan@oulu.fi, Marja.Matinmikko@oulu.fi.}\vspace{-0.3in}}
\maketitle
\begin{abstract}
This paper addresses the following question which is of interest in
designing efficient exclusive-use spectrum licenses sold through spectrum
auctions. Given a system model in which customer demand, revenue, and
bids of wireless operators are characterized by stochastic processes
and an operator is interested in joining the market only if its expected
revenue is above a threshold and the lease duration is below a threshold,
what is the optimal lease duration which maximizes the net customer
demand served by the wireless operators? Increasing or decreasing
lease duration has many competing effects; while shorter
lease duration may increase the efficiency of spectrum allocation,
longer lease duration may increase market competition by incentivizing
more operators to enter the market. We formulate this problem as a
two-stage Stackelberg game consisting of the regulator and the wireless
operators and design efficient algorithms to find the Stackelberg equilibrium
of the entire game. These algorithms can also be used to find the
Stackelberg equilibrium under some generalizations of our model. Using these
algorithms, we obtain important numerical results and insights that
characterize how the optimal lease duration varies with respect to
market parameters in order to maximize the spectrum utilization. A few
of our numerical results are non-intuitive as they suggest that increasing
market competition may not necessarily improve spectrum utilization.
To the best of our knowledge, this paper presents the first mathematical
approach to optimize the lease duration of spectrum licenses.
\end{abstract}

\begin{IEEEkeywords}
Spectrum license, spectrum auctions, lease duration, spectrum utilization,
Stackelberg game, Nash equilibrium\vspace{-1.5em}
\end{IEEEkeywords}

\section{Introduction\label{sec:Introduction}}

\noindent With the rapid growth of wireless services and devices,
wireless data traffic is increasing. Cisco's forecast \cite{cisco2018cisco}
shows a 6-fold increase in global data traffic from 2017 to 2022.
There is only a finite amount of wireless spectrum that can be used
to support the growing wireless data traffic. There are various reports
\cite{underutilization1,underutilization2} that show that many licensed
spectrum channels are underutilized, leading to inefficient use of
the spectrum. It is widely accepted that the legacy policy of static spectrum
allocations is a major cause of inefficient spectrum utilization \cite{survey1}.
Long-term spectrum leases can also lead to spectrum hoarding \cite{redesigning}.
Long-term spectrum leases are likely to have higher license fees. Higher
license fees lead to a lower number of wireless operators in the
market and can also lead to collusion among wireless operators \cite{entrybarrier1}.
This can reduce market competition and hence may lead to inefficient
spectrum utilization. For the recently proposed Citizens Broadband
Radio Service band \cite{fcc2015}, the lease duration of Priority
Access Licenses (PAL) is an important topic of debate. Since potential
wireless operators prefer different lease duration of PALs, it has
changed multiple times over the last few years of debate; $1$ year
in $2015$ \cite{fcc2015}, $3$ years in $2016$ \cite{fcc2016},
$10$ years in $2017$ \cite{fcc2017}. In spite of the importance
of lease duration, there is no formal study to optimize lease duration
(except for our previous work \cite{optimalleaseDySPAN}).

In this paper, we present a mathematical model to capture the effect
of lease duration on spectrum utilization when channels are allocated
for \textit{exclusive use}. Our model can be summarized as follows.
\textit{First}, the customer demand, the revenue of an operator and
its bids are modeled as statistically correlated stochastic processes.
\textit{Second}, spectrum utilization is equal to the net customer
demand served by the operators over a long time horizon. \textit{Third},
the revenue of an operator and its valuation of channel is solely
dependent on the amount of customer demand it can serve using the
channel; the more the customer demand served by the operator, the
higher its revenue and valuation of the channel. \textit{Fourth},
an operator will join the market only if the lease duration is below
a threshold so that it can afford the licensing fees and if the lease
duration is such that its expected revenue is above a threshold so
that it can generate return sufficient return on its investments.

Based on our system model, we formulate an optimization problem whose
objective is to maximize spectrum utilization. The optimization problem
manifests itself as a two-stage Stackelberg game consisting of the
regulator and the wireless operators. The optimization problem, in
essence, has only one scalar decision variable, the lease duration.
Shorter lease duration increases the frequency of spectrum auctions.
Therefore, there is frequent re-allocation of channels to those operators
who values it the most. This leads to more efficient allocation of
spectrum and hence improves spectrum utilization. On the other hand,
longer lease duration, in general, ensures that the operators get
their desired return on investment. This incentivizes more operators
to join the market and hence lead to more competition which in turn
improves spectrum utilization. However, if the lease duration is too
long, some of the operators may not join the market because they cannot
afford license fees \cite{fcc2017}. These opposing factors suggest
that the optimization problem should have a non-trivial solution.

There have been several active areas of research related to spectrum
licenses, such as pricing \cite{licensepricing}, auction design \cite{licenseauction},
flexible licensing \cite{licenseflexible}, enforcement \cite{licenseenforcement},
etc. But, to the best of our knowledge, a mathematical treatment of
the impact of lease duration of spectrum licenses has been only considered
in \cite{lease_duration_main}. In \cite{lease_duration_main}, the
authors took a data-driven approach and concluded that lease duration
has no significant impact on spectrum market competition. But there
are several works like \cite{survey1,redesigning,entrybarrier1} that
suggest otherwise and also data-driven approaches cannot be generalized,
especially for extrapolation. Furthermore, higher market competition
may not necessarily imply higher spectrum utilization as we show in
section \ref{sec:Numerical-Results}. Other than \cite{lease_duration_main},
we found no paper that mathematical studies the impact of lease duration
even in other synergistic areas such as electricity markets and cloud
computing.

However, there are a few works in the spectrum sharing literature that
consider the effect of certain ``duration aspects'' on the overall
performance of the system. The work in \cite{main2Contract} considers
a market of only two service providers with a common customer base.
Time is divided into intervals. At the beginning of every interval,
an auction is conducted which redistributes the available bandwidth
based on the bids of individual service providers. The ratio of the
customer demand reaching each service provider is governed by evolutionary
game theory. The authors use simulations to conclude that shorter
allocation interval corresponds to better spectrum utilization. In
\cite{hybridpricing5}, the authors model various factors that a secondary
service provider considers when buying spectrum resources from primary
service providers. The authors design a utility function for the secondary
service provider that suggest that longer contract duration is better.
In \cite{leaseduration}, the primary user leases its bandwidth to
secondary users for a fraction of time in exchange for cooperation
(relaying). If the fraction of time is too small, it will not compensate
for the overall cost of transmission (including relaying), and hence
the secondary users may not agree to cooperate. For opportunistic
spectrum use, optimal spectrum sensing time is an area which received
wide attention from the spectrum community \cite{sensingduration0}.
There are few works in economic journals like \cite{economiccontractduration3}
that consider the problem of optimizing contract duration for welfare
analysis. The fundamental idea governing these works is a tradeoff
between opportunity cost and transaction cost. The definitions of
transaction cost and opportunity cost change with the market setting,
like housing property market \cite{economiccontractduration1}, priority
service market \cite{economiccontractduration2}, etc.

In Section~\ref{subsec:System-Model}, we present a system model
to study the effect of lease duration on spectrum utilization. Our
model captures important properties of the effect of lease duration
and market competition on spectrum utilization and an operator's expected
revenue. We also discuss how some of these properties are affected
by bidding accuracy, the statistical correlation between an operator's
bid and its revenue. These properties are discussed in Section~\ref{subsec:Characteristics}.
Up to our knowledge, this is the first system model that enables the
mathematical analysis of optimal lease duration. This constitutes
the first contribution of the paper. The work done in this paper extends
to other system models as long as it satisfies the properties discussed
in Section~\ref{subsec:Characteristics}. Few of these generalizations
are hypothesized in Section~\ref{subsec:Characteristics} as well.

In Section~\ref{subsec:Optimization-Problem}, we capture the interaction
between a regulator and the operators as a two-stage Stackelberg game
with \textit{incomplete information}. In the first stage, the regulator
sets the lease duration to optimize spectrum utilization. In the second
stage (subgame), the operators decide whether to enter the market
or not based on the lease duration set by the regulator in the first
stage.  Our model admits an unique subgame Nash equilibrium (NE) and
hence finding the Stackelberg Equilibrium of the two-stage Stackelberg game reduces to solving the
optimization problem of the first stage which has only one scalar
decision variable, the lease duration. Yet, the optimization problem
is not trivial because it is reminiscent of combinatorial optimization.
To elaborate, the debate over lease duration of PALs shows that it
may not be possible to choose a lease duration which interests all
the operators. In fact, a lease duration which interests all the operators,
even if it exists, may not lead to the optimal spectrum utilization.
Hence, in certain sense, we want to find the optimal set of interested
operators which is a combinatorial optimization problem. The formulation
of the Stackelberg game is the second contribution of the paper.

In Section~\ref{sec:Optimal-Solution}, we design algorithms to solve
the optimization problem of the first stage for two scenarios: (i)
homogeneous market with complete information, (ii) heterogeneous market
with incomplete information. Since the optimization problem has a
combinatorial nature, the number of candidate sets of interested operators
may be exponential in the number of operators in the market. The design
of an \textit{efficient} optimization algorithm relies on the result
that, with lease duration as the decision variable, the number of
candidate sets of interested operators is \textit{polynomial} in the
number of operators in the market. Designing an efficient optimization
algorithm for the first stage game is the third contribution of the
paper. The final contribution is the numerical results presented in
Section~\ref{sec:Numerical-Results}. We use our optimization algorithm
to numerically study the variation of optimal characteristics, i.e.,
optimal lease duration and optimal value of the objective function,
as a function of market parameters. We also study how bidding accuracy
and incomplete information decreases spectrum utilization. Few of
our numerical results are non-intuitive as they suggest that increasing
market competition may not necessarily improve spectrum utilization.\vspace{-0.5em}

\section{Problem Formulation\label{sec:Problem-Formulation}}

We present our system model in Section~\ref{subsec:System-Model}
and also introduce the revenue function and the objective function,
which capture the revenue of an operator and spectrum utilization,
respectively. The expressions of the revenue and the objective function
are derived in Section~\ref{subsec:Derivations}. The properties
of the objective and the revenue function are discussed in Section~\ref{subsec:Characteristics}
which reveal their practical relevance. We also discuss few generalizations
of our system model in Section~\ref{subsec:Characteristics}. Table~\ref{notations}
lists frequently used notations while other notations are standard.\vspace{-1.5em}

\noindent 
\begin{table}[t]
\caption{A table of frequently used notations. \label{notations}}

\begin{centering}
\begin{tabular}{|>{\centering}m{1.1cm}|m{6.9cm}|}
\hline 
\textbf{Notation} & \textbf{Description}\tabularnewline
\hline 
$\mathbb{Z}^{+}$ & Set of positive integers.\tabularnewline
\hline 
$\left\lceil x\right\rceil $ & Ceiling Function.\tabularnewline
\hline 
$T$ & Lease duration.\tabularnewline
\hline 
$N$ & Number of operators.\tabularnewline
\hline 
$M$ & Number of channels.\tabularnewline
\hline 
$x_{k}\left(t\right)$ & Revenue of the $k^{th}$ operator in $t^{th}$ time slot.\tabularnewline
\hline 
$Y_{k}\left(c,T\right)$ & Net revenue of the $k^{th}$ operator in $c^{th}$ epoch if lease
duration is $T$.\tabularnewline
\hline 
$\widehat{Y}_{k}\left(c,T\right)$ & Bid of the $k^{th}$ operator in $c^{th}$ epoch if lease duration
is $T$.\tabularnewline
\hline 
$\mu_{k}$ , $\widehat{\mu}_{k}$ & True and estimated mean respectively of the revenue process of the
$k^{th}$ operator.\tabularnewline
\hline 
$\sigma_{k}$ , $\widehat{\sigma}_{k}$ & True and estimated standard deviation respectively of the revenue
process of the $k^{th}$ operator.\tabularnewline
\hline 
$a_{k}$ , $\widehat{a}_{k}$ & True and estimated autocorrelation coefficient respectively of the
revenue process of the $k^{th}$ operator.\tabularnewline
\hline 
$\rho_{k}$ , $\widehat{\rho}_{k}$ & True and estimated bid correlation coefficient respectively of the
$k^{th}$ operator.\tabularnewline
\hline 
$\lambda_{k}$ , $\widehat{\lambda}_{k}$ & True and estimated minimum expected revenue (MER) requirement respectively
of the $k^{th}$ operator.\tabularnewline
\hline 
$\Lambda_{k}$ , $\widehat{\Lambda}_{k}$ & True and estimated maximum lease duration respectively above which
the $k^{th}$ operator cannot afford a channel.\tabularnewline
\hline 
$\xi_{k}$ , $\widehat{\xi}_{k}$ & Tuples representing the true and estimated parameters of the $k^{th}$
operator resp. We have, $\xi_{k}=\left(\mu_{k},\sigma_{k},a_{k},\rho_{k},\lambda_{k},\Lambda_{k}\right)$
and $\widehat{\xi}_{k}=\left(\widehat{\mu}_{k},\widehat{\sigma}_{k},\widehat{a}_{k},\widehat{\rho}_{k},\widehat{\lambda}_{k},\widehat{\Lambda}_{k}\right)$.\tabularnewline
\hline 
$\mathcal{S}$ & Set of interested operators.\tabularnewline
\hline 
$s$ & Number of interested operators. We have $s=\left|\mathcal{S}\right|$.\tabularnewline
\hline 
$\widetilde{S}$ & Set of interested operators according to the regulator.\tabularnewline
\hline 
$\mathcal{S}_{k}^{L}$ & Largest set of interested operators who \textit{may} join the market
according to the $k^{th}$ operator.\tabularnewline
\hline 
$\widetilde{S}^{L}$ & \noindent Largest set of interested operators according to the regulator.\tabularnewline
\hline 
$\mathcal{R}_{k}\left(\mathcal{S},T\right)$ & Revenue function of the $k^{th}$ operator.\tabularnewline
\hline 
$\widehat{\mathcal{R}}_{k}\left(\mathcal{S},T\right)$ & Revenue function of the $k^{th}$ operator as perceived by itself.\tabularnewline
\hline 
$\widetilde{\mathcal{R}}_{k}\left(\mathcal{S},T\right)$ & Revenue function of the $k^{th}$ operator as perceived by the regulator.\tabularnewline
\hline 
$U\left(\mathcal{S},T\right)$ & Objective function as a function of set of interested operators $\mathcal{S}$
and lease duration $T$.\tabularnewline
\hline 
$\widetilde{U}\left(\mathcal{S},T\right)$ & Objective function as perceived by the regulator as a function of
set of interested operators $\mathcal{S}$ and lease duration $T$.\tabularnewline
\hline 
$U\left(T\right)$ & Objective function as a function of lease duration $T$.\tabularnewline
\hline 
$\widetilde{U}\left(T\right)$ & Objective function as perceived by the regulator as a function of
lease duration $T$.\tabularnewline
\hline 
$\mathcal{R}\left(s,T\right)$ & Revenue function of an operator for a market that is homogeneous in
$\mu_{k}$, $\sigma_{k}$, $a_{k}$ and $\rho_{k}$.\tabularnewline
\hline 
$U\left(s,T\right)$ & Objective function as a function of number of interested operators
$s$ and lease duration $T$. It only applies for a market that is
homogeneous in $\mu_{k}$, $\sigma_{k}$, $a_{k}$ and $\rho_{k}$.\tabularnewline
\hline 
\end{tabular}
\par\end{centering}
\vspace{-1.0em}
\end{table}

\subsection{System Model\label{subsec:System-Model}}

We consider a time slotted model where $t\in\mathbb{Z}^{+}$ is a
time slot. Let $T\in\mathbb{Z}^{+}$ denote the lease duration. The
word \textit{epoch} denotes a lease duration. Hence, the time slots
corresponding to the $c^{th}$ epoch are $t\in\left[\left(c-1\right)T+1\,,\,cT\right]$
where $c\in\mathbb{Z}^{+}$. There are $N$ operators indexed $k=1,2,\ldots,N$.
Let $\mathcal{S}\subseteq\left\{ 1,2,\ldots,N\right\} $ be the set
of operators who are interested in entering the market. In our model,
the number of interested operators, $s=\left|\mathcal{S}\right|$,
is the measure of market competition \cite{lease_duration_main}.
There are $M$ channels indexed $m=1,2,\ldots,M$ which are to be
allocated to the operators in set $\mathcal{S}$, at the beginning
of every epoch, for \textit{exclusive use.} Similar to prior works
like \cite{zhou2008ebay,wu2012strategy}, these channels are assumed
to be identical. Our model assumes spectrum cap of one, i.e. an operator
is allocated \textit{at most} one channel in every epoch.

Let $x_{k}\left(t\right)$ denonte the revenue of the $k^{th}$ operator
at time slot $t$ if it is allocated a channel. The revenue of an
operator is $0$ if it is not allocated a channel. We model the revenue
$x_{k}\left(t\right)$ as a first order Gaussian Autoregressive (AR)
process. Modeling time-series with AR models is a common practice
in academic literature \cite{autoregressive1,autoregressive2}. Federal
Communication Commission report \cite{fcc2016} expresses the need
for ``periodic, market-based reassignment of channels in response
to changes in local conditions and operator needs.'' A first order
AR process is a\textit{ simple} stochastic process capturing autocorrelation
among time series data. We can model fast (slow) ``changes in local
conditions and operator needs'' by setting a lower (higher) autocorrelation
among $x_{k}\left(t\right)$. Mathematically,

\vspace{-0.75em}

\begin{equation}
x_{k}\left(t+1\right)=a_{k}x_{k}\left(t\right)+\varepsilon_{k}\left(t\right)\:;\:\forall t\geq1\label{eq:2.1.1}
\end{equation}

\noindent where $a_{k}\in\left[0,1\right)$ is the autocorrelation
coefficient, $\varepsilon_{k}\left(t\right)$ is an iid Gaussian random
process with mean $\mu_{k}^{\varepsilon}$ and standard deviation
$\sigma_{k}^{\varepsilon}$, i.e. $\varepsilon_{k}\left(t\right)\sim\mathcal{N}\left(\mu_{k}^{\varepsilon},\sigma_{k}^{\varepsilon}\right)\,,\,\forall t$,
and $x_{k}\left(1\right)$ is a gaussian random variable with mean
$\mu_{k}$ and standard deviation $\sigma_{k}$, i.e. $x_{k}\left(1\right)\sim\mathcal{N}\left(\mu_{k},\sigma_{k}\right)$
where

\vspace{-0.5em}

\begin{equation}
\mu_{k}=\frac{\mu_{k}^{\varepsilon}}{1-a_{k}}\;;\;\sigma_{k}=\frac{\sigma_{k}^{\varepsilon}}{\sqrt{1-a_{k}^{2}}}\label{eq:2.1.2}
\end{equation}

It can be shown that $x_{k}\left(t\right)$ is a \textit{stationary}
Gaussian random process \cite{chatfield} with mean $\mu_{k}$ and
standard deviation $\sigma_{k}$, i.e. $x_{k}\left(t\right)\sim\mathcal{N}\left(\mu_{k},\sigma_{k}\right)\,,\,\forall t$.
It should be noted that $x_{k}\left(t\right)$, as given by (\ref{eq:2.1.1}),
can become negative for some $t$. This however is not practical because
revenue is always positive. We can reduce the probability of $x_{k}\left(t\right)$
becoming negative by setting a low coefficient of variation $\frac{\sigma_{k}}{\mu_{k}}$.
Mathematically, $P\left[x_{k}\left(t\right)<0\right]=\frac{1}{2}\left(1+\text{erf}\left(-\frac{\mu_{k}}{\sqrt{2}\sigma_{k}}\right)\right)$.
So if $\frac{\sigma_{k}}{\mu_{k}}\leq0.5$, $P\left[x_{k}\left(t\right)<0\right]\leq0.02$.
This model is similar to other approaches for modeling non-negative
quantities by Gaussian processes for ease of analysis, e.g. \cite{gaussian1,gaussian2}.
\begin{prop}
\label{prop:NetRevenue}Let $Y_{k}\left(c,T\right)=\stackrel[t=\left(c-1\right)T+1]{cT}{\sum}x_{k}\left(t\right)$
be the net revenue of the $k^{th}$ operator in $c^{th}$ epoch if
it is allocated a channel and the lease duration is $T$. Then, $Y_{k}\left(c,T\right)$
is a gaussian random variable with mean, $\widetilde{\mu}_{k}\left(T\right)$,
and standard deviation deviation, $\widetilde{\sigma}_{k}\left(T\right)$,
where\vspace{-1.5em}

\begin{eqnarray}
\widetilde{\mu}_{k}\left(T\right) & = & \mu_{k}T\label{eq:2.1.3}\\
\widetilde{\sigma}_{k}\left(T\right) & = & \frac{\sqrt{T-a_{k}\left(2-2a_{k}^{T}+a_{k}T\right)}}{\left(1-a_{k}\right)}\sigma_{k}\label{eq:2.1.4}
\end{eqnarray}

Mathematically, $Y_{k}\left(c,T\right)\sim\mathcal{N}\left(\widetilde{\mu}_{k}\left(T\right),\widetilde{\sigma}_{k}^{2}\left(T\right)\right)\,;\,\forall c$.
\end{prop}
\begin{IEEEproof}
\noindent Please refer to Appendix~A for the proof.
\end{IEEEproof}
\noindent 
Channels are allocated through auctions in every epoch. There are
$M$ channels to be allocated and the spectrum cap is one. In a given
epoch, the regulator allocates the channels to the wireless operators
with the $M$ highest bids in that epoch. Let $\widehat{Y}_{k}\left(c,T\right)$
denote the bid of the $k^{th}$ operator in $c^{th}$ epoch if the
lease duration is $T$. Our model assumes truthful spectrum auctions.
Therefore, the bid of an operator is equal to its valuation of a channel.
An operator's valuation of a channel is equal to its revenue in an
epoch if it is allocated the channel. Our model does not account for
the strategic value of a spectrum as discussed in \cite{valuation_strategic}
which arises because ``markets are not fully competitive, and there
is value in controlling access to that market.'' To this end we have,
$\widehat{Y}_{k}\left(c,T\right)=Y_{k}\left(c,T\right)$. But this
is true only if during bidding, an operator exactly knows the true
net revenue it will earn in that epoch. In reality, the bid $\widehat{Y}_{k}\left(c,T\right)$
is only has an estimate of the true revenue $Y_{k}\left(c,T\right)$.
In our model, $\widehat{Y}_{k}\left(c,T\right)$ and $Y_{k}\left(c,T\right)$
assumes the following joint probability distribution,\vspace{-1.0em}
\begin{equation}
\begin{bmatrix}Y_{k}\left(c,T\right)\\
\widehat{Y}_{k}\left(c,T\right)
\end{bmatrix}\sim\mathcal{N}\left(\begin{bmatrix}\widetilde{\mu}_{k}\left(T\right)\\
\widetilde{\mu}_{k}\left(T\right)
\end{bmatrix},\begin{bmatrix}\widetilde{\sigma}_{k}^{2}\left(T\right) & \rho_{k}\widetilde{\sigma}_{k}^{2}\left(T\right)\\
\rho_{k}\widetilde{\sigma}_{k}^{2}\left(T\right) & \widetilde{\sigma}_{k}^{2}\left(T\right)
\end{bmatrix}\right)\,;\,\forall c\label{eq:2.1.5}
\end{equation}
where $\rho_{k}\in\left[0,1\right)$ is the correlation
coefficient between bid $\widehat{Y}_{k}\left(c,T\right)$ and true
revenue $Y_{k}\left(c,T\right)$. A higher $\rho_{k}$ implies a higher
accuracy of bidding estimate. Also note that in (\ref{eq:2.1.5}),
the bid $\widehat{Y}_{k}\left(c,T\right)$ has the same marginal distribution
as $Y_{k}\left(c,T\right)$, 
\begin{equation}
\widehat{Y}_{k}\left(c,T\right)\sim\mathcal{N}\left(\widetilde{\mu}_{k}\left(T\right),\widetilde{\sigma}_{k}^{2}\left(T\right)\right)\,;\,\forall c\label{eq:2.1.6}
\end{equation}

In our model, an operator generates revenue solely by serving customer
demand. Let $d_{k}\left(t\right)$ denonte the customer demand served
by the $k^{th}$ operator at time slot $t$ if it is allocated a channel.
An operator's revenue in time slot $t$ is $x_{k}\left(t\right)=p_{k}\left(t\right)d_{k}\left(t\right)$.
In a competitive market, the price $p_{k}\left(t\right)$ charged
by an operator to serve a unit of customer demand cannot vary significantly
with operators. Otherwise, the operator may suffer a significant loss
of its market share. Hence, our model assumes that $p_{k}\left(t\right)=p\left(t\right)\,,\,\forall k$.
Similar results are shown in \cite{berry2}. In fact, the famous Bertrand
and Cournot competition models \cite{bertrand,cournot} suggest that
with two or more operators, the market reaches perfect competition
and all operators sell at the same price. So we have, $x_{k}\left(t\right)=p\left(t\right)d_{k}\left(t\right)$.
This implies that, in our model, an operator who is generating more revenue
is also utilizing the spectrum better as it is serving more customer
demand. The results presented in this paper may not hold if $p_{k}\left(t\right)$
varies significantly among operators, e.g. when operators have strategic
valuation of spectrum because such markets are not competitive.

An operator has to invest in infrastructure development to enter the
market and further invest to lease a channel. Since the cost of leasing
a channel generally increases with lease duration, some operators
cannot afford to lease a channel if the lease duration is too high
\cite{fcc2017}. This is captured in our model using $\Lambda_{k}$,
the maximum lease duration above which the $k^{th}$ operator cannot
afford a channel. In order to get return on infrastructure development
cost and the cost of leasing a channel, the $k^{th}$ operator wants
to make a minimum expected revenue (MER) $\lambda_{k}$ in an epoch.
The $k^{th}$ operator is interested in entering the market iff the
lease duration is less than $\Lambda_{k}$ and its expected revenue
in an epoch is greater than $\lambda_{k}$.

If $k\in\mathcal{S}$, then $\mathcal{R}_{k}\left(\mathcal{S},T\right)$
is the \textit{revenue function} of the $k^{th}$ operator and it
denotes its expected revenue in an epoch if the set of interested
operators is $\mathcal{S}$ and the lease duration is $T$. In our
model, the \textit{objective function} is $U\left(\mathcal{S},T\right)$
which is proportional to the spectrum utilization when the set of
interested operators is $\mathcal{S}$ and the lease duration is $T$.
We derive expressions for $\mathcal{R}_{k}\left(\mathcal{S},T\right)$
and $U\left(\mathcal{S},T\right)$ in the next section.\vspace{-0.5em}

\subsection{Analytical Expressions of Revenue and Objective Function\label{subsec:Derivations}}

We start by introducing few notions and notations which are required
for the derivation of revenue and objective function. Let $w_{c}^{m}$
denote the index of the operator who is allocated the $m^{th}$ channel
in $c^{th}$ epoch. Without any loss of generality let us assume that
$w_{c}^{m}$ is decided by the following rule: $m^{th}$ channel is
allocated to the operator having the $m^{th}$ highest value of $\widehat{Y}_{k}\left(c,T\right)$
in the $c^{th}$ epoch. The number of channels being allocated is
$\widetilde{M}=\min\left(M,s\right)$. The revenue function $\mathcal{R}_{k}\left(\mathcal{S},T\right)$
can be expressed as\vspace{-0.5em}
\[
\mathcal{R}_{k}\left(\mathcal{S},T\right)=\stackrel[m=1]{\widetilde{M}}{\sum}E\left[Y_{k}\left(c,T\right)|w_{c}^{m}=k\right]P\left[w_{c}^{m}=k\right]
\]
\begin{equation}
\quad\quad+0\cdot P\left[\stackrel[m=1]{\widetilde{M}}{{\textstyle \bigcap}}w_{c}^{m}\neq k\right]\label{eq:2.2.2.1}
\end{equation}

\vspace{-1.25em}

\begin{equation}
\quad\quad\quad\quad\quad=\stackrel[m=1]{\widetilde{M}}{\sum}E\left[Y_{k}\left(1,T\right)|w_{1}^{m}=k\right]P\left[w_{1}^{m}=k\right]\label{eq:2.2.2}
\end{equation}

\vspace{-0.25em}

In (\ref{eq:2.2.2.1}), $P\left[w_{c}^{m}=k\right]$ is the probability
that the $k^{th}$ operator is allocated the $m^{th}$ channel in
the $c^{th}$ epoch in which case its net expected revenue is $E\left[Y_{k}\left(c,T\right)|w_{c}^{m}=k\right]$.
$P\left[\stackrel[m=1]{\widetilde{M}}{{\textstyle \bigcap}}w_{c}^{m}\neq k\right]$
is the probability that the $k^{th}$ operator is not allocated a
channel in the $c^{th}$ epoch in which case its revenue is $0$.
In (\ref{eq:2.2.2.1}), $w_{c}^{m}$ is dependent on the random variable
$\widehat{Y}_{k}\left(c,T\right)$. This shows that $Y_{k}\left(c,T\right)$
and $\widehat{Y}_{k}\left(c,T\right)$ are the only random variables
in (\ref{eq:2.2.2.1}). Hence, the expectation in (\ref{eq:2.2.2.1})
is over $Y_{k}\left(c,T\right)$ and $\widehat{Y}_{k}\left(c,T\right)$.
Based on Proposition \ref{prop:NetRevenue} and (\ref{eq:2.1.6}),
statistical properties of $Y_{k}\left(c,T\right)$ and $\widehat{Y}_{k}\left(c,T\right)$
are not dependent on epoch $c$. Therefore, the expectation in (\ref{eq:2.2.2.1})
and hence the revenue function does not depend on $c$. This means
that we can simply substitute $c=1$ to get (\ref{eq:2.2.2}). Please
note that $\mathcal{R}_{k}\left(\mathcal{S},T\right)$ as given by
(\ref{eq:2.2.2}) is a function of $\mu_{k}$, $\sigma_{k}$, $a_{k}$
and $\rho_{k}$ because the statistical properties of $Y_{k}\left(c,T\right)$
and $\widehat{Y}_{k}\left(c,T\right)$ if governed by (\ref{eq:2.1.5})
which in turn depends on $\mu_{k}$, $\sigma_{k}$, $a_{k}$ and $\rho_{k}$.
Equation \ref{eq:2.2.2} is enough for the remaining discussion in
the paper. However, we have derived a more explicit equation to calculate
$\mathcal{R}_{k}\left(\mathcal{S},T\right)$ in Appendix~B. This equation involves numerical integration.

In this paper, we consider a scenario where the regulator wants to
maximize the expected spectrum utilization. As discussed in Section
\ref{subsec:System-Model}, in our model, an operator who generates
more revenue also utilizes the spectrum better. Therefore, maximizing
expected spectrum utilization is equivalent to maximizing the net
expected revenue $V$ in optimization horizon $\mathcal{T}\gg T$.
Assume that $\mathcal{T}$ is a multiple of $T$, i.e $\mathcal{T}=CT$
where $C\in\mathbb{Z}^{+}$. We have,\vspace{-0.5em}

\begin{equation}
{\textstyle V=\stackrel[c=1]{C}{\sum}E\left[\stackrel[m=1]{\widetilde{M}}{\sum}Y_{w_{c}^{m}}\left(c,T\right)\right]}\label{eq:2.2.9}
\end{equation}

In (\ref{eq:2.2.9}), $E\left[\stackrel[m=1]{\widetilde{M}}{\sum}Y_{w_{c}^{m}}\left(c,T\right)\right]$
denotes the net expected revenue in the $c^{th}$ epoch over all the
$\mbox{{\ensuremath{\widetilde{M}}=\ensuremath{\min\left(M,s\right)}}}$
allocated channels. Similar to (\ref{eq:2.2.2.1}), the expectation
in (\ref{eq:2.2.9}) is over random variables $Y_{k}\left(c,T\right)$
and $\widehat{Y}_{k}\left(c,T\right)$ and hence the term $E\left[\stackrel[m=1]{\widetilde{M}}{\sum}Y_{w_{c}^{m}}\left(c,T\right)\right]$
is not dependent on epoch $c$. In other words, the net expected revenue
is equal in all epochs. Hence, (\ref{eq:2.2.9}) can be simplified
to
\begin{equation}
{\textstyle V=CE\left[\stackrel[m=1]{\widetilde{M}}{\sum}Y_{w_{1}^{m}}\left(1,T\right)\right]=\frac{\mathcal{T}}{T}E\left[\stackrel[m=1]{\widetilde{M}}{\sum}Y_{w_{1}^{m}}\left(1,T\right)\right]}\label{eq:2.2.10}
\end{equation}

Maximizing $V$ in (\ref{eq:2.2.10}) is equivalent to maximizing
$\frac{E\left[\stackrel[m=1]{\widetilde{M}}{\sum}Y_{w_{1}^{m}}\left(1,T\right)\right]}{T}$.
This holds even if $\mathcal{T}$ is not a multiple of $T$ provided
$\mathcal{T}\gg T$. Finally, the regulator wants to maximize\vspace{-1.0em}
\begin{equation}
U\left(\mathcal{S},T\right)=\frac{E\left[\stackrel[m=1]{\widetilde{M}}{\sum}Y_{w_{1}^{m}}\left(1,T\right)\right]}{T}\quad\quad\quad\quad\quad\quad\quad\quad\quad\quad\quad\quad\quad\label{eq:2.2.14}
\end{equation}

\vspace{-1.5em}

\begin{equation}
\quad\quad\quad=\frac{\stackrel[m=1]{\widetilde{M}}{\sum}\underset{k\in\mathcal{S}}{\sum}E\left[Y_{k}\left(1,T\right)|w_{1}^{m}=k\right]P\left[w_{1}^{m}=k\right]}{T}\label{eq:2.2.12}
\end{equation}

\vspace{-1.0em}

\begin{equation}
=\frac{1}{T}\underset{k\in\mathcal{S}}{\sum}\mathcal{R}_{k}\left(\mathcal{S},T\right)\quad\quad\quad\quad\quad\quad\quad\quad\quad\quad\label{eq:2.2.13}
\end{equation}

Equation \ref{eq:2.2.12} is obtained by first applying \textit{linearity
of expectation} and then applying \textit{law of iterated expectation}
over all possible $w_{1}^{m}$ in (\ref{eq:2.2.14}). To obtain (\ref{eq:2.2.13}),
we change the order of summation and then observe that $\mathcal{R}_{k}\left(\mathcal{S},T\right)$
is equal to $\stackrel[m=1]{\widetilde{M}}{\sum}E\left[Y_{k}\left(1,T\right)|w_{1}^{m}=k\right]P\left[w_{1}^{m}=k\right]$
(refer to (\ref{eq:2.2.2})).

We end this section by defining two new notations. Let $\mathcal{R}\left(s,T\right)$
and $U\left(s,T\right)$ denote the revenue and the objective function
respectively if the market is homogeneous in $\mu_{k}$, $\sigma_{k}$,
$a_{k}$ and $\rho_{k}$, i.e. $\mu_{k}=\mu$, $\sigma_{k}=\sigma$,
$a_{k}=a$ and $\rho_{k}=\rho\,,\,\forall k$. In such a market, the
revenue and the objective function only depend on the number of interested
operators. Also, the revenue function is the same for all the operators.
The formula for $\mathcal{R}\left(s,T\right)$ is derived in Appendix~C. The objective function $U\left(s,T\right)$
is obtained by substituting $\mathcal{R}_{k}\left(\mathcal{S},T\right)=\mathcal{R}\left(s,T\right)\,,\,\forall k$
in (\ref{eq:2.2.13}) which yields,\vspace{-0.5em}

\begin{equation}
U\left(s,T\right)=\frac{s}{T}\mathcal{R}\left(s,T\right)\label{eq:2.p.3}
\end{equation}

\subsection{Properties of the Revenue and the Objective Function\label{subsec:Characteristics}}

In this section, we discuss few properties of the revenue and the
objective function that are crucial for formulating and solving the
Stackelberg game in Section~\ref{subsec:Optimization-Problem} and
\ref{sec:Optimal-Solution}. We also discuss a few generalizations
of the system model proposed in Section~\ref{subsec:System-Model}
under which these properties of the revenue and the objective function
should remain valid. $\mathcal{R}_{k}\left(\mathcal{S},T\right)$
and $U\left(\mathcal{S},T\right)$ have the following properties.

\begin{property} \label{RevenueFuncProp1}
$\mathcal{R}_{k}\left(\mathcal{S},T\right)$ is unimodal in $T$ with a maximum.
\end{property}

\begin{property} \label{RevenueFuncProp2}
$\mathcal{R}_{k}\left(\mathcal{S},T\right)$ is monotonic decreasing in $\mathcal{S}$, i.e. $\mathcal{R}_{k}\left(\mathcal{S},T\right)\geq \mathcal{R}_{k}\left(\mathcal{S}\bigcup\left\{ a\right\} ,T\right)$ where $a\notin\mathcal{S}$.
\end{property}

\begin{property} \label{ObjFuncProp1}
$U\left(\mathcal{S},T\right)$ is monotonic increasing in $T$ or it is unimodal in $T$ with a minimum.
\end{property}

Consider an operator indexed $a$, where $a\notin\mathcal{S}$, whose
bid correlation coefficient is $\rho_{a}$.

\begin{property} \label{ObjFuncProp2}
As $\rho_{a}\rightarrow1$, $U\left(\mathcal{S}\bigcup\left\{ a\right\} ,T\right)\geq U\left(\mathcal{S},T\right)$. As $\rho_{a}$ decreases, $U\left(\mathcal{S}\bigcup\left\{ a\right\} ,T\right)$ decreases.
\end{property}

\begin{figure}[t]
\begin{centering}
\includegraphics[scale=0.55]{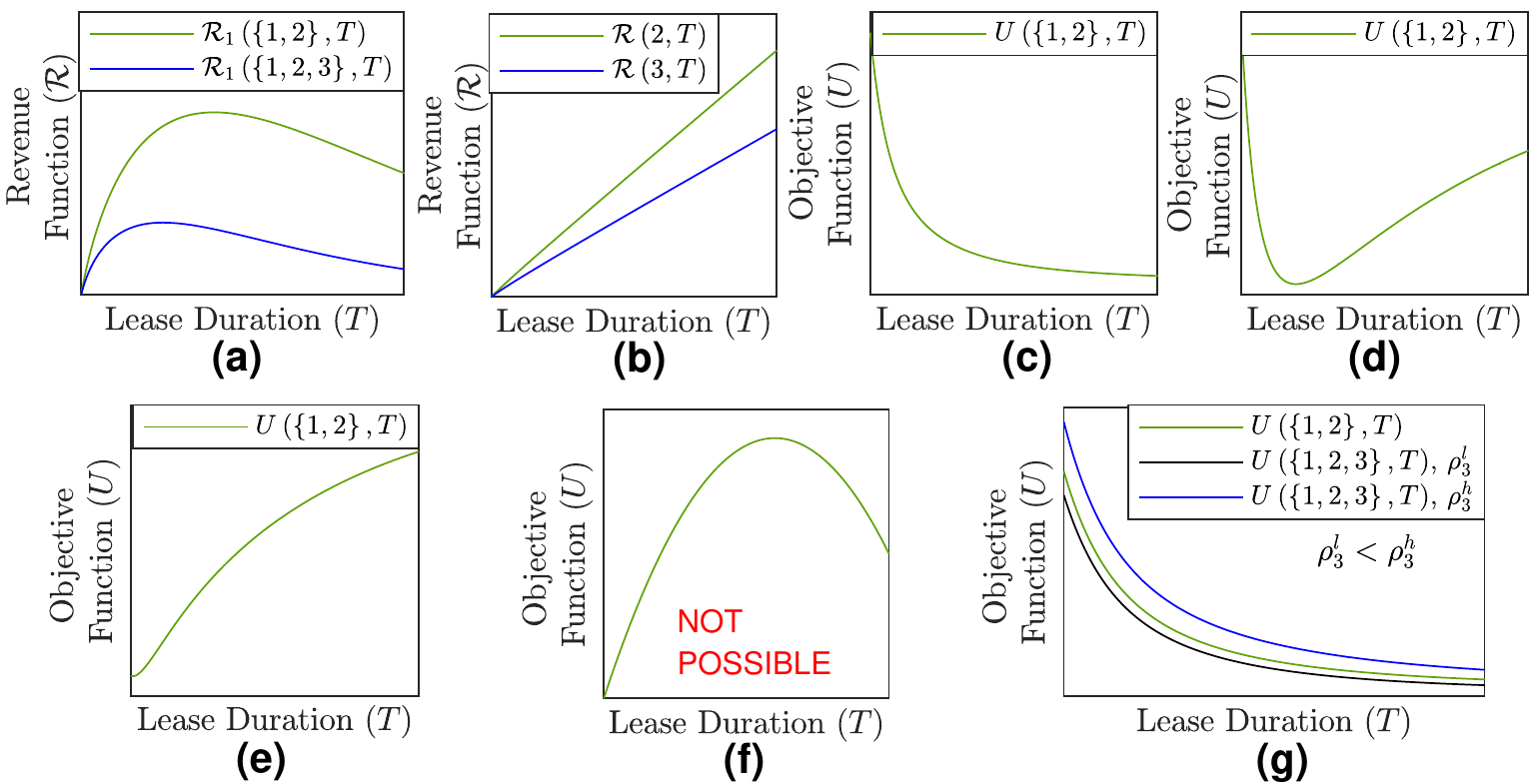}
\par\end{centering}
\caption{(a) A typical trend of $\mathcal{R}_{k}\left(\mathcal{S},T\right)$
with respect to $T$ and $\mathcal{S}$ in a heterogeneous market.
(b) A typical trend of $\mathcal{R}\left(s,T\right)$ with respect
to $T$ and $s$ in a homogeneous market. (c, d, and e) A typical
trend of $U\left(\mathcal{S},T\right)$ with respect to $T$ in a
homogeneous market (c) and in a heterogeneous market with high $\rho_{k}$
(c), mid-range $\rho_{k}$ (d) and low $\rho_{k}$ (e). (f) A trend
of $U\left(\mathcal{S},T\right)$ which is not possible in any market
scenario. (g) A typical trend of $U\left(\mathcal{S},T\right)$ with
respect to $\mathcal{S}$. The black and the blue curves corresponds
to objective function, $U\left(\left\{ 1,2,3\right\} ,T\right)$,
if bid correlation coefficient of operator $3$ is $\rho_{3}^{l}$
and $\rho_{3}^{h}$ respectively where $\rho_{3}^{h}>\rho_{3}^{l}$.\vspace{-1.25em}}
\label{func_prop_fig}
\end{figure}

Figure~\ref{func_prop_fig} is a pictorial representation of these
properties. According to Property~\ref{RevenueFuncProp1}, the revenue
function is unimodal in $T$ with a maximum. This is shown in Figure~\ref{func_prop_fig}.a
where both the blue and the green curves first increase with $T$
and then start decreasing after a certain value of lease duration.
This can be qualitatively explained as follows. The increase of revenue
function with $T$ simply happens because an operator can earn more
revenue if it has the channel for a longer duration. However, the
decrease in revenue function with $T$ is \textit{non-intuitive}.
This can be explained as follows. The coefficient of variation of
bids $\widehat{Y}_{k}\left(c,T\right)$ is $\frac{\widetilde{\sigma}_{k}\left(T\right)}{\widetilde{\mu}_{k}\left(T\right)}$
(refer to (\ref{eq:2.1.6}), (\ref{eq:2.1.3}) and (\ref{eq:2.1.4})).
As $T$ increases, coefficient of variation of $\widehat{Y}_{k}\left(c,T\right)$,
$\frac{\widetilde{\sigma}_{k}\left(T\right)}{\widetilde{\mu}_{k}\left(T\right)}$,
tends to zero and hence $\widehat{Y}_{k}\left(c,T\right)\rightarrow\widetilde{\mu}_{k}\left(T\right)=\mu_{k}T$.
In our model, those operators with high bids $\widehat{Y}_{k}\left(c,T\right)$
are allocated channels in the $c^{th}$ epoch. Since $\mbox{\ensuremath{\widehat{Y}_{k}\left(c,T\right)}}$
is approximately equal to $\mu_{k}T$ for large $T$, operators with
low $\mu_{k}$ are not likely to be allocated channels as $T$ increases.
This leads to a decrease in their revenue function. This result shows
that not all the operators would prefer a long lease duration. For
a heterogeneous market, we could only verify Property~\ref{RevenueFuncProp1}
numerically. But for a homogeneous market, we could rigorously prove
that revenue function $\mathcal{R}\left(s,T\right)$ is monotonic
increasing in $T$ (special case of unimodal function with maximum
at infinity). This is shown in Figure~\ref{func_prop_fig}.b where
both the blue and the green curves increase as $T$ increases. Please
refer to Appendix~D for the proof.

\noindent 

\vspace{-1.0em}

According to Property~\ref{RevenueFuncProp2}, the revenue function
decreases as the set of interested operators $\mathcal{S}$ grows.
This is shown in Figure~\ref{func_prop_fig}.a and \ref{func_prop_fig}.b
where the blue curve is below the green curve. As $\mathcal{S}$ grows,
an operator competes with more operators to get a channel. Hence,
the probability of an operator being allocated a channel decreases
which in turn decreases its revenue function. We have verified Property~\ref{RevenueFuncProp2}
numerically.

According to Property~\ref{ObjFuncProp1}, the objective function
$U\left(\mathcal{S},T\right)$ is unimodal in $T$ with a minimum
or it is monotonic increasing in $T$. This happens because of two
competing causes. \textit{First}, as lease duration increases, the
regulator is allocating the channels to the operators with the best
spectrum utilization less often. This reduces the efficiency of spectrum
allocation and hence spectrum utilization decreases.\textit{ Second},
due to bidding inaccuracy an operator may place a lower bid and hence
not allocated a channel even though it has a high net revenue in an
epoch (and hence a higher spectrum utilization). The chances of such
inefficient spectrum allocation increases as lease duration decreases
because spectrum auctions occurs more frequently and hence more the
chances of such erroneous spectrum allocation. As bidding coefficient
$\rho_{k}$ decreases, the bidding inaccuracy increases making the
second cause more dominating than the first one. Therefore, the objective
function is monotonic non-increasing (special case of unimodal function
with minimum at infinity) in $T$ if $\rho_{k}$ is high (Figure \ref{func_prop_fig}.c),
unimodal in $T$ with a minimum if $\rho_{k}$ is mid-range (Figure
\ref{func_prop_fig}.d) and monotonic non-decreasing in $T$ if $\rho_{k}$
is low (Figure \ref{func_prop_fig}.e). However, the objective function
will never be unimodal in $T$ with a maximum (Figure \ref{func_prop_fig}.f).
For a heterogeneous market, we could only verify Property~\ref{ObjFuncProp1}
numerically. But for a homogeneous market, we could rigorously prove
that objective function $U\left(s,T\right)$ is monotonic decreasing
in $T$ (Figure \ref{func_prop_fig}.c) for any value of $\rho_{k}$.
Please refer to Appendix~D for the proof.

Property~\ref{ObjFuncProp2} discusses the effect of a set of interested
operators on objective function and how it changes with bid correlation
coefficient. According to Property~\ref{ObjFuncProp2}, as $\rho_{a}\rightarrow1$,
$U\left(\mathcal{S}\bigcup\left\{ a\right\} ,T\right)\geq U\left(\mathcal{S},T\right)$;
but as $\rho_{a}$ decreases, $U\left(\mathcal{S}\bigcup\left\{ a\right\} ,T\right)$
decreases as well. According to our model, market competition increases
as the set of interested operators grows from $\mathcal{S}$ to $\mathcal{S}\bigcup\left\{ a\right\} $. Qualitatively, property~\ref{ObjFuncProp2} states that increasing market
competition may have two outcomes: (a) Increase in spectrum utilization
if bid correlation is high: as $\rho_{a}\rightarrow1$, there is a value of $\rho_{a}$ above which $U\left(\mathcal{S}\bigcup\left\{ a\right\} ,T\right)\geq U\left(\mathcal{S},T\right)$ implying that the increase in competition due to addition of operator $a$ lead to an increase in spectrum utilization. This is shown in Figure \ref{func_prop_fig}.g where the blue curve is above the green curve. (b) Decrease in spectrum utilization if bid correlation is low: as $\rho_{a}$ decreases, the
bid of operator $a$ is not a good estimate of its true net revenue
which in turn is proportional to operator $a$'s spectrum utilization.
Therefore, with decrease in $\rho_{a}$, there is a higher probability
of erroneous spectrum allocation to operator $a$ when its spectrum
utilization is low. In other words, spectrum allocation becomes less
efficient with decrease in $\rho_{a}$ which in turn decreases objective
function $U\left(\mathcal{S}\bigcup\left\{ a\right\} ,T\right)$. In some cases, the decrease in spectrum utilization may be large enough that $U\left(\mathcal{S}\bigcup\left\{ a\right\} ,T\right)< U\left(\mathcal{S},T\right)$. This is non-intuitive as it suggests that increasing market competition may not necessarily improve spectrum
utilization. This is shown in Figure \ref{func_prop_fig}.g where the black curve is below the green curve. For a heterogeneous market, we could only verify Property~\ref{ObjFuncProp2} numerically. But for a homogeneous market, we could rigorously prove that objective function $U\left(s,T\right)$ is monotonic increasing in $s$ for any value of $\rho_{k}$. Please refer to Appendix~D for the proof.

We end this section by stressing that the results in the subsequent
sections remain valid as long as Properties~\ref{RevenueFuncProp1}-\ref{ObjFuncProp1}
hold. We believe that these properties are\textit{ robust} and hold
under various generalizations of our system model. Two such generalizations are as follows. \textit{First}, we can relax our system model such that an operator can be allocated more than one channel. \textit{Second}, we can generalize the revenue process given by (\ref{eq:2.1.1})  to other stochastic processes. An interesting generalization is to consider \textit{cross-correlation} among operators' revenue processes. Such generalizations may not guarantee closed-form expressions of the revenue function, in which case, it can be estimated using \textit{Monte-Carlo
simulations}.\vspace{-1.0em}

\section{Stackelberg Game Formulation and Solution\label{sec:Optimal-Solution}}

In this section, we formulate the problem as a Stackelberg game in in Section~\ref{subsec:Optimization-Problem}. We solve Stage-2 of the Stackelberg game and use the result to formulate Stage-1 of the Stackelberg game as an optimization problem $OP1$. We then design  algorithms to solve $OP1$ for two market settings: (a) homogeneous market with complete information in Section~\ref{subsec:Homogeneous} (Proposition~\ref{prop:optimal-homogeneous}), and (b) heterogeneous market with incomplete information in Section~\ref{subsec:Heterogeneous} (Algorithm~\ref{opalgo_heterogeneous}).\vspace{-1.0em}

\subsection{Stackelberg game formulation\label{subsec:Optimization-Problem}}

In this section, we formulate the optimization problem from the regulator's
and the operators' perspective. The optimization problem manifests
itself in the form of a two-stage Stackelberg game. In Stage-1, the
regulator sets the lease duration to maximize spectrum utilization.
The payoff of an operator depends on the lease duration. In Stage-2,
the operators decide whether to join the market or not depending on
the decision which maximizes its payoff. To make decisions the regulator
needs information about the operators and the operators need information
about other operators in the market. The $k^{th}$ operator can be
completely characterized by six parameters which can be represented
using the tuple $\xi_{k}=\left(\mu_{k},\sigma_{k},a_{k},\rho_{k},\lambda_{k},\Lambda_{k}\right)$.
Only the $k^{th}$ operator knows the true value of these six parameters.
To model \textit{incomplete information games}, we assume the regulator
and other operators in the market only has a point estimate of the
$k^{th}$ operator's parameters \cite{flexauc}. Let the estimate
be $\widehat{\xi}_{k}=\left(\widehat{\mu}_{k},\widehat{\sigma}_{k},\widehat{a}_{k},\widehat{\rho}_{k},\widehat{\lambda}_{k},\widehat{\Lambda}_{k}\right)$.
Please note that for simplicity, we have assumed that the entire market
has one common estimate of $k^{th}$ operator's parameters. This can
be easily generalized where the regulator and the operators have different
estimates of $k^{th}$ operator's parameters.

Stackelberg equilibrium of a Stackelberg game can be found using backward
induction \cite{mas1995microeconomic}. To apply backward induction,
we start with Stage-2 and analyze the operators' decision strategy
given a lease duration. Then we solve for Stage-1 where the regulator
decides the lease duration knowing the possible response of the operators
to the lease duration. Consider Stage-2 of the game, also referred
as the subgame. The outcome of this process is the function $\mathcal{S}\left(T\right)$
which characterizes the set of interested operators as a function
of lease duration $T$. An operators decision to enter the market
depends on its revenue function $\mathcal{R}_{k}\left(\mathcal{S},T\right)$.
$\mathcal{R}_{k}\left(\mathcal{S},T\right)$ is the \textit{true}
revenue function of the $k^{th}$ operator. To compute $\mathcal{R}_{k}\left(\mathcal{S},T\right)$,
the $k^{th}$ operator needs to know $\xi_{j}$ of all the operators
in the market which it does not. Let $\widehat{\mathcal{R}}_{k}\left(\mathcal{S},T\right)$
denotes the revenue function of the $k^{th}$ operator as perceived
by the $k^{th}$ operator due to incomplete information scenario.
To compute $\widehat{\mathcal{R}}_{k}\left(\mathcal{S},T\right)$,
the $k^{th}$ operator uses its true parameters $\xi_{k}$ and estimates
$\widehat{\xi}_{j}$, where $j\neq k$, of other operator's parameters.
If $T>\Lambda_{k}$, then the $k^{th}$ operator will definitely not
enter the market. If $T\leq\Lambda_{k}$, then the payoff of the $k^{th}$
operator is\vspace{-0.75em}

\begin{equation}
\pi_{k}\left(\mathcal{X}\right)=\widehat{\mathcal{R}}_{k}\left(\mathcal{X},T\right)-\lambda_{k}\label{eq:2.3.1}
\end{equation}

\noindent if it enters the market where $\mathcal{X}$ is the set
of operators who decided to enter the market and $k\in\mathcal{X}$.
Payoff of the $k^{th}$ operator is $0$ if it does not enter the
market. With (\ref{eq:2.3.1}) as the payoff function, the subgame
can have \textit{multiple} pure strategy Nash equilibria. For example,
consider a complete information game which is a subset of incomplete
information game with $\widehat{\xi}_{k}=\xi_{k}\,;\,\forall k$.
There is $M=1$ channel and $N=2$ operators. Parameters
$\xi_{k}$ of these operators are $\mu_{k}=\mu\,,\,\lambda_{k}=\lambda\,,\,\Lambda_{k}=\infty\,;\,k=1,2$. If the lease duration $T=\frac{\lambda}{\mu}$, then   $\widehat{\mathcal{R}}_{k}\left(\left\{ k\right\} ,T\right)=\mu T=\lambda\,;\,k=1,2$ and $\widehat{\mathcal{R}}_{k}\left(\left\{ 1,2\right\} ,T\right)<\mu T=\lambda\,;\,k=1,2$ (due to Property~\ref{RevenueFuncProp2}). Hence, there are two pure
strategy Nash equilibria; Operator $1$ enters the market while Operator
$2$ does not and vice-versa.

If the subgame has multiple pure strategy Nash equilibria, then Stage-1
will have multiple optimal solutions for lease duration each corresponding
to one of the NE of Stage-1. In other words, for a given set of market
parameters, there can be multiple optimal lease durations. In order
to simplify the analysis, we consider a setting where the subgame
has a unique pure strategy NE. In one such setting, an operator is
interested in maximizing its minimum payoff. The obtained NE is called
\textit{Max-Min NE} and has been considered in previous works like
\cite{maxmin1}. According to Property~\ref{RevenueFuncProp2}, payoff
decreases as the set $\mathcal{X}$ increases. So the minimum payoff
corresponds to the largest $\mathcal{X}$. The largest $\mathcal{X}$
is composed of those operators who \textit{may} join the market. The
$k^{th}$ operator will definitely not join the market if either $T>\Lambda_{k}$
or $\mu_{k}T<\lambda_{k}$. The inequality $T>\Lambda_{k}$
is obvious. To appreciate the inequality $\mu_{k}T<\lambda_{k}$,
note that the maximum payoff of the $k^{th}$ operator is $\mu_{k}T-\lambda_{k}$
which happens when it is alone in the market, i.e. $\mathcal{X}=\left\{ k\right\} $
and hence $\widehat{\mathcal{R}}_{k}\left(\left\{ k\right\} ,T\right)=\mu_{k}T$
. This is due to Property~\ref{RevenueFuncProp2}. If $\mu_{k}T<\lambda_{k}$,
then the payoff of the $k^{th}$ operator is negative if it enters
the market, irrespective of the decision of other operators. Hence,
its \textit{dominant strategy} is not to enter the market. To conclude,
the $k^{th}$ operator may join the market if and only if $T\leq\Lambda_{k}$
and $\mu_{k}T\geq\lambda_{k}$. But the $k^{th}$ operator does not
know the true value of $\mu_{j}$, $\lambda_{j}$ and $\Lambda_{j}$
if $j\neq k$. Therefore, the largest set of interested operators
who \textit{may} join the market, according to the $k^{th}$ operator,
for lease duration $T$ is\vspace{-1.2em}

\begin{eqnarray}
\mathcal{S}_{k}^{L}\left(T\right) & = & \left\{ \vphantom{\widehat{\Lambda}_{j}}j\::\:j=k\,,\,T\leq\Lambda_{j}\,,\,\mu_{j}T\geq\lambda_{j}\quad\text{{or}}\right.\nonumber \\
 &  & \left.\qquad\:j\neq k\,,\,T\leq\widehat{\Lambda}_{j}\,,\,\widehat{\mu}_{j}T\geq\widehat{\lambda}_{j}\right\} \label{eq:2.3.2}
\end{eqnarray}

To this end we can conclude that according to the $k^{th}$ operator,
its minimum payoff is $\widehat{\mathcal{R}}_{k}\left(\mathcal{S}_{k}^{L}\left(T\right),T\right)-\lambda_{k}$.
This leads to the following proposition.
\begin{prop}
\label{prop:UniqueSubGameNE}The subgame has a unique Max-Min Nash
equilibrium which is given by the set of interested operators\vspace{-1.0em}

\begin{equation}
\mathcal{S}\left(T\right)=\left\{ k\::\:T\leq\Lambda_{k}\,,\,\widehat{\mathcal{R}}_{k}\left(\mathcal{S}_{k}^{L}\left(T\right),T\right)\geq\lambda_{k}\right\} \label{eq:2.3.3}
\end{equation}
\end{prop}
\noindent Equation \ref{eq:2.3.3} is the solution (true) of the subgame.
With slight abuse of notation let\vspace{-0.5em}
\begin{equation}
U\left(T\right)=U\left(\mathcal{S}\left(T\right),T\right)\label{eq:2.3.4}
\end{equation}
$U\left(T\right)$ is the \textit{true} objective function because
it depends on the true solution of the subgame, $\mathcal{S}\left(T\right)$,
and also because computation of $U\left(\mathcal{S},T\right)$ is
based only on the true parameters $\xi_{j}$. In Stage-1, the regulator
wants to maximize $U\left(T\right)$. Hence, to calculate $U\left(T\right)$,
the regulator needs to know $\xi_{j}$ of all the operators. But the
regulator does not know $\xi_{j}$ of \textit{any} operator; it only
knows the estimates $\widehat{\xi}_{j}$. Let $\widetilde{\mathcal{R}}_{k}\left(\mathcal{S},T\right)$
and $\widetilde{U}\left(\mathcal{S},T\right)=\frac{1}{T}\underset{k\in S}{\sum}\widetilde{\mathcal{R}}_{k}\left(S,T\right)$
denote the revenue function of the $k^{th}$ operator and the objective
function as perceived by the regulator respectively. Unlike $\widehat{\mathcal{R}}_{k}\left(\mathcal{S},T\right)$,
$\widetilde{\mathcal{R}}_{k}\left(\mathcal{S},T\right)$ is calculated
\textit{only} based on estimates $\widehat{\xi}_{j}$. The largest
set of interested operators as perceived by the regulator, $\widetilde{\mathcal{S}}^{L}\left(T\right)$,
and the perceived set of interested operators, $\widetilde{S}\left(T\right)$,
is given by\vspace{-0.5em}

\begin{equation}
\widetilde{\mathcal{S}}^{L}\left(T\right)=\left\{ k\,:\,T\leq\widehat{\Lambda}_{k}\,,\,\widehat{\mu}_{k}T\geq\widehat{\lambda}_{k}\right\} \label{eq:2.3.5}
\end{equation}

\vspace{-1.0em}

\begin{equation}
\widetilde{S}\left(T\right)=\left\{ k\,:\,T\leq\widehat{\Lambda}_{k}\,,\,\widetilde{\mathcal{R}}_{k}\left(\widetilde{\mathcal{S}}^{L}\left(T\right),T\right)\geq\widehat{\lambda}_{k}\right\} \label{eq:2.3.6}
\end{equation}

Note that $\widetilde{\mathcal{S}}\left(T\right)\subseteq\widetilde{\mathcal{S}}^{L}\left(T\right)$
because $\widetilde{\mathcal{R}}_{k}\left(\widetilde{\mathcal{S}}^{L}\left(T\right),T\right)\leq\widetilde{\mathcal{R}}_{k}\left(\left\{ k\right\} ,T\right)=\widehat{\mu}_{k}T$
(Property~\ref{RevenueFuncProp2}). With slight abuse of notation
let $\widetilde{U}\left(T\right)=\widetilde{U}\left(\widetilde{\mathcal{S}}\left(T\right),T\right)$
be the perceived objective function. In Stage-1, the regulator solves
the following optimization problem\vspace{-1.0em}

\[
OP1\begin{cases}
\underset{T\in\mathbb{Z}^{+}}{\max}\quad\widetilde{U}\left(T\right)=\widetilde{U}\left(\widetilde{\mathcal{S}}\left(T\right),T\right)\end{cases}
\]

\noindent 
\begin{figure}[t]
\begin{centering}
\includegraphics[scale=0.73]{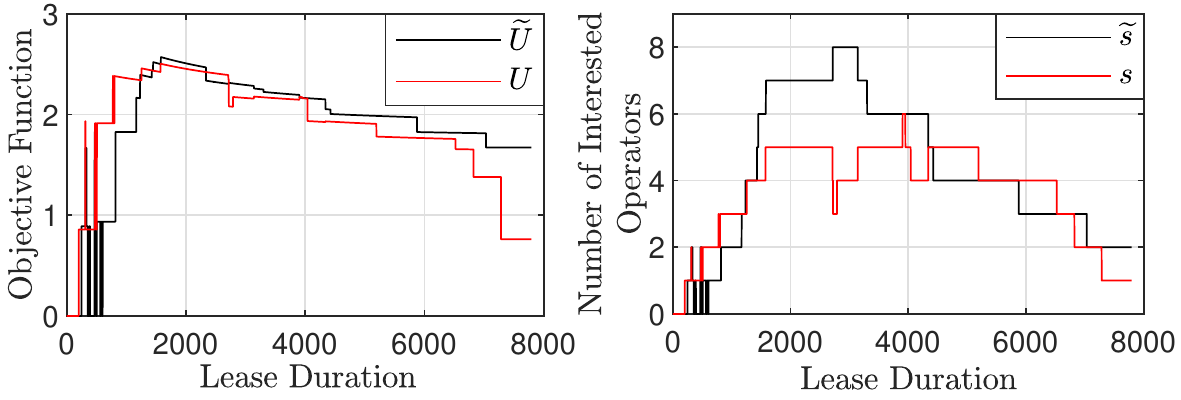}
\par\end{centering}
\caption{A typical plot of objective function (true) $U\left(T\right)$, perceived
objective function $\widetilde{U}\left(T\right)$, number of interested
operators (true) $s\left(T\right)=\left|\mathcal{S}\left(T\right)\right|$,
and the perceived number of interested operators $\widetilde{s}\left(T\right)=\left|\widetilde{\mathcal{S}}\left(T\right)\right|$.
For this plot, $N=10$ , $M=2$ and the true parameters $\xi_{k}$
are chosen uniformly at random. The estimated parameters $\widehat{\xi}_{k}$
are chosen uniformly at random such that they are within $\pm25\%$
of the true parameters.\vspace{-1.0em}}

\label{sim_typical_obj_func}
\end{figure}

\vspace{-1.0em}

The regulator chooses lease duration $T$ to maximize the perceived
objective function $\widetilde{U}\left(T\right)$. The perceived objective
function and the objective function (true) $U\left(T\right)$ may
not be equal as shown in Figure~\ref{sim_typical_obj_func}. Therefore,
a lease duration which maximizes the perceived objective function
may not maximize the objective function. This leads to sub-optimal
spectrum utilization due to incomplete information games.

The perceived set of interested operators, $\widetilde{\mathcal{S}}\left(T\right)$,
can be \textit{implicitly} controlled by choosing a suitable $T$.
A significant part of solving $OP1$ is to find an $\widetilde{\mathcal{S}}\left(T\right)$
that maximizes $\widetilde{U}\left(T\right)$. The number of combinations
of $\widetilde{\mathcal{S}}\left(T\right)$ can be exponential in
$N$. Therefore, $OP1$ is reminiscent of combinatorial optimization
which makes it difficult to solve even though it is a scalar optimization
problem in $T$. A typical plot of objective function $\widetilde{U}\left(T\right)$
and the perceived number of interested operators $\widetilde{s}\left(T\right)=\left|\widetilde{\mathcal{S}}\left(T\right)\right|$
is shown in Figure~\ref{sim_typical_obj_func} (black curve). The
\textit{discontinuous} and \textit{non-smooth} nature of $\widetilde{U}\left(T\right)$
is another reason why it is difficult to solve $OP1$.

Figure~\ref{sim_typical_obj_func} also shows that the optimal lease
duration is\textit{ non-trivial}. If the lease duration is too low,
MER of many operators are not satisfied (Property~\ref{RevenueFuncProp1})
and hence $\widetilde{\mathcal{S}}\left(T\right)$ is small. Therefore,
$\widetilde{U}\left(T\right)$ is low (high) according to Property~\ref{ObjFuncProp2}
if bid correlation if high (low). If the lease duration is too high
and bid correlation is high (low), $\widetilde{U}\left(T\right)$
is low (high) for one of the two reasons. \textit{First}, due to Property~\ref{ObjFuncProp1}.
This is because the objective function is monotonic decreasing (increasing)
in lease duration if bid correlation is high (low). \textit{Second},
$\widetilde{\mathcal{S}}\left(T\right)$ is small either because the
operators cannot afford a channel with long lease duration or because the revenue function decreases for a higher value of lease duration as suggested by Property~\ref{RevenueFuncProp1}.
Hence, the optimal lease duration is neither too high nor too low
as shown in Figure~\ref{sim_typical_obj_func}.\vspace{-1.0em}

\subsection{Stackelberg game solution: Homogeneous Market with Complete Information\label{subsec:Homogeneous}}

For complete information games, $\widehat{\xi}_{k}=\xi_{k}\,;\,\forall k$,
and for homogeneous market, $\xi_{k}=\xi\,;\,\forall k$. Let, $\xi=\left(\mu,\sigma,a,\rho,\lambda,\Lambda\right)$.
Since $\widehat{\xi}_{k}=\xi_{k}$, the revenue function of the $k^{th}$
operator as perceived by the $k^{th}$ operator, $\widehat{\mathcal{R}}_{k}\left(\mathcal{S},T\right)$,
and as perceived by the regulator ,$\widetilde{\mathcal{R}}_{k}\left(\mathcal{S},T\right)$,
are equal to the revenue function (true) $\mathcal{R}_{k}\left(\mathcal{S},T\right)$.
Similarly, the perceived objective function, $\widetilde{U}\left(T\right)$,
is equal to the objective function (true), $U\left(T\right)$. Also,
since the market is homogeneous in $\xi_{k}$, the revenue is same
for all the operators, i.e. $\mathcal{R}_{k}\left(\mathcal{S},T\right)=\mathcal{R}\left(s,T\right)\,;\,\forall k$.
The following proposition can be used to calculate the optimal lease
duration, $T^{*}$, and the optimal value of the objective function,
$U^{*}$, for a homogeneous market with complete information.
\begin{prop}
\label{prop:optimal-homogeneous}Let $\theta$ be the solution to
$\mathcal{R}\left(N,\theta\right)=\lambda$ and $\left\lceil \cdot\right\rceil $
be the ceiling function. If $\left\lceil \theta\right\rceil \leq\Lambda$,
then $T^{*}=\left\lceil \theta\right\rceil $ and $U^{*}=\frac{N}{\left\lceil \theta\right\rceil }\mathcal{R}\left(N,\left\lceil \theta\right\rceil \right)$.
However, if $\left\lceil \theta\right\rceil >\Lambda$, then $U^{*}=0$
and $T^{*}$ can be set to any value.
\end{prop}
\begin{IEEEproof}
Please refer to Appendix~E for the proof.
\end{IEEEproof}
Intuitively, Proposition \ref{prop:optimal-homogeneous} can be understood
as follows. In a homogeneous market, either all or none of the operators
are interested in joining the market. If none of the operators are
interested in joining the market, then the objective function is zero
which is trivial. Hence, to have $U^{*}>0$, $T^{*}$ should be such
that all the operators are interested in joining the market. If all
the $N$ operators join the market, then the revenue function of an
operator is $\mathcal{R}\left(N,T^{*}\right)$. Also, for operators
to be interested in joining the market, $T^{*}$ must also satisfy
$\mathcal{R}\left(N,T^{*}\right)\geq\lambda$ and $T^{*}\leq\Lambda$
(refer to (\ref{eq:2.3.6})). As discussed in Section~\ref{subsec:Characteristics},
the revenue function of a homogeneous market is monotonic increasing
in lease duration. Hence, the solution to $\mathcal{R}\left(N,T^{*}\right)\geq\lambda$
is $T^{*}\geq\left\lceil \theta\right\rceil $ where $\theta$ if
the solution to $\mathcal{R}\left(N,\theta\right)=\lambda$. The ceiling
function $\left\lceil \cdot\right\rceil $ is needed because we consider
a time slotted model. To this end we conclude that, $U^{*}>0$ if
and only if $T^{*}$ satisfies $\left\lceil \theta\right\rceil \leq T^{*}\leq\Lambda$.
If $\left\lceil \theta\right\rceil >\Lambda$, then there does not
exists a $T^{*}$ that satisfies $\left\lceil \theta\right\rceil \leq T^{*}\leq\Lambda$.
Hence, $U^{*}=0$ and $T^{*}$ can be set to any value. If $\left\lceil \theta\right\rceil \leq\Lambda$,
then $T^{*}=\left\lceil \theta\right\rceil $ maximizes the objective
function because for a homogeneous market, objective function is monotonic
decreasing in lease duration (refer to Section~\ref{subsec:Characteristics}).
For $T^{*}=\left\lceil \theta\right\rceil $, all the operators are
interested in joining the market. Hence, according to (\ref{eq:2.p.3}),
$U^{*}=\frac{N}{\left\lceil \theta\right\rceil }\mathcal{R}\left(N,\left\lceil \theta\right\rceil \right)$.
This completes the explanation of Proposition \ref{prop:optimal-homogeneous}.

Finally, to solve $OP1$ for a homogeneous market with complete information,
we have to compute $\theta$. Since, $\mathcal{R}\left(N,\theta\right)$
is monotonic increasing in $\theta$, the equation $\mathcal{R}\left(N,\theta\right)=\lambda$
can be solved using \textit{binary search} or \textit{Newton-Raphson}
method.

\subsection{Stackelberg game solution: Heterogeneous Market with Incomplete Information\label{subsec:Heterogeneous}}
\noindent \SetInd{0.1em}{0.75em}
\SetAlgoHangIndent{0.0em}
\begin{algorithm}[t]

	\DontPrintSemicolon
	 
	\SetKwInput{Input}{Input}
	\SetKwInput{Output}{Output}

	\Input{$N$, $M$, $\widehat{\xi}_{k}\,;\,k=1,\ldots,N$}

	\Output{$\widetilde{T}^{*}$, $\widetilde{U}^{*}$, $\widetilde{\mathcal{S}}^{*}$}

	\justfy Initialize an empty list $Q^{L}$ whose elements are ordered pairs $\left(T,k\right)$ where $T$ denotes lease duration and $k$ denotes operator index\;

	\justfy \For{$k\leftarrow 1$ \KwTo $N$}
	{
		\justfy Append $\left(\left\lceil \frac{\widehat{\lambda}_{k}}{\widehat{\mu}_{k}}\right\rceil,k\right)$ and $\left(\widehat{\Lambda}_{k}+1,k\right)$ onto $Q^{L}$\;
	}

	\justfy Sort $Q^{L}$ in ascending order of lease duration\;
	
	\justfy Set $\widetilde{U}^{*}=0$ and $\mathcal{X}^{L}_{0}=\emptyset$\;

	\justfy \For{$i\leftarrow 1$ \KwTo $\left|Q^{L}\right|$}
	{

		\justfy \textbf{if} $Q^{L}_{i}.k \in \mathcal{X}^{L}_{i-1}$ \textbf{then} set $\mathcal{X}^{L}_{i}=\mathcal{X}^{L}_{i-1}-\left\{ Q^{L}_{i}.k\right\} $; \textbf{else} set $\mathcal{X}^{L}_{i}=\mathcal{X}^{L}_{i-1}\bigcup\left\{ Q^{L}_{i}.k\right\} $\;

		\justfy \If{$Q^{L}_{i}.T < Q^{L}_{i+1}.T$ \textbf{or} $i=\left|Q^{L}\right|$}
		{
	
			\justfy $\mathcal{X}_{i}^{L}$ is one of the sets in $\mathcal{F}^{L}$. Set $\theta_{i}^{L}=Q^{L}_{i}.T$ and $\Theta_{i}^{L}=Q^{L}_{i+1}.T-1$\; 

			\justfy Initialize an empty list $Q$ whose elements are ordered pairs $\left(T,k\right)$ where $T$ denotes lease duration and $k$ denotes operator index\;
	
			\justfy \For{$k$ \textbf{in} $\mathcal{X}^{L}_{i}$}
			{
				\justfy \label{bounds_line} Find $\gamma_{k}$ and $\Gamma_{k}$, the minimum and the maximum lease duration resp. in the interval $\left[\theta_{i}^{L},\Theta_{i}^{L}\right]$ s.t. $\widetilde{\mathcal{R}}_{k}\left(\mathcal{X}_{i}^{L},T\right)\geq \widehat{\lambda}_{k}\,;\,\forall T\in\left[\gamma_{k},\Gamma_{k}\right]$\;

				\justfy \If{$\gamma_{k}$ and $\Gamma_{k}$ \textit{exists}}
				{
					\justfy Append $\left(\gamma_{k},k\right)$ onto $Q$

					\justfy \textbf{if} $\Gamma_{k}<\Theta_{i}^{L}$ \textbf{then} append $\left(\Gamma_{k}+1,k\right)$ onto $Q$
				}
			}

			\justfy Sort $Q$ in ascending order of lease duration\;

			\justfy Set $\mathcal{X}_{0}=\emptyset$\;

			\justfy \For{$j\leftarrow 1$ \KwTo $\left|Q\right|$}
			{

				\justfy \textbf{if} $Q_{j}.k \in \mathcal{X}_{j-1}$ \textbf{then} set $\mathcal{X}_{j}=\mathcal{X}_{j-1}-\left\{ Q_{j}.k\right\} $; \textbf{else} set $\mathcal{X}_{j}=\mathcal{X}_{j-1}\bigcup\left\{ Q_{j}.k\right\}$\;

				\justfy \If{($Q_{j}.T < Q_{j+1}.T$ \textbf{or} $j=\left|Q\right|$)}
				{
	
					\justfy $\mathcal{X}_{j}$ is one of the sets in $\mathcal{F}$. Set $\theta_{j}=Q_{j}.T$ and $\Theta_{j}=Q_{j+1}.T-1$\;

					\justfy \If{$\widetilde{U}\left(\mathcal{X}_{j}\,,\,\theta_{j}\right)>\widetilde{U}^{*}$}
					{
						\justfy Set $\widetilde{T}^{*}=\theta_{j}$ , $\widetilde{U}^{*}=\widetilde{U}\left(\mathcal{X}_{j}\,,\,\theta_{j}\right)$ , $\widetilde{\mathcal{S}}^{*}=\mathcal{X}_{j}$\;
					}

					\justfy \If{$\widetilde{U}\left(\mathcal{X}_{j}\,,\,\Theta_{j}\right)>\widetilde{U}^{*}$}
					{
						\justfy Set $\widetilde{T}^{*}=\Theta_{j}$ , $\widetilde{U}^{*}=\widetilde{U}\left(\mathcal{X}_{j}\,,\,\Theta_{j}\right)$ , $\widetilde{\mathcal{S}}^{*}=\mathcal{X}_{j}$\;
					}
				}
			}
		}
	}
	\caption{Optimization algorithm to solve $OP1$ for a heterogeneous market with incomplete information}
	\label{opalgo_heterogeneous}
\end{algorithm}

\begin{figure}[t]
\begin{centering}
\includegraphics[scale=0.48]{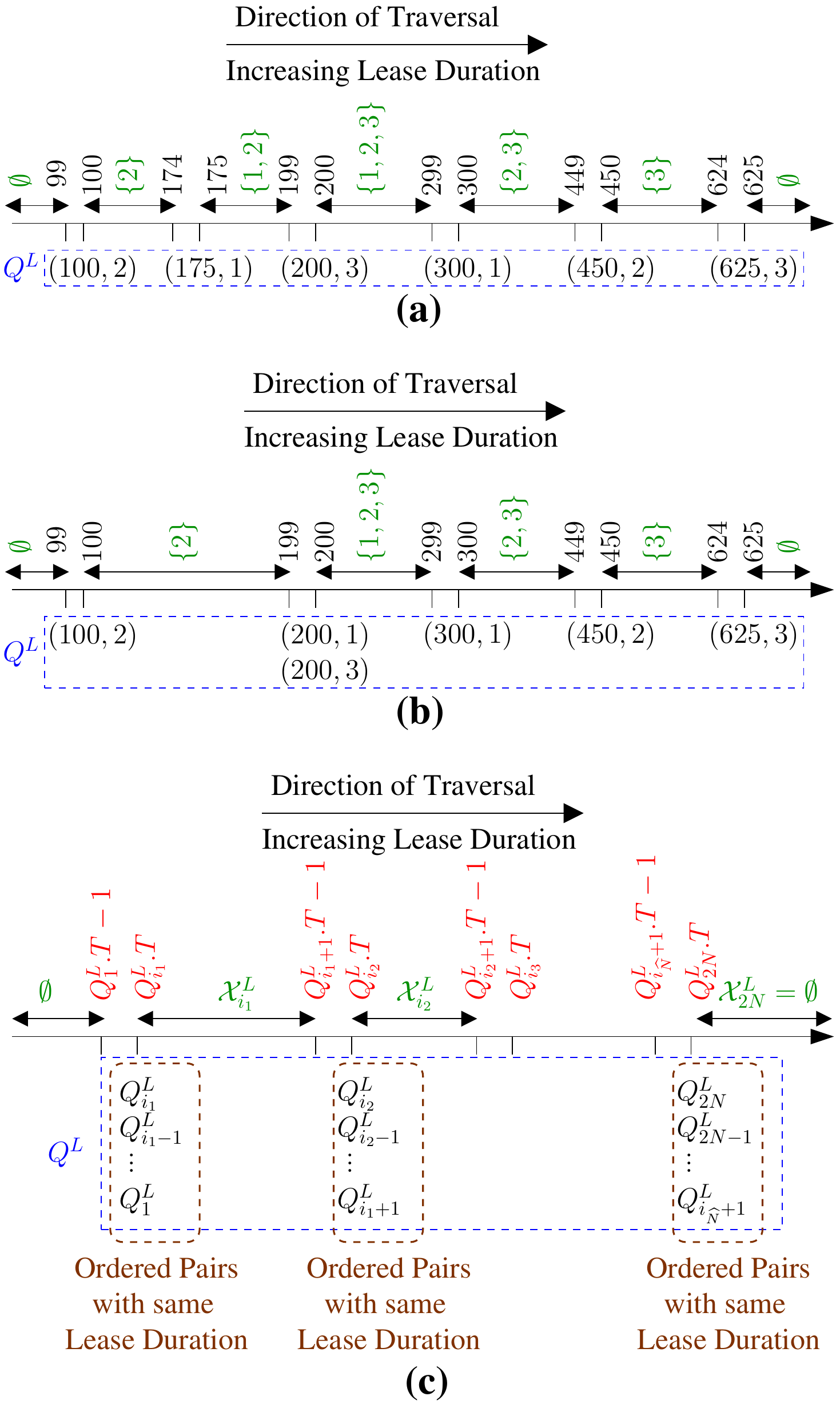}
\par\end{centering}
\caption{(a) Figure showing sorted list $Q^{L}$ (in blue) and the sets in
$\mathcal{F}^{L}$ (in green) for Example \ref{exa:Example1}. (b)
Figure showing sorted list $Q^{L}$ (in blue) and the sets in $\mathcal{F}^{L}$
(in green) for Example \ref{exa:Example2}. (c) A \textit{generic}
pictorial representation showing the sorted list $Q^{L}$ (in blue),
the sets in $\mathcal{F}^{L}$ (in green) and the interval of lease
duration corresponding to the sets in $\mathcal{F}^{L}$ (in red).}

\label{example1_fig}\vspace{-1.0em}
\end{figure}

\vspace{-1.5em}Algorithm~\ref{opalgo_heterogeneous} is a psuedocode
to solve $OP1$ for a heterogeneous market with incomplete information. Proposition \ref{prop:optimal-homogeneous} which solves $OP1$ for a homogeneous market with complete information is a special case of Algorithm~\ref{opalgo_heterogeneous}. The main difficulty about solving $OP1$ is the change in $\widetilde{\mathcal{S}}\left(T\right)$
with change in $T$. This leads to discontinuities in the objective
function of $OP1$. Algorithm~\ref{opalgo_heterogeneous} solves
this issue by dividing the entire positive real axis which represents
the lease duration into intervals such that $\widetilde{\mathcal{S}}\left(T\right)$
does not change within these intervals. The optimal lease duration
within these intervals will lie in its boundaries because of Property~\ref{ObjFuncProp1}.
Finally, the optimal lease duration can be found by comparing the
maximum lease duration within each of these intervals. In the rest
of this section, we will device an efficient approach to find these
intervals. We will approach this in steps. \textit{First}, we will
convert $OP1$ into a combinatorial optimization problem $OP2$. By
doing so, we formalize the idea discussed in this paragraph.\textit{
Second}, we will discuss how to divide the entire positive real axis
into intervals such that $\widetilde{\mathcal{S}}^{L}\left(T\right)$
does not change within these intervals. This is required because $\widetilde{\mathcal{S}}\left(T\right)$
is a function of $\widetilde{\mathcal{S}}^{L}\left(T\right)$ (refer
to (\ref{eq:2.3.6})). Therefore, to find the intervals corresponding
to $\widetilde{\mathcal{S}}\left(T\right)$, we have to first find
the intervals corresponding to $\widetilde{\mathcal{S}}^{L}\left(T\right)$.
The process of finding the intervals corresponding to $\widetilde{\mathcal{S}}^{L}\left(T\right)$
will be exemplified using Example 1, Example 2 and Figure \ref{example1_fig}.
Finally, we discuss how to find the intervals corresponding to $\widetilde{\mathcal{S}}\left(T\right)$
which is very similar to finding the intervals corresponding to $\widetilde{\mathcal{S}}^{L}\left(T\right)$.

Let $\mathcal{F}$ be a family of sets containing all possible perceived
sets of interested operators. Mathematically, $\mathcal{F}=\left\{ \overline{\mathcal{S}}\::\:\left(\exists\,T\in\mathbb{Z}^{+}\right)\left[\widetilde{\mathcal{S}}\left(T\right)=\overline{\mathcal{S}}\right]\right\} $.
For a given $\overline{\mathcal{S}}\in\mathcal{F}$, there can be
several values of $T$ satisfying $\widetilde{\mathcal{S}}\left(T\right)=\overline{\mathcal{S}}$.
Let $T_{m}\left(\overline{\mathcal{S}}\right)$ and $T_{M}\left(\overline{\mathcal{S}}\right)$
denote the minimum and the maximum $T$ respectively satisfying $\widetilde{\mathcal{S}}\left(T\right)=\overline{\mathcal{S}}$.
According to Property~\ref{ObjFuncProp1}, either $T_{m}\left(\overline{\mathcal{S}}\right)$
or $T_{M}\left(\overline{\mathcal{S}}\right)$ maximizes $OP1$ if
$\widetilde{\mathcal{S}}\left(T\right)=\overline{\mathcal{S}}$. Based
on this discussion, $OP1$ is equivalent to the following optimization
problem\vspace{-1.2em}

\[
OP2\begin{cases}
\underset{\overline{\mathcal{S}}\in\mathcal{F}}{\max}\;\widetilde{U}\left(\overline{\mathcal{S}}\right)=\max\left(\widetilde{U}\left(\overline{\mathcal{S}},T_{m}\left(\overline{\mathcal{S}}\right)\right),\widetilde{U}\left(\overline{\mathcal{S}},T_{M}\left(\overline{\mathcal{S}}\right)\right)\right)\end{cases}
\]

$OP2$ is a combinatorial optimization problem in $\overline{\mathcal{S}}$.
Let $\widetilde{T}^{*}$ be the optimal solution of $OP1$. If $\widetilde{\mathcal{S}}^{*}$
is an optimal solution of $OP2$, then $\widetilde{T}^{*}$ is either
$T_{m}\left(\widetilde{\mathcal{S}}^{*}\right)$ or $T_{M}\left(\widetilde{\mathcal{S}}^{*}\right)$,
whichever maximizes $\widetilde{U}\left(\widetilde{\mathcal{S}}^{*},T\right)$.
In Algorithm~\ref{opalgo_heterogeneous}, we find $\widetilde{\mathcal{S}}^{*}$
(and hence $\widetilde{T}^{*}$) by iterating over all $\overline{\mathcal{S}}\in\mathcal{F}$
to find the one which maximizes $\widetilde{U}\left(\overline{\mathcal{S}}\right)$.
To do so, we need a constructive method to find all the sets in $\mathcal{F}$.
In the rest of the section, we discuss the steps involved in finding
all the sets in $\mathcal{F}$ and the corresponding line number of
Algorithm~\ref{opalgo_heterogeneous} which implements that step.
One of the outputs of Algorithm~\ref{opalgo_heterogeneous} is $\widetilde{U}^{*}$,
the optimal value of the perceived objective function. But the value
of the objective function (true) corresponding to optimal solution
of $OP1$, $\widetilde{T}^{*}$, is given by (\ref{eq:2.3.4}) and
is equal to\vspace{-1.0em}

\begin{equation}
U^{*}=U\left(\mathcal{S}\left(\widetilde{T}^{*}\right),\widetilde{T}^{*}\right)\label{eq:3.2.1}
\end{equation}

In order to find all the sets in $\mathcal{F}$, we have to first
find all the sets in $\mathcal{F}^{L}=\left\{ \overline{\mathcal{S}}\::\:\left(\exists\,T\in\mathbb{Z}^{+}\right)\left[\widetilde{\mathcal{S}}^{L}\left(T\right)=\overline{\mathcal{S}}\right]\right\} $.
$\mathcal{F}^{L}$ is a family of sets containing all possible \textit{largest}
sets of interested operators as perceived by the regulator. According
to (\ref{eq:2.3.5}), $k\in\widetilde{\mathcal{S}}^{L}\left(T\right)$
\textit{if and only if} $T\geq\left\lceil \frac{\widehat{\lambda}_{k}}{\widehat{\mu}_{k}}\right\rceil $
and $T<\widehat{\Lambda}_{k}+1$. The ceiling function $\left\lceil \cdot\right\rceil $
is needed because we consider a time slotted model. Consider the ordered
pairs $\left(\left\lceil \frac{\widehat{\lambda}_{k}}{\widehat{\mu}_{k}}\right\rceil ,k\right)$
and $\left(\widehat{\Lambda}_{k}+1,k\right)$ where the first element
is lease duration and the second element is operator index. A list
$Q^{L}$ contains such ordered pairs corresponding to all the $N$
operators (\textit{line 1-3}). The size of $Q^{L}$ is $\left|Q^{L}\right|=2N$
as there are $2$ ordered pairs corresponding to each of the $N$
operators. Let $Q_{i}^{L}$ be the $i^{th}$ element of $Q^{L}$.
We use the dot ($\cdot$) operator to access the lease duration and
the operator index of the elements of $Q^{L}$. In other words, $Q_{i}^{L}.T$
and $Q_{i}^{L}.k$ denote the lease duration and the operator index
respectively corresponding to ordered pair $Q_{i}^{L}$. All the sets
in $\mathcal{F}^{L}$ can be found using the following steps:
\begin{enumerate}
\item[(A1)] Sort $Q^{L}$ in \textit{ascending order} of lease duration (\textit{line
4}). Traverse the sorted list $Q^{L}$ from $i=1$ to $\left|Q^{L}\right|$
and repeat steps (A2) to (A4) in every iteration (\textit{line 6}).
Let $\mathcal{X}_{i}^{L}$ be the largest set of interested operators
as perceived by the regulator which is obtained in the $i^{th}$ iteration.
Set $\mathcal{X}_{0}^{L}=\emptyset$ (\textit{line 5}). Let $i$ be
the current iteration.
\item[(A2)] If the operator with index $Q_{i}^{L}.k$ is \textit{not} in set
$\mathcal{X}_{i-1}^{L}$, \textit{add} $Q_{i}^{L}.k$ to $\mathcal{X}_{i-1}^{L}$
to get $\mathcal{X}_{i}^{L}$. Else if the operator with index $Q_{i}^{L}.k$
is in set $\mathcal{X}_{i-1}^{L}$, \textit{remove} $Q_{i}^{L}.k$
from $\mathcal{X}_{i-1}^{L}$ to get $\mathcal{X}_{i}^{L}$. This
is implemented in \textit{line 7}.
\item[(A3)] If $Q_{i}^{L}.T<Q_{i+1}^{L}.T$ \textit{or} $i=\left|Q^{L}\right|$,
then the obtained $\mathcal{X}_{i}^{L}$ in step (A2) is one of the
sets in $\mathcal{F}^{L}$. $\widetilde{\mathcal{S}}^{L}\left(T\right)=\mathcal{X}_{i}^{L}$
for all $\mbox{\ensuremath{T\in\left[Q_{i}^{L}.T\,,\,Q_{i+1}^{L}.T-1\right]}}$.
\item[(A4)] If $Q_{i}^{L}.T=Q_{i+1}^{L}.T$, then the obtained $\mathcal{X}_{i}^{L}$
in step (A2) is \textit{not} one of the sets in $\mathcal{F}^{L}$.
This is because operators $Q_{i}^{L}.k$ and $Q_{i+1}^{L}.k$ corresponding
to ordered pairs $Q_{i}^{L}$ and $Q_{i+1}^{L}$ respectively must
update $\mathcal{X}_{i-1}^{L}$ simultaneously as both these ordered
pairs have the same lease duration.
\end{enumerate}

The \textit{if statement} in \textit{line 8} implements steps (A3)
and (A4). If $Q_{i}^{L}.T=Q_{i+1}^{L}.T$, then the if statement in
line 8 is\textit{ false} and the algorithm simply loops to the next
iteration without considering the obtained $\mathcal{X}_{i}^{L}$
in line 7 as one of the sets in $\mathcal{F}^{L}$. The following
examples exemplifies the working of steps (A1) to (A4).

\noindent \begin{example}
\label{exa:Example1} Consider $N=3$ operators with $\widehat{\mu}_{k}=1\,;\,\forall k$, $\widehat{\lambda}_{k}$'s are $\left[175,100,200\right]$ and $\widehat{\Lambda}_{k}$'s are $\left[300,450,625\right]$. The sorted list $Q^{L}$ for this example is shown in Figure~\ref{example1_fig}.a. As we traverse Figure~\ref{example1_fig}.a from left to right, $\widetilde{\mathcal{S}}^{L}\left(T\right)$ is $\emptyset$ if $T \leq 99$, $\left\{ 2\right\} $ if $T\in\left[100,174\right]$, $\left\{ 1,2\right\} $ if $T\in\left[175,199\right]$, $\left\{ 1,2,3\right\} $ if $T\in\left[200,299\right]$, $\left\{ 2,3\right\} $ if $T\in\left[300,449\right]$, $\left\{ 3\right\} $ if $T\in\left[450,624\right]$ and $\emptyset$ if $T\geq625$. Hence, $\mathcal{F}^{L}$ consists of the sets $\left\{ 2\right\} $, $\left\{ 1,2\right\} $, $\left\{ 1,2,3\right\} $, $\left\{ 2,3\right\} $, $\left\{ 3\right\} $, and $\emptyset$. \vspace{-1.0em}
\end{example}

\noindent \begin{example}
\label{exa:Example2} This example demonstrates the importance of step (A4) by considering mutiple ordered pairs with same lease duration. The setting is the same as Example \ref{exa:Example1} except that $\widehat{\lambda}_{k}$'s are $\left[200,100,200\right]$. The sorted list $Q^{L}$ for this example is shown in Figure~\ref{example1_fig}.b. As we traverse Figure~\ref{example1_fig}.b from left to right, $\widetilde{\mathcal{S}}^{L}\left(T\right)$ is $\emptyset$ if $T \leq 99$,  $\left\{ 2\right\} $ if $T\in\left[100,199\right]$, $\left\{ 1,2,3\right\} $ if  $T\in\left[200,299\right]$, $\left\{ 2,3\right\} $ if  $T\in\left[300,449\right]$, $\left\{ 3\right\} $ if  $T\in\left[450,624\right]$ and $\emptyset$ if $T\geq625$. Hence, $\mathcal{F}^{L}$ consists of the sets $\left\{ 2\right\} $, $\left\{ 1,2,3\right\} $, $\left\{ 2,3\right\} $, $\left\{ 3\right\} $, and $\emptyset$. Unlike Example \ref{exa:Example1}, $\left\{ 1,2\right\} \notin\mathcal{F}^{L}$ since ordered pairs $\left(200,1\right)$ and $\left(200,3\right)$ have the same lease duration.
\end{example}

\noindent 

\vspace{-1.5em}

Figures~\ref{example1_fig}.a and \ref{example1_fig}.b show that
steps (A1) to (A4) divides the set of positive integers into contiguous
intervals of lease duration. Each interval has its corresponding $\mathcal{X}_{i}^{L}$.
A general setup is shown in Figure~\ref{example1_fig}.c. Let $i_{1}$,
$i_{2}$, $\ldots$ , $2N$, where $\mbox{\ensuremath{i_{1}<i_{2}<\cdots<2N}}$,
denote all the iterations such that $Q_{i}^{L}.T<Q_{i+1}^{L}.T$ (or
$i=\left|Q^{L}\right|=2N$). Refering to step (A3), $\mathcal{F}^{L}$
consists of the sets $\mathcal{X}_{i_{1}}^{L}$, $\mathcal{X}_{i_{2}}^{L}$,
$\ldots$ , $\mathcal{X}_{2N}^{L}$. Each of these sets are associated
with a corresponding interval of lease duration. As shown in Figure~\ref{example1_fig}.c,
$\widetilde{\mathcal{S}}^{L}\left(T\right)$ is equal to $\mathcal{X}_{i_{1}}^{L}$
in the interval $\left[Q_{i_{1}}^{L}.T\,,\,Q_{i_{1}+1}^{L}.T-1\right]$,
$\mathcal{X}_{i_{2}}^{L}$ in the interval $\left[Q_{i_{2}}^{L}.T\,,\,Q_{i_{2}+1}^{L}.T-1\right]$
etc. These intervals are non-overlapping and their union spans the
entire set of positive integers. We want to design an algorithm to
find all the sets in $\mathcal{F}$. $\mathcal{F}$ contains all the
sets $\widetilde{\mathcal{S}}\left(T\right)$ as $T$ varies in the
set of positive integers. This problem is equivalent to finding all
the sets $\widetilde{\mathcal{S}}\left(T\right)$ as $T$ varies in
each one of these intervals. The equivalence is due to the fact that
the union of these intervals spans the entire set of positive integers.

\begin{figure}[t]
\begin{centering}
\includegraphics[scale=0.48]{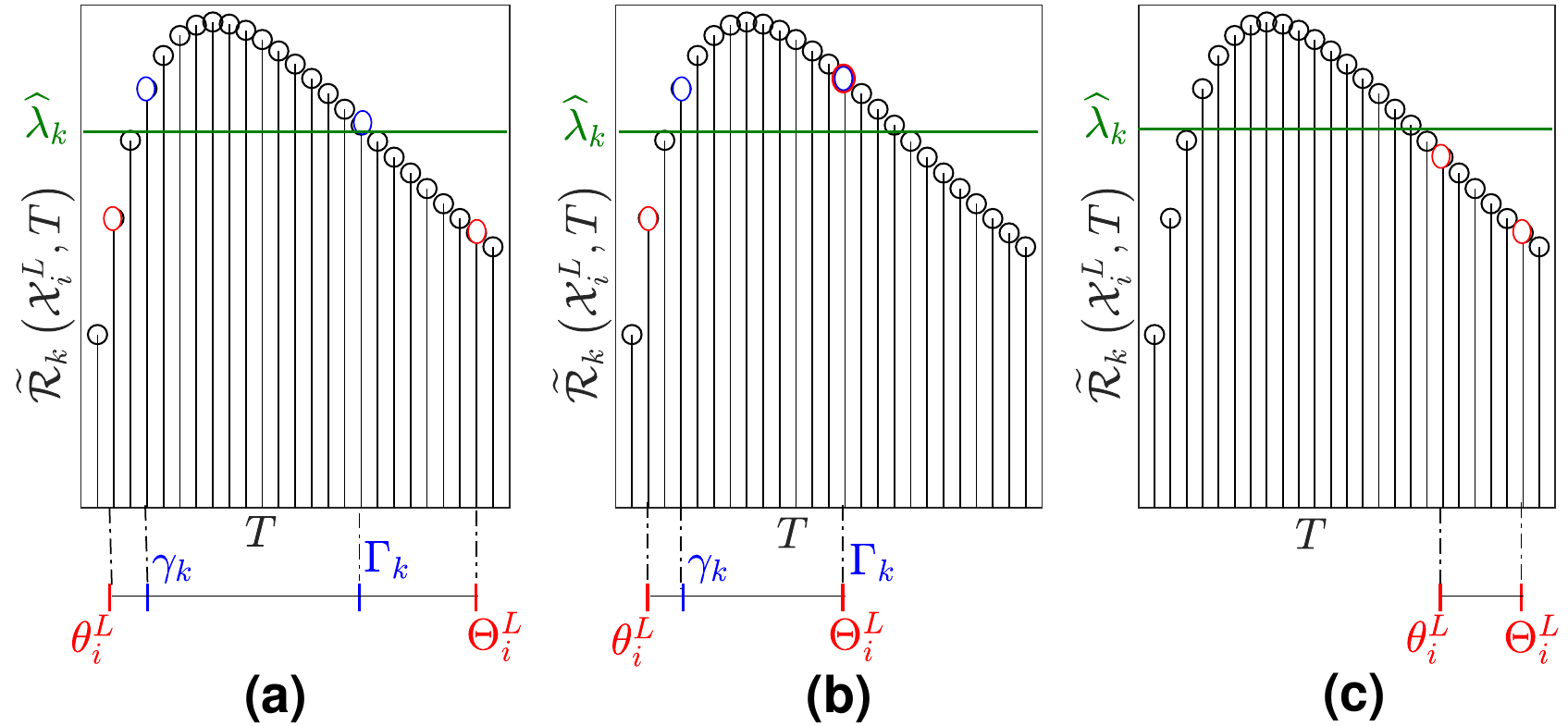}
\par\end{centering}
\caption{Graph of unimodal function $\widetilde{\mathcal{R}}_{k}\left(\mathcal{X}_{i}^{L},T\right)$
depicting an example when (a) $\gamma_{k}$ and $\Gamma_{k}$ exist
and $\Gamma_{k}<\Theta_{i}^{L}$. (b) $\gamma_{k}$ and $\Gamma_{k}$
exist and $\Gamma_{k}=\Theta_{i}^{L}$. (c) $\gamma_{k}$ and $\Gamma_{k}$
do not exist.}

\label{unimodal_figure}\vspace{-1.0em}
\end{figure}

Let $\left[\theta_{i}^{L},\Theta_{i}^{L}\right]$ be one such interval
where $\mbox{{\ensuremath{Q_{i}^{L}}.T<\ensuremath{Q_{i+1}^{L}}.T}}$,
$\mbox{\ensuremath{\theta_{i}^{L}=Q_{i}^{L}.T}}$, and $\mbox{\ensuremath{\Theta_{i}^{L}=Q_{i+1}^{L}.T-1}}$
(\textit{line 9}). We have $\mbox{\ensuremath{\widetilde{\mathcal{S}}^{L}\left(T\right)=\mathcal{X}_{i}^{L}\,;\,\forall T\in\left[\theta_{i}^{L},\Theta_{i}^{L}\right]}}$.
In the interval $\left[\theta_{i}^{L},\Theta_{i}^{L}\right]$, the
perceived set of interested operators $\widetilde{\mathcal{S}}\left(T\right)\subseteq\mathcal{X}_{i}^{L}$.
If $k\in\mathcal{X}_{i}^{L}$, then $T\leq\widehat{\Lambda}_{k}$
(refer to (\ref{eq:2.3.5})). Hence, for the interval $\left[\theta_{i}^{L},\Theta_{i}^{L}\right]$,
(\ref{eq:2.3.6}) is equivalent to $\mbox{\ensuremath{\widetilde{\mathcal{S}}\left(T\right)=\left\{ k\in\mathcal{X}_{i}^{L}\,:\,\widetilde{\mathcal{R}}_{k}\left(\mathcal{X}_{i}^{L},T\right)\geq\widehat{\lambda}_{k}\right\} }}$.
According to Property~\ref{RevenueFuncProp1}, $\widetilde{\mathcal{R}}_{k}\left(\mathcal{X}_{i}^{L},T\right)$
is unimodal in $T$. Therefore, the solution to $\widetilde{\mathcal{R}}_{k}\left(\mathcal{X}_{i}^{L},T\right)\geq\widehat{\lambda}_{k}$
in the interval $\left[\theta_{i}^{L},\Theta_{i}^{L}\right]$ is also
an interval $\left[\gamma_{k},\Gamma_{k}\right]$. $\gamma_{k}$ and
$\Gamma_{k}$ are the minimum and the maximum lease duration satisfying
$\mbox{\ensuremath{\theta_{i}^{L}\leq\gamma_{k}\leq\Gamma_{k}\leq\Theta_{i}^{L}}}$
such that $\mbox{\ensuremath{\widetilde{\mathcal{R}}_{k}\left(\mathcal{X}_{i}^{L},T\right)\geq\widehat{\lambda}_{k}\,;\,\forall T\in\left[\gamma_{k},\Gamma_{k}\right]}}$.
If $T\in\left[\theta_{i}^{L},\Theta_{i}^{L}\right]$, there are three
possible cases:
\begin{enumerate}
\item[(B1)] \textit{$\gamma_{k}$ and $\Gamma_{k}$ exist and $\Gamma_{k}<\Theta_{i}^{L}$:}
One such example is shown in Figure~\ref{unimodal_figure}.a. In
this case, the $k^{th}$ operator is associated with two ordered pairs
$\left(\gamma_{k},k\right)$ and $\left(\Gamma_{k}+1,k\right)$ implying
that $k\in\widetilde{\mathcal{S}}\left(T\right)$ \textit{if and only
if} $T\geq\gamma_{k}$ and $T<\Gamma_{k}+1$.
\item[(B2)] \textit{$\gamma_{k}$ and $\Gamma_{k}$ exist and $\Gamma_{k}=\Theta_{i}^{L}$:}
One such example is shown in Figure~\ref{unimodal_figure}.b. In
this case, the $k^{th}$ operator is associated with an ordered pair
$\left(\gamma_{k},k\right)$ implying that $k\in\widetilde{\mathcal{S}}\left(T\right)$
\textit{if and only if} $T\geq\gamma_{k}$.
\item[(B3)] \textit{$\gamma_{k}$ and $\Gamma_{k}$ do not exist:} One such example
is shown in Figure~\ref{unimodal_figure}.c. In this case, the $k^{th}$
operator is not associated with any ordered pair because $k\notin\widetilde{\mathcal{S}}\left(T\right)$
for all $T$.
\end{enumerate}

Consider a list $Q$ containing the ordered pairs associated with
all the operators in $\mathcal{X}_{i}^{L}$. List $Q$ is constructed
in \textit{lines 11-15}. This involves computation of $\gamma_{k}$
and $\Gamma_{k}$ in \textit{line 12} followed by accounting for cases
(B1) to (B3) in \textit{lines 13-15}. $\gamma_{k}$ and $\Gamma_{k}$
can be computed as follows. First, we find the maximum of $\widetilde{\mathcal{R}}_{k}\left(\mathcal{X}_{i}^{L},T\right)$
in the interval $\left[\theta_{i}^{L},\Theta_{i}^{L}\right]$ time
using \textit{fibonnaci search} \cite{fibonnacisearch}. Let $\widehat{\Theta}$
be the maxima of $\widetilde{\mathcal{R}}_{k}\left(\mathcal{X}_{i}^{L},T\right)$
in the interval $\left[\theta_{i}^{L},\Theta_{i}^{L}\right]$. Second,
$\gamma_{k}$ ($\Gamma_{k}$, resp.) can be found by solving the equation
$\mbox{\ensuremath{\widetilde{\mathcal{R}}_{k}\left(\mathcal{X}_{i}^{L},T\right)=\widehat{\lambda}_{k}}}$
in the interval $\left[\theta_{i}^{L},\widehat{\Theta}\right]$ ($\left[\widehat{\Theta},\Theta_{i}^{L}\right]$,
resp.) using \textit{binary search}. This strategy to compute $\gamma_{k}$
and $\Gamma_{k}$ requires $\mathcal{O}\left(\log_{2}\left(\Theta_{i}^{L}-\theta_{i}^{L}\right)\right)$
computations of $\widetilde{\mathcal{R}}_{k}\left(\mathcal{X}_{i}^{L},T\right)$
for various values of $T$.

\noindent 

\vspace{-1.0em}

To find all the sets $\widetilde{\mathcal{S}}\left(T\right)$ as $T$
varies in the interval $\left[\theta_{i}^{L},\Theta_{i}^{L}\right]$,
we simply apply steps (A1) to (A4) to list $Q$. This is implemented
in \textit{lines 16-20}. Let $Q_{j}$ be the $j^{th}$ element of
the sorted list $E$. If $Q_{j}.T<Q_{j+1}.T$, then $\mathcal{X}_{j}$
is one of the sets in $\mathcal{F}$. We have, $\mbox{\ensuremath{\widetilde{\mathcal{S}}\left(T\right)=\mathcal{X}_{j}\,;\,T\in\left[\theta_{j},\Theta_{j}\right]}}$
where $\theta_{j}=Q_{j}.T$ and $\Theta_{j}=Q_{j+1}.T-1$ . Therefore,
the objective function in the interval $\left[\theta_{j},\Theta_{j}\right]$
is $\widetilde{U}\left(\mathcal{X}_{j},T\right)$ which is maximum
for $T=\texttt{\ensuremath{\theta_{j}}}$ or $T=\Theta_{j}$ (Property~\ref{ObjFuncProp1}).
We can find the optimal lease duration in the interval $\left[\theta_{i}^{L},\Theta_{i}^{L}\right]$
by iterating over all $\mathcal{X}_{j}$ such that $Q_{j}.T<Q_{j+1}.T$.
Finally, we can find the optimal lease duration $\widetilde{T}^{*}$
by repeating the same procedure for all such intervals $\left[\theta_{i}^{L},\Theta_{i}^{L}\right]$
that satisfies $Q_{i}^{L}.T<Q_{i+1}^{L}.T$. These steps are implemented
in \textit{lines 21-25}.

We end this section by discussing the time complexity of Algorithm~\ref{opalgo_heterogeneous}
and comparing it with a bruteforce approach to solve $OP1$. Lines
12 and 22 are the most computationally demanding steps of Algorithm~\ref{opalgo_heterogeneous}
as it involves numerical integration to evaluate the revenue function
$\widetilde{\mathcal{R}}_{k}\left(\mathcal{S},T\right)$. All other
computations is absorbed (up to a constant factor) by the time taken
for evaluating the revenue function. Let $\widehat{\Lambda}^{L}=\underset{1\leq k\leq N}{\max}\widehat{\Lambda}_{k}$,
the maximum lease duration above which none of the operators can afford
a channel.
\begin{prop}
\label{prop:time-complexity}Time complexity of Algorithm~\ref{opalgo_heterogeneous}
is $\mathcal{O}\left(N^{2}\log_{2}\left(\widehat{\Lambda}^{L}\right)+N^{3}\right)$.
\end{prop}
\begin{IEEEproof}
Please refer to Appendix~F for the proof.
\end{IEEEproof}
A bruteforce approach to solve $OP1$ involves iterating from $T=1$
to $\widehat{\Lambda}^{L}$ to find the $T$ which maximizes $\widetilde{U}\left(T\right)$.
This is because for $T>\widehat{\Lambda}^{L}$, $\widetilde{\mathcal{S}}\left(T\right)=\emptyset$
and hence $\widetilde{U}\left(T\right)=0$. To evaluate $\widetilde{U}\left(T\right)$,
we need $\mathcal{O}\left(N\right)$ computations of $\widetilde{\mathcal{R}}_{k}\left(\mathcal{S},T\right)$
to find $\widetilde{\mathcal{S}}\left(T\right)$ (refer to (\ref{eq:2.3.3}))
and finally $\widetilde{U}\left(T\right)$ (refer to (\ref{eq:2.2.13})).
Therefore, time complexity of the bruteforce approach is $\mathcal{O}\left(N\widehat{\Lambda}^{L}\right)$.
In practice, $\widehat{\Lambda}^{L}$ is much larger compared to $N$.
Hence, time complexity of bruteforce approach, $\mathcal{O}\left(N\widehat{\Lambda}^{L}\right)$,
is much larger compared to time complexity of Algorithm~\ref{opalgo_heterogeneous},
$\mathcal{O}\left(N^{2}\log_{2}\left(\widehat{\Lambda}^{L}\right)+N^{3}\right)$.

\section{Numerical Results\label{sec:Numerical-Results}}

In Sections \ref{subsec:Optimal-Trends} to \ref{subsec:Suboptimal},
we use the optimization algorithms from Section~\ref{sec:Optimal-Solution}
to numerically explore the effect of true market parameters $\xi_{k}$
on $\widetilde{T}^{*}$, $U^{*}=U\left(\mathcal{S}\left(\widetilde{T}^{*}\right),\widetilde{T}^{*}\right)$
and $s^{*}=\left|\mathcal{S}\left(\widetilde{T}^{*}\right)\right|$
for complete information games. Recall that $\widetilde{T}^{*}$ is
the optimal lease duration corresponding to the perceived objective
function, $U^{*}$ is the value of the objective function (true) corresponding
to $\widetilde{T}^{*}$ and $s^{*}$ is the number of interested operators
corresponding to $\widetilde{T}^{*}$ (refer to (\ref{eq:3.2.1})).
For complete information games, the perceived objective function,
$\widetilde{U}\left(T\right)$, is equal to the objective function
(true), $U\left(T\right)$. Hence, $\widetilde{T}^{*}=T^{*}$ here
$T^{*}$ is the true optimal lease duration corresponding to $U\left(T\right)$. In Section \ref{subsec:Effect-Incomplete}, we discuss how incomplete information games leads to sub-optimal solutions.

One of the market parameters in $\xi_{k}$ is the autocorrelation
coefficient $a_{k}$. Instead of $a_{k}$, we use time constant $\tau_{k}$
where $a_{k}=\exp\left(-\frac{1}{\tau_{k}}\right)$. A higher time
constant implies higher autocorrelation. Throughout this section,
number of operators $N=10$ and number of channels $M=2$.\vspace{-0.5em}

\subsection{Optimal Trends\label{subsec:Optimal-Trends}}

\noindent 
\begin{figure}[t]
\begin{centering}
\includegraphics[scale=0.57]{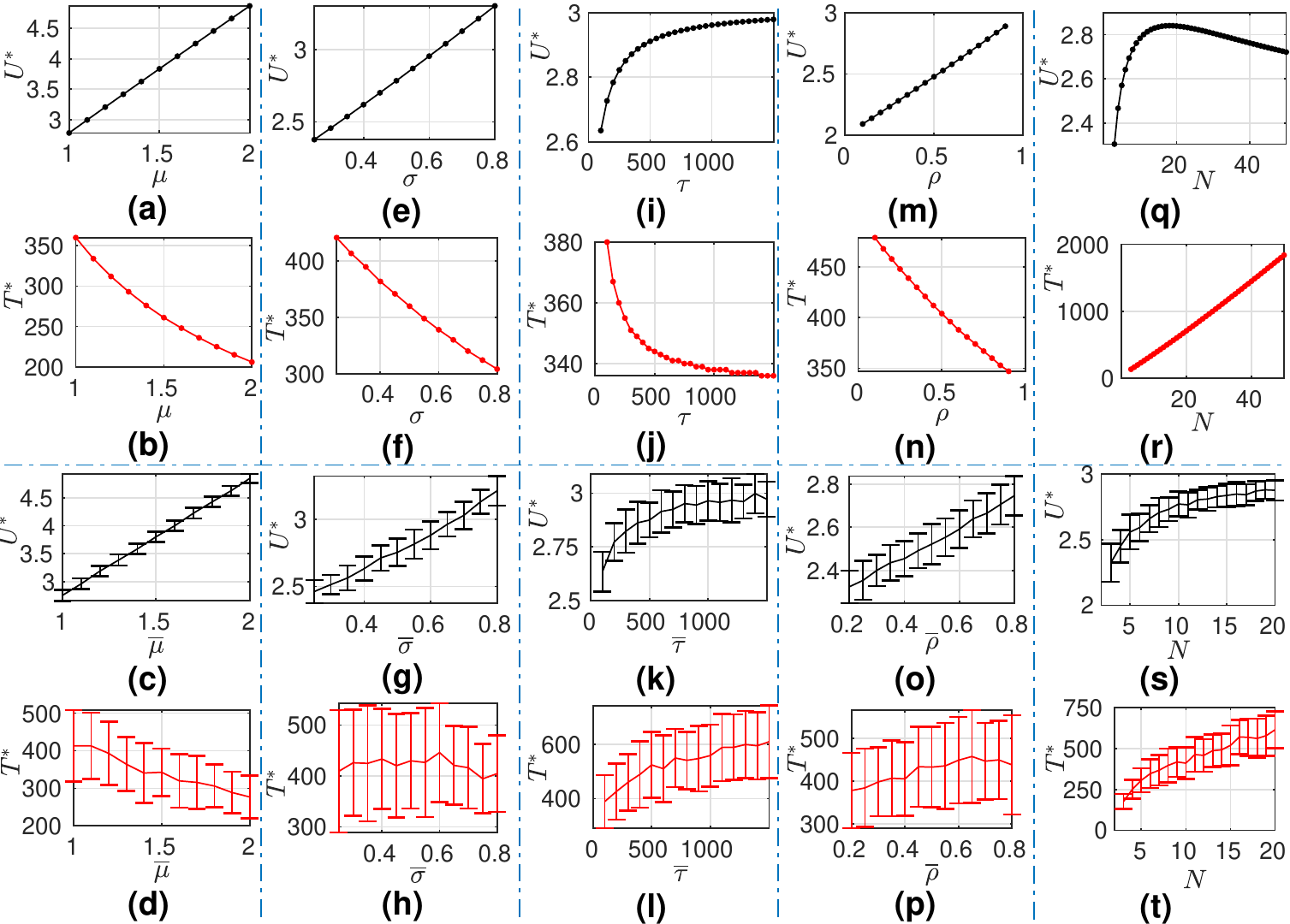}
\par\end{centering}
\caption{Plots showing $T^{*}$ and $U^{*}$ as a function of mean (a, b, c,
and d), standard deviation (e, f, g, and h), time constant (i, j,
k, and l), MER (m, n, o, and p), and number of operators (q, r, s,
and t). The top two and the bottom two rows correspond to homogeneous
and heterogeneous markets respectively. For every mean of the market
parameters (x-axis) in a heterogeneous market, we averaged $T^{*}$
and $U^{*}$ over $100$ instances of market parameters.\vspace{-1.0em}}
\label{sim_homogeneous}
\end{figure}

\vspace{-1.0em}

The trends of $T^{*}$ and $U^{*}$ as a function of number of operators
$N$ and parameters $\mu_{k}$, $\sigma_{k}$, $\tau_{k}$, and $\rho_{k}$
are discussed in this subsection. Throughout this subsection, $\Lambda_{k}=\infty\,;\,\forall k$.
We consider both homogeneous and heterogeneous market. The default
parameters for homogeneous market are $\mu=1$, $\sigma=0.5$, $\tau=100$,
$\rho=0.8$ and $\lambda=100$. We vary one parameter at a time while
holding the other parameters at their default values. We solve for
$T^{*}$ and $U^{*}$ as we vary the parameters and plot the result
in Figure~\ref{sim_homogeneous}.

For heterogeneous markets, we randomly choose the values for the market
parameters from an uniform distribution. The default uniform distributions
are $\mu_{k}\sim\mathcal{U}\left(0.8,1.2\right)$, $\sigma_{k}\sim\mathcal{U}\left(0.4,0.6\right)$,
$\tau_{k}\sim\mathcal{U}\left(50,150\right)$, $\rho_{k}\sim\mathcal{U}\left(0.7,0.9\right)$,
and $\lambda_{k}\sim\mathcal{U}\left(50,150\right)\,;\,\forall k$.
Each of these distributions are associated with a mean and a range,
e.g. the mean of $\mu_{k}$ is $\overline{\mu}=\frac{0.8+1.2}{2}=1$
and the range is $\left(1.2-0.8\right)=0.4$. The range of these distributions
remains the same throughout this section; only the mean is varied.
One of these distributions is varied at a time while holding the other
distributions at their default value. For every distribution, we generate
$100$ instances of market parameters sampled from the five distributions.
We solve for $T^{*}$ and $U^{*}$ for each of the $100$ instances
and plot the result as errorbar graphs in Figure~\ref{sim_homogeneous}.
The errobar graphs show the sample mean and standard deviation of
$T^{*}$ and $U^{*}$.

We will now explain the effect of various parameters on $T^{*}$ and
$U^{*}$. These explanations will rely on Properties \ref{RevenueFuncProp1}
- \ref{ObjFuncProp2}. Also, recall that special cases of Properties
\ref{ObjFuncProp1} and \ref{ObjFuncProp2} holds for homogeneous
market, i.e. objective function is monotonic decreasing in $T$ (Property
\ref{ObjFuncProp1}) and monotonic increasing in $s$ (Property \ref{ObjFuncProp2})
for a homogeneous market.

\textit{Effect of mean:} In a homogeneous market, as $\mu$ increases,
an operator's revenue per time slot increases. Therefore, it will
take less time to generate the MER $\lambda$. Hence, $T^{*}$ decreases
as shown in Figure~\ref{sim_homogeneous}.b. With decrease in $T^{*}$,
$U^{*}$ increases according to Property~\ref{ObjFuncProp1}. This
is shown in Figure~\ref{sim_homogeneous}.a. Similar trends hold
for heterogeneous market. As the mean of $\mu_{k}$, $\overline{\mu}$,
increases, the sample mean of $U^{*}$ increases while the sample
mean of $T^{*}$ decreases. This is shown in Figures~\ref{sim_homogeneous}.c.
and \ref{sim_homogeneous}.d.

\textit{Effect of standard deviation:} In a homogeneous market, $T^{*}$
decreases with increase in \textit{$\sigma$ }as shown in Figure~\ref{sim_homogeneous}.f.
This can be explained as follows. As $\sigma$ increases, an operator's
revenue fluctuates more around the mean. These fluctuations can lead
to a revenue which is either greater or lower than the mean. If an
operator is allocated a channel, there is a higher probability that
the revenue is greater than the mean. This is due to the allocation
policy which, in general, ensures that the operator who is allocated
a channel has a high revenue. This suggests that the revenue function
increases with $\sigma$. Since the revenue function increases, an
operator takes less time to generate its MER. Hence, $T^{*}$ decreases
with increase in \textit{$\sigma$. }As $T^{*}$ decreases,\textit{
}$U^{*}$ increases due to Property~\ref{ObjFuncProp1}. This is
shown in Figure~\ref{sim_homogeneous}.e. For a heterogeneous market,
the sample mean of $U^{*}$ increases with increase in mean of $\sigma_{k}$,
$\overline{\sigma}$. This is shown in Figure~\ref{sim_homogeneous}.g
and its is similar to homogeneous market. However, unlike a homogeneous
market, the sample mean of $T^{*}$ remains almost the same with increase
in $\overline{\sigma}$ as shown in Figure~\ref{sim_homogeneous}.h.
This happens due to a cyclic effect which can be explained as follows.
Similar to homogeneous market, with increase in $\overline{\sigma}$,
the revenue function increases. But as the revenue function increases,
more operators are interested in entering the market which in turn
decreases the revenue function (Property~\ref{RevenueFuncProp2}).
These two competing factors negates the impact of $\overline{\sigma}$
on the revenue function and hence on $T^{*}$.

\textit{Effect of time constant: }Consider the homogeneous market
first. Autocorrelation defines the self-similarity of a random process.
As autocorrelation increases, an operator with higher revenue at current
time slot will have higher revenue at a later time slot. Therefore,
with increase in time constant $\tau$ (and hence autocorrelation),
the revenue function increases. As the revenue function increases,
an operator takes less time to generate its MER. Hence, $T^{*}$ decreases
with increase in \textit{$\tau$. }As $T^{*}$ decreases,\textit{
}$U^{*}$ increases due to Property~\ref{ObjFuncProp1}. This is
shown in Figure~\ref{sim_homogeneous}.i and \ref{sim_homogeneous}.j.
For a heterogeneous market, the sample mean of $U^{*}$ increases
with increase in mean of $\tau_{k}$, $\overline{\tau}$. This trend
is shown in Figure~\ref{sim_homogeneous}.k and its is similar to
a homogeneous market. However, unlike a homogeneous market, the sample
mean of $T^{*}$ increases with increase in $\overline{\tau}$ as
shown in Figure~\ref{sim_homogeneous}.l. This is due to the same
cyclic effect mentioned while explaining the effect of standard deviation.
But in this case, the effect of the increase in number of interested
operator is more dominant. As a result, the revenue function decreases
with increase in $\overline{\tau}$. Since the revenue function decreases,
the sample mean of $T^{*}$ increases because an operator takes more
time to generate its MER.

\textit{Effect of bid correlation coefficient:} Consider the homogeneous
market first. As bid correlation coefficient increases, an operator
with a high bid is more likely to generate a higher revenue. Since
channels are allocated to operators with high bids, we can equivalently
say that if an operator is allocated a channel, then its revenue increases
with increase in bid correlation coefficient. Therefore, it will take
less time to generate the MER $\lambda$. Hence, $T^{*}$ decreases
with increase in \textit{$\rho$.} With decrease in $T^{*}$, $U^{*}$
increases according to Property~\ref{ObjFuncProp1}. This is shown
in Figure~\ref{sim_homogeneous}.m and Figure~\ref{sim_homogeneous}.n.
For a heterogeneous market, the sample mean of $U^{*}$ increases
with increase in mean of $\rho_{k}$, $\overline{\rho}$. This is
shown in Figure~\ref{sim_homogeneous}.o and its is similar to a
homogeneous market. However, unlike a homogeneous market, the sample
mean of $T^{*}$ increases with increase in $\overline{\rho}$ as
shown in Figure~\ref{sim_homogeneous}.p. This is due to the same
cyclic effect mentioned while explaining the effect of time constant.

\textit{Effect of number of operators: }Consider the homogeneous market
first. As the number of operators increases, the probability that
a given operator is allocated a channel decreases. To compensate for
this decrease in probability, an operator has to generate more revenue
when it is allocated a channel in order to satisfy its MER. Hence,
$T^{*}$ increases as shown in Figure~\ref{sim_homogeneous}.r. Now
we will explain the effect of $N$ on $U^{*}$. As $N$ increases,
$U^{*}$ increases according to Property~\ref{ObjFuncProp2}. However,
with increase in $N$, $T^{*}$ increases which leads to decrease
in $U^{*}$ according to Property~\ref{ObjFuncProp1}. Because of
these two competing factors, $U^{*}$ first increases and then decreases
with increase in $N$ as shown in Figure~\ref{sim_homogeneous}.q.
Recall that in our model, the number of interested operators is a
measure of market competition. Then this numerical study shows that
too much competition may not necessarily improve spectrum utilization.

For heterogeneous market, we sampled $\mu_{k}$, $\sigma_{k}$, $\tau_{k}$,
$\rho_{k}$ and $\lambda_{k}$ from their default uniform distributions.
As $N$ increases, the sample mean of $T^{*}$ and $U^{*}$ increases.
This is shown in Figure~\ref{sim_homogeneous}.s and \ref{sim_homogeneous}.t.
These trends are similar to homogeneous market. Similar to homogeneous
market, we expect the sample mean of $U^{*}$ to start decreasing
if $N$ is above a threshold. But, we could not verify the same. This
is because as $N$ increases, computing the revenue function for a
heterogeneous market becomes computationally expensive which in turn
makes Algorithm~\ref{opalgo_heterogeneous} computationally expensive.
This problem does not exists for homogeneous market because the expression
for revenue function is simpler for homogeneous market. Please refer
to the proof of Propositions 5 and equation 60 in the appendix to
appreciate the relative complexity of the revenue function for a heterogeneous
and a homogeneous market.\vspace{-0.5em}

\subsection{Comparison with an intuitive algorithm\label{subsec:Suboptimal}}

\noindent 
\begin{figure}[t]
\begin{centering}
\includegraphics[scale=0.54]{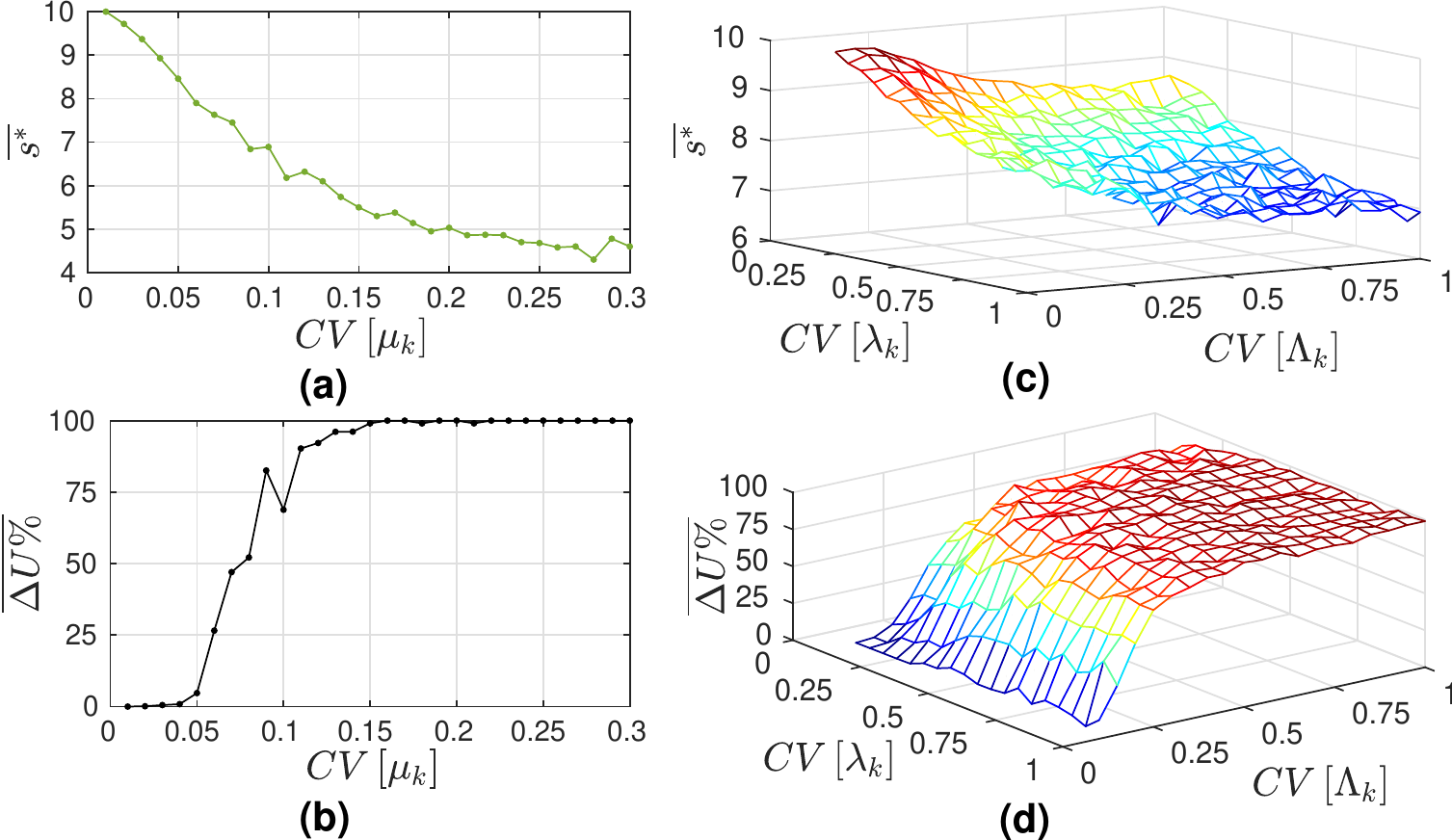}
\par\end{centering}
\caption{Plots comparing the performance of Algorithm~\ref{opalgo_heterogeneous}
and SUBOP as the market becomes more heterogeneous in Mean $\mu_{k}$
(a, b), MER $\lambda_{k}$ and maximum affordable lease duration $\Lambda_{k}$
(c, d). In (a, b), for each coefficient of variation $CV\left[\mu_{k}\right]$,
$\overline{s^{*}}$ and $\overline{\Delta U_{\%}}$ have been averaged
over $100$ instance of $\mu_{k}$. In (c, d), for each pair of coefficient
of variations $\left(CV\left[\lambda_{k}\right],CV\left[\Lambda_{k}\right]\right)$,
$\overline{s^{*}}$ and $\overline{\Delta U_{\%}}$ have been averaged
over $100$ instance of $\lambda_{k}$ and $\Lambda_{k}$.\vspace{-1.5em}}
\label{sim_subop}
\end{figure}

\vspace{-1.0em}

In this section, we compare the performance of Algorithm~\ref{opalgo_heterogeneous}
with an intuitive, but sub-optimal, algorithm SUBOP which maximizes the objective function
by setting a lease duration that satisfies all the $N$ operators
in terms of affordability $\Lambda_{k}$ and MER $\lambda_{k}$. Through
this comparison we exemplify that as the market becomes more heterogeneous,
it is not optimal to satisfy all the operators even if it is possible.

We start by describing SUBOP. Define $\mathcal{S}_{N}=\left\{ 1,2,\ldots,N\right\} $.
Since SUBOP has to satisfy all the $N$ operators, the $k^{th}$ operator
is satisfied if the lease duration satisfies $T\leq\Lambda_{k}$ and
$\mathcal{R}_{k}\left(\mathcal{S}_{N},T\right)\geq\lambda_{k}$. The
solution to $\mathcal{R}_{k}\left(\mathcal{S}_{N},T\right)\geq\lambda_{k}$
is a range $\left[\gamma_{k},\Gamma_{k}\right]$ (refer to line~\ref{bounds_line}
of Algorithm~\ref{opalgo_heterogeneous}). But $T\leq\Lambda_{k}$
and hence the range has to be modified as $\left[\widetilde{\gamma}_{k},\widetilde{\Gamma}_{k}\right]$
where $\widetilde{\gamma}_{k}=\gamma_{k}$ and $\widetilde{\Gamma}_{k}=\min\left(\Gamma_{k},\Lambda_{k}\right)$.
The $k^{th}$ operator is interested in entering the market \textit{iff}
$T\in\left[\widetilde{\gamma}_{k},\widetilde{\Gamma}_{k}\right]$.
The range of lease duration that satisfies all the $N$ operators
is $\left[\widetilde{\theta},\widetilde{\Theta}\right]$ where $\widetilde{\theta}=\underset{k}{\max}\:\widetilde{\gamma}_{k}$
and $\widetilde{\Theta}=\underset{k}{\min}\:\widetilde{\Gamma}_{k}$.
If $\widetilde{\Theta}<\widetilde{\theta}$, then there is no lease
duration that satisfies all the operators and hence the value of the
objective function is $U_{S}^{*}=0$ where the subscript $S$ implies
sub-optimal. If $\widetilde{\Theta}\geq\widetilde{\theta}$, then either
$T=\widetilde{\theta}$ or $T=\widetilde{\Theta}$ maximizes the objective
function (Property~\ref{ObjFuncProp1}) subjected to $\mathcal{S}=\mathcal{S}_{N}$.
Accordingly, the value of the objective function is $U_{S}^{*}=\max\left(U\left(\mathcal{S}_{N},\widetilde{\theta}\right),U\left(\mathcal{S}_{N},\widetilde{\Theta}\right)\right)$.

We first compare the performance of Algorithm~\ref{opalgo_heterogeneous}
with SUBOP as the market becomes more heterogeneous in mean $\mu_{k}$.
To compare the algorithms, lets define $\mbox{{\ensuremath{\Delta U_{\%}}=\ensuremath{\frac{U^{*}-U_{S}^{*}}{U_{S}^{*}}\times}100}}$,
the percentage increase in $U^{*}$ compared to $U_{s}^{*}$. The
setup is homogeneous in all market parameters but $\mu_{k}$. We have
$\sigma_{k}=0.5$, $\tau_{k}=100$, $\lambda_{k}=100$, and $\Lambda_{k}=\infty\,;\,\forall k$.
The mean $\mu_{k}$ is sampled from a truncated gaussian distribution
with mean $1$, coefficient of variation $CV\left[\mu_{k}\right]$
and the truncation bounds are $0.5$ and $1.5$. As $CV\left[\mu_{k}\right]$
increases, the gaussian distribution spreads out more and hence there
is a wider range of $\mu_{k}$ making the market more heterogeneous.
As shown in Figure~\ref{sim_subop}.a, expected optimal number of
interested operators $\overline{s^{*}}$ decreases with increase in
$CV\left[\mu_{k}\right]$. This is because as the market becomes more
heterogeneous in $\mu_{k}$, the revenue function becomes unimodal
in nature (Property~\ref{RevenueFuncProp1}). This suggests that
there may not be a lease duration that satisfies MER of all the operators.
Even if it is possible to satisfy lease duration of all the operators,
such lease durations may too large which may significantly decrease
the objective function according to Property~\ref{ObjFuncProp1}
(assuming bid correlation coefficients of the operators are high).
It is also possible that some of the operators have low bid correlation
coefficient. If they enter the market, objective function function
can decrease (Property~\ref{ObjFuncProp2}). Therefore, it may not
be optimal to satisfy those operators whose bid correlation coefficient
is low. But since SUBOP tries to satisfy all the operators, its performance
compared to Algorithm~\ref{opalgo_heterogeneous} decreases as the
market becomes more heterogeneous in $\mu_{k}$. This is shown in
Figure~\ref{sim_subop}.b. where $\overline{\Delta U_{\%}}$ increases
with $CV\left[\mu_{k}\right]$.

Similarly, we compare the performance of Algorithm~\ref{opalgo_heterogeneous}
with SUBOP as the market becomes more heterogeneous in $\lambda_{k}$
and $\Lambda_{k}$. The setup is homogeneous in all market parameters
but $\lambda_{k}$ and $\Lambda_{k}$. We have $\mu_{k}=1$, $\sigma_{k}=0.5$
and $\tau_{k}=100\,;\,\forall k$. $\lambda_{k}$ is sampled from
a truncated gaussian distribution with mean $500$, coefficient of
variation $CV\left[\lambda_{k}\right]$ and the truncation bounds
are $100$ and $900$. $\Lambda_{k}$ is sampled from a truncated
gaussian distribution with mean $5000$, coefficient of variation
$CV\left[\Lambda_{k}\right]$ and the truncation bounds are $900$
and $9100$. As $CV\left[\lambda_{k}\right]$ and $CV\left[\Lambda_{k}\right]$
increases, there is a wider of $\lambda_{k}$ and $\Lambda_{k}$.
As shown in Figure~\ref{sim_subop}.c, expected optimal number of
interested operators $\overline{s^{*}}$ decreases with increase in
$CV\left[\lambda_{k}\right]$ and $CV\left[\Lambda_{k}\right]$. This
is because as the market becomes more heterogeneous in $\lambda_{k}$
and $\Lambda_{k}$, it is possible that a lease duration that satisfies
MER of one operator is not affordable by another operator. Hence,
there may not exist a lease duration that satisfies all the operator.
Even if it is possible to satisfy lease duration of all the operators,
such lease durations may be too large because few of the operators
have high MER. Setting such a large lease duration may not be optimal
according to Property~\ref{ObjFuncProp1}. But since SUBOP tries
to satisfy all the operators, its performance compared to Algorithm~\ref{opalgo_heterogeneous}
decreases as the market becomes more heterogeneous in $\lambda_{k}$
and $\Lambda_{k}$. This is shown in Figure~\ref{sim_subop}.d.

\subsection{Discontinuity in Optimal Trends\label{subsec:Discontinuity}}

This numerical result deals with complete information games. We demonstrate
certain interesting discontinuities in optimal trends as MER of operators
changes in a heterogeneous market. Since we are considering complete
information games, we have $\widehat{\xi}_{k}=\xi_{k}\,;\,\forall k$,
and hence $\mathcal{S}_{k}^{L}\left(T\right)$ is same for all $k$'s
(refer to (16)). So we have, $\mathcal{S}_{k}^{L}\left(T\right)=\mathcal{S}^{L}\left(T\right)\,;\,\forall k$.
Let $s^{L*}=\left|\mathcal{S}^{L}\left(T^{*}\right)\right|$ denote
the largest number of interested operators corresponding to optimal
lease duration. The numerical setup is as follows. The market is homogeneous
in all parameters but $\lambda_{k}$. We have $\mu_{k}=\mu=1\,;\,\forall k$,
$\sigma_{k}=\sigma=0.5\,;\,\forall k$, $\tau_{k}=\tau=100\,;\,\forall k$
and $\Lambda_{k}=\infty\,;\,\forall k$. MER of the first $8$ operators
are $100$ while the $9^{th}$ and the $10^{th}$ operator has MER
$\overline{\lambda}\geq100$. As our setup is homogeneous in $\mu_{k}$,
$\sigma_{k}$ and $\tau_{k}$, the revenue function of all the operators
is $\mathcal{R}\left(s,T\right)$ and the objective function is $U\left(s,T\right)$
(refer to Section~II-C). Recall that $U\left(s,T\right)$ is monotonic
decreasing in $T$ (special case of Property~3) and monotonic increasing
in $s$ (special case of Property~4) while $\mathcal{R}\left(s,T\right)$
is monotonic increasing in $T$ (special case of Property~1). Consider
if the market consists of only the first $8$ operators. This market
is completely homogeneous and the optimal lease duration is the solution
to $\mathcal{R}\left(8,T\right)=100$ (refer to Section~III-A) which
is equal to $306$. Now consider the market with all the $10$ operators.
The optimal lease duration of this market must be at least $306$
because the MER of the first $8$ operators must be satisfied for
optimality. We explore how $U^{*}$, $T^{*}$, $s^{*}$ and $s^{L*}$
vary with $\overline{\lambda}$.

\begin{figure}[t]
\begin{centering}
\includegraphics[scale=0.6]{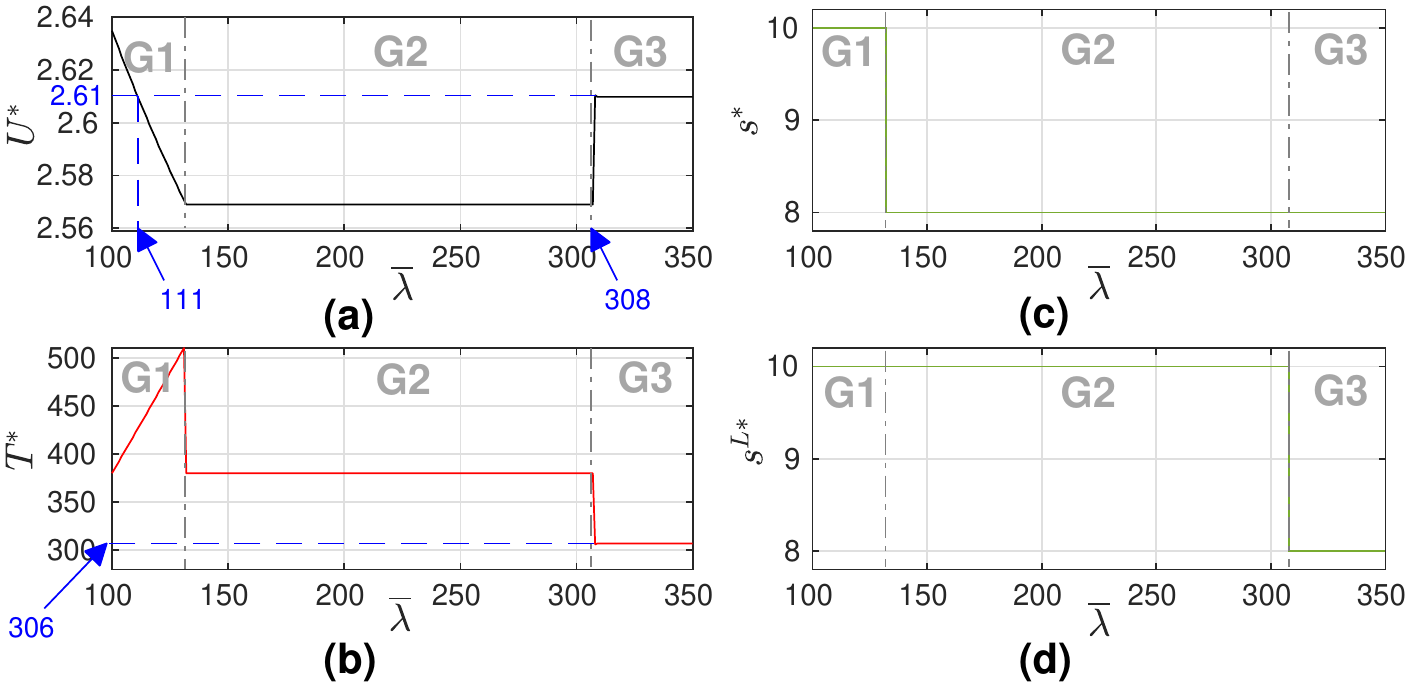}
\par\end{centering}
\caption{A figure demonstrating the discontinuous nature of $U^{*}$, $T^{*}$,
$s^{*}$ and $s^{L*}$ as a function of MER $\overline{\lambda}$.\vspace{-1.0em}}
\label{sim_implicitcontrol}
\end{figure}

As $\overline{\lambda}$ varies, there are three discontinuities in
$U^{*}$, $T^{*}$, $s^{*}$ and $s^{L*}$ as shown in Figure~\ref{sim_implicitcontrol}.
Therefore, we divide our explanation into three regions. As mentioned
in the previous paragraph, $T^{*}\geq306$. In \textit{region G1},
$\overline{\lambda}\leq\mu T^{*}\leq306$, implying that the $9^{th}$
and the $10^{th}$ operator \textit{may} join the market along with
the other $8$ operators. Hence, $s^{L*}=\left|\mathcal{S}^{L}\left(T^{*}\right)\right|=10$
(refer to (16)). Therefore, the minimum value of the revenue function
is $\mathcal{R}\left(10,T\right)$ which decides whether an operator
is interested in entering the market (refer to Section~II-D). There
are two possible candidates for optimal lease duration. \textit{First,}
the lease duration is $T_{1}$ that satisfies $\mathcal{R}\left(10,T_{1}\right)=100$.
In this case, only the first $8$ operators are interested in entering
the market and hence the value of the objective function is $U\left(8,T_{1}\right)$.
\textit{Second,} the lease duration is $T_{2}$ that satisfies $\mathcal{R}\left(10,T_{2}\right)=\overline{\lambda}$.
Since $\overline{\lambda}\geq100$, all the $10$ operators are interested
in entering the market and hence the value of the objective function
is $U\left(10,T_{2}\right)$. Definitely, $T_{2}\geq T_{1}$ because
$\overline{\lambda}\geq100$ and the $\mathcal{R}\left(s,T\right)$
is monotonic increasing in $T$. In region G1, $T_{2}$ is not much
larger compared to $T_{1}$ because $\overline{\lambda}$ is very
close to $100$. Hence, according to Property~4, $U\left(10,T_{2}\right)>U\left(8,T_{1}\right)$.
Therefore, $T^{*}=T_{2}$, $s^{*}=10$ and $U^{*}=U\left(10,T_{2}\right)$.
This is shown in Figure~\ref{sim_implicitcontrol}.a, \ref{sim_implicitcontrol}.b
and \ref{sim_implicitcontrol}.c. As $\overline{\lambda}$ increases,
$T^{*}=T_{2}$ increases and hence $U^{*}=U\left(10,T_{2}\right)$
decreases (Property~3).

In \textit{region G2}, $\overline{\lambda}$ is much more than $100$
and hence $T_{2}$ is much larger compared to $T_{1}$. Hence, according
to Property~3, $U\left(10,T_{2}\right)<U\left(8,T_{1}\right)$. Therefore,
$T^{*}=T_{1}$, $s^{*}=8$ and $U^{*}=U\left(8,T_{1}\right)$. This
is shown in Figure~\ref{sim_implicitcontrol}.a, \ref{sim_implicitcontrol}.b
and \ref{sim_implicitcontrol}.c. As $T_{1}$ is a constant, $T^{*}$
and $U^{*}$ are constants in region G2. In\textit{ region G3}, $\overline{\lambda}>306$
as shown in Figure~\ref{sim_implicitcontrol}.a. Since it was not
optimal to satisfy MER of the $9^{th}$ and the $10^{th}$ operator
in region G2, it is not optimal to satisfy their MER in region G3.
This is because $\overline{\lambda}$ in region G3 is greater compared
to region G2. But we must satisfy the MER of first $8$ operators.
Say that the lease duration is $306$. It is the least lease duration
that satisfies MER of the first $8$ operators. Also, it does not
satisfy MER of the $9^{th}$ and the $10^{th}$ operator becasuse
$\overline{\lambda}>306$. Infact, if lease duration is $306$, the
\textit{dominant strategy} of the $9^{th}$ and the $10^{th}$ operator
is not to enter the market and hence $s^{L*}=8$ as shown in Figure~\ref{sim_implicitcontrol}.d.
So we conlude that $T^{*}=306$ in region G3 and the corresponding
$U^{*}=U\left(8,306\right)=2.61$. This is shown in Figure~\ref{sim_implicitcontrol}.a
and \ref{sim_implicitcontrol}.b.

We conclude this section with two critical observations. If the market
consisted of only the first $8$ operators, then $\mbox{{\ensuremath{U^{*}}=2.61}}$.
With the $9^{th}$ and the $10^{th}$ operator in the market, $U^{*}$
is less than $2.61$ if $\overline{\lambda}$ lies in the interval
$\left(111,308\right)$. This is shown in Figure~\ref{sim_implicitcontrol}.a.
Based on this observation, we can conclude the following. \textit{First,}
too much competition may not necessarily lead to better spectrum utilization.
In region G1, even though all the $10$ operators are interested in
entering the market, $U^{*}$ is less than $2.61$ if $\overline{\lambda}>111$.
\textit{Second,} as a thumb rule, the spectrum utilization is higher
if the MER of the operators are low and hence leading to a lower lease
duration or if the MER is high enough that it does not affect the
decision of the other operators to enter the market. To appreciate
the last point, note that in region G2, the first $8$ operators enter
the market only if their MER is satisfied even if all $10$ operator
enters the market. So, even though the $9^{th}$ and the $10^{th}$
operators did not enter the market, these two operators affected the
decision of the first $8$ operators.

\subsection{Effect of Incomplete Information\label{subsec:Effect-Incomplete}}

\noindent 
\begin{figure}[t]
\begin{centering}
\includegraphics[scale=0.58]{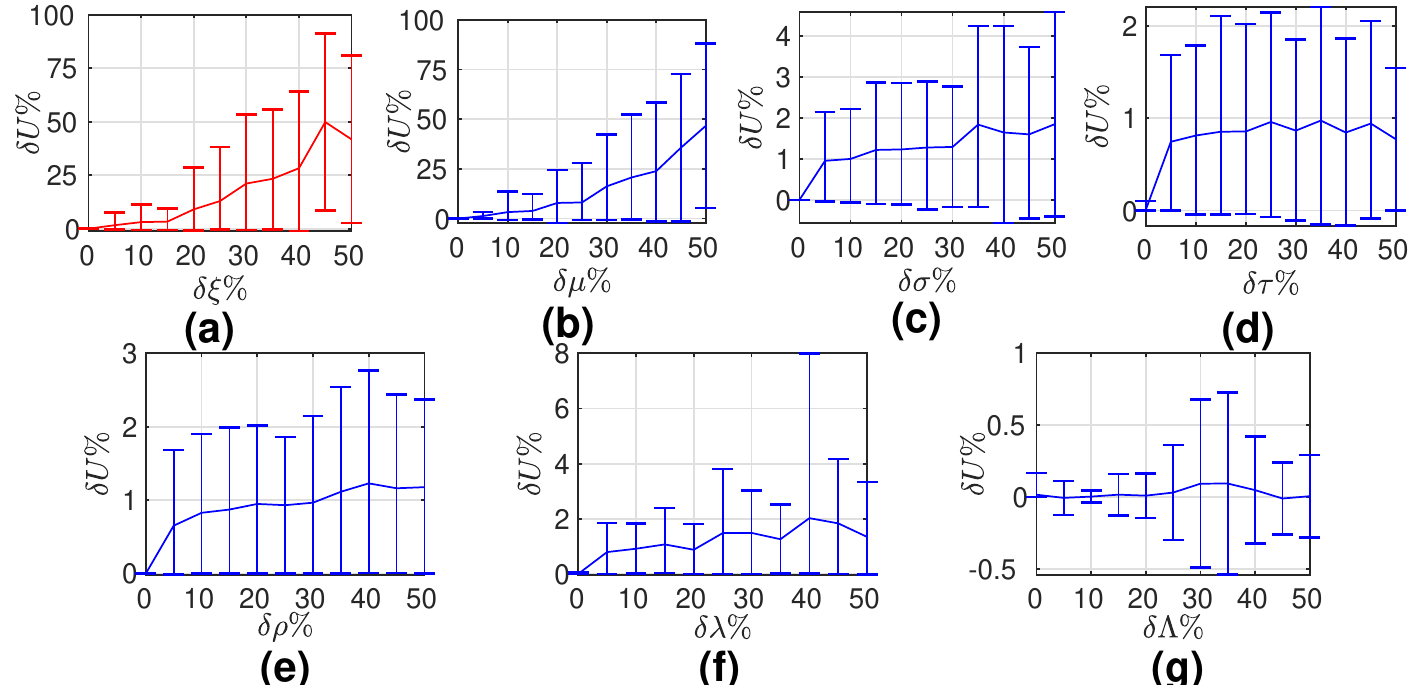}
\par\end{centering}
\caption{Plots of $\delta U\%$, the relative change in the optimal value of
the objective function due to incomplete information, when there is
a error in (a) all the parameters in $\widehat{\xi}_{k}$, (b) $\widehat{\mu}_{k}$,
(c) $\widehat{\sigma}_{k}$, (d) $\widehat{\tau}_{k}$, (e) $\widehat{\rho}_{k}$,
(f) $\widehat{\lambda}_{k}$, and (g) $\widehat{\Lambda}_{k}$.\vspace{-1.0em}}
\label{sim_incomplete_info}
\end{figure}

\vspace{-1.0em}

Our final numerical study analyzes how the deviation of the estimated
market parameters, $\widehat{\xi}_{k}$, from the true market parameters,
$\xi_{k}$, leads to sub-optimal spectrum utilization. Recall that
$\widetilde{T}^{*}$ is the optimal lease duration corresponding to
estimated market parameters, $\widehat{\xi}_{k}$. $\widetilde{T}^{*}$
is one of the outputs of Algorithm 1 of the main paper. Let
$T^{*}$ be the optimal lease duration corresponding to true market
parameters, $\xi_{k}$. In other words, $\widetilde{T}^{*}=T^{*}$
when $\widehat{\xi}_{k}=\xi_{k}$. The value of the objective function
(true) corresponding to $\widetilde{T}^{*}$ and $T^{*}$ are $U\left(\mathcal{S}\left(\widetilde{T}^{*}\right),\widetilde{T}^{*}\right)$
and $U\left(\mathcal{S}\left(T^{*}\right),T^{*}\right)$ respectively
(refer to (21) of the main paper). Let,\vspace{-1.0em}

\[
\delta U\%=\frac{U\left(\mathcal{S}\left(T^{*}\right),T^{*}\right)-U\left(\mathcal{S}\left(\widetilde{T}^{*}\right),\widetilde{T}^{*}\right)}{U\left(\mathcal{S}\left(T^{*}\right),T^{*}\right)}\cdot100
\]

$\delta U\%$ is the relative change in the optimal value of the objective
function (spectrum utilization) due to incomplete information. We
randomly choose true market parameters from uniform distributions:
$\mu_{k}\sim\mathcal{U}\left(0.8,1.2\right)$, $\sigma_{k}\sim\mathcal{U}\left(0.4,0.6\right)$,
$\tau_{k}\sim\mathcal{U}\left(150,250\right)$, $\rho_{k}\sim\mathcal{U}\left(0.5,0.7\right)$,
$\lambda_{k}\sim\mathcal{U}\left(50,150\right)$, and $\Lambda_{k}\sim\mathcal{U}\left(500,2000\right)\,;\,\forall k$.
Consider the true parameter $\mu_{k}$. The estimated parameter $\widehat{\mu}_{k}$
is choosen uniformly at random in the interval $\left[\mu_{k}-\frac{\delta\mu\%}{100}\cdot\mu_{k},\mu_{k}+\frac{\delta\mu\%}{100}\cdot\mu_{k}\right]$
where $\delta\mu\%$ is the error window associated with $\mu_{k}$.
Similarly, we have error windows $\delta\sigma\%$, $\delta\tau\%$,
$\delta\rho\%$, $\delta\lambda\%$ and $\delta\Lambda\%$ associated
with $\sigma_{k}$, $\tau_{k}$, $\rho_{k}$, $\lambda_{k}$ and $\Lambda_{k}$
respectively. In Figure~\ref{sim_incomplete_info}, we plot errorbar
graphs to study how $\delta U\%$ vary with error window. In Figure~\ref{sim_incomplete_info}.a,
all the six error windows are varied while in the remaining six plots,
one of the error windows is varied while the remaining five error
windows are set to zero. For each value of error window, we sample
over $100$ random instances of true and estimated market parameters
to generate the errorbar graphs.

As expected, the sample mean of $\delta U\%$ increases with error
window. Comparing Figure~\ref{sim_incomplete_info}.a with the remaining
six graphs, we can conclude that most of the reduction in spectrum
utilization happens due to error in $\mu_{k}$. Finally, we point
to a non-intuitive result which can be observed by zooming into Figure~\ref{sim_incomplete_info}.
There are some cases where $\delta U\%$ is less than zero. In other
words, it is possible that in an incomplete information scenario,
spectrum utilization is more than complete information scenario. This
can happen when an operator makes an erroneous decision of joining
the market (not joining the market) due to error in $\widehat{\xi}_{k}$.
This can improve spectrum utilization if the true bid correlation
coefficient is high (low).

\section{Conclusion\label{sec:Conclusion}}

The duration of a spectrum lease is a critical parameter that influences
the efficiency of spectrum utilization. The main contribution of this
paper is a mathematical model that is used to find the lease duration
which maximizes spectrum utilization. This model captures the effects
of lease duration on spectrum utilization for a market where an operators'
revenue is a measure of its spectrum utilization. Based on the system
model, we formulate a Stackelberg game with lease duration as one
of the decision variables. We also design algorithms to find the Stackelberg equilibrium and hence find the optimal lease
duration. Using these algorithms, we find several numerical trends
that show how lease duration should change with respect to various
market parameters in order to maximize spectrum utilization.

There are several possible avenues for extending this work, including: (a) Generalization
of our system model to capture the transaction costs associated with
re-allocation of channels. (b) Including variance in our system model
to capture risk aversion of the operators. (c) Second price
auctions to capture the variable market-dependent price of a spectrum
lease.\balance

\appendices{}

\section{Proof of Proposition 1\label{sec:Proof-of-Proposition1}}

We want to find the probability density function (pdf ) of the random
variable $Y_{k}\left(c,T\right)=\stackrel[t=\left(c-1\right)T+1]{cT}{\sum}x_{k}\left(t\right)$.
Since $x_{k}\left(t\right)$ is a stationary process, the pdf of $Y_{k}\left(c,T\right)$
is same for all epochs. Therefore, we can simply derive the pdf of
$Y_{k}\left(1,T\right)=\stackrel[t=1]{T}{\sum}x_{k}\left(t\right)$.
Now, $Y_{k}\left(1,T\right)$ is gaussian because $x_{k}\left(t\right)$
is gaussian and sum of gaussian random variable is always gaussian.
Since $Y_{k}\left(1,T\right)$ is gaussian, its pdf is completely
characterized by its mean $\widetilde{\mu}_{k}\left(T\right)$ and
standard deviation $\widetilde{\sigma}_{k}\left(T\right)$. Based
on this discussion we can conclude that 
\begin{equation}
Y_{k}\left(c,T\right)\sim\mathcal{N}\left(\widetilde{\mu}_{k}\left(T\right),\widetilde{\sigma}_{k}^{2}\left(T\right)\right)\,;\,\forall c\label{eq:2.1}
\end{equation}

Now all we have to do is to find expressions for $\widetilde{\mu}_{k}\left(T\right)$
and $\widetilde{\sigma}_{k}\left(T\right)$. We have,\vspace{-1.0em}

\begin{equation}
\widetilde{\mu}_{k}\left(T\right)=E\left[\stackrel[t=1]{T}{\sum}x_{k}\left(t\right)\right]=\stackrel[t=1]{T}{\sum}E\left[x_{k}\left(t\right)\right]=\mu_{k}T\label{eq:2.2}
\end{equation}

For a first order AR process as governed by (1), $x_{k}\left(t\right)$
can be expressed as

\vspace{-1.0em}

\begin{equation}
x_{k}\left(t\right)=a_{k}^{t}x_{k}\left(0\right)+\stackrel[v=0]{t-1}{\sum}a_{k}^{t-1-v}\varepsilon_{k}\left(v\right)\label{eq:2.3}
\end{equation}

Equation \ref{eq:2.3} can be easily proved using mathematical induction.
Also,

\vspace{-1.0em}

\begin{eqnarray}
\stackrel[t=1]{T}{\sum}x_{k}\left(t\right) & = & \stackrel[t=1]{T}{\sum}\left(a_{k}^{t}x_{k}\left(0\right)+\stackrel[v=0]{t-1}{\sum}a_{k}^{t-1-v}\varepsilon_{k}\left(v\right)\right)\label{eq:2.4}\\
 & = & x_{k}\left(0\right)\stackrel[t=1]{T}{\sum}a_{k}^{t}+\stackrel[t=1]{T}{\sum}\stackrel[v=0]{t-1}{\sum}a_{k}^{t-1-v}\varepsilon_{k}\left(v\right)\nonumber \\
 & = & x_{k}\left(0\right)\stackrel[t=1]{T}{\sum}a_{k}^{t}+\stackrel[v=0]{T-1}{\sum}\stackrel[t=v+1]{T}{\sum}a_{k}^{t-1-v}\varepsilon_{k}\left(v\right)\label{eq:2.5}\\
 & = & x_{k}\left(0\right)\frac{a_{k}-a_{k}^{T+1}}{1-a_{k}}+\stackrel[v=0]{T-1}{\sum}\frac{1-a_{k}^{T-v}}{1-a_{k}}\varepsilon_{k}\left(v\right)\;\;\;\;\;\;\label{eq:2.6}
\end{eqnarray}

Equation \ref{eq:2.4} is obtained using (\ref{eq:2.3}). Equation
\ref{eq:2.5} is obtained by changing the order of summation. Now,
$\widetilde{\sigma}_{k}\left(T\right)=\sqrt{\text{Var}\left[\stackrel[t=1]{T}{\sum}x_{k}\left(t\right)\right]}$
where,

\begin{eqnarray}
 &  & \text{Var}\left[\stackrel[t=1]{T}{\sum}x_{k}\left(t\right)\right]\nonumber \\
 & = & \text{Var}\left[x_{k}\left(0\right)\frac{a_{k}-a_{k}^{T+1}}{1-a_{k}}+\stackrel[v=0]{T-1}{\sum}\frac{1-a_{k}^{T-v}}{1-a_{k}}\varepsilon_{i}\left(v\right)\right]\label{eq:2.7}\\
 & = & \left(\frac{a_{k}-a_{k}^{T+1}}{1-a_{k}}\right)^{2}\text{Var}\left[x_{k}\left(0\right)\right]\nonumber \\
 &  & \qquad+\stackrel[v=0]{T-1}{\sum}\left(\frac{1-a_{k}^{T-v}}{1-a_{k}}\right)^{2}\text{Var}\left[\varepsilon_{k}\left(v\right)\right]\label{eq:2.8}\\
 & = & \left(\frac{a_{k}-a_{k}^{T+1}}{1-a_{k}}\right)^{2}\sigma_{k}^{2}+\stackrel[v=0]{T-1}{\sum}\left(\frac{1-a_{k}^{T-v}}{1-a_{k}}\right)^{2}\left(\sigma_{k}^{\varepsilon}\right)^{2}\label{eq:2.9}\\
 & = & \frac{T-a_{k}\left(2-2a_{k}^{T}+a_{k}T\right)}{\left(1-a_{k}\right)^{2}}\cdot\frac{\left(\sigma_{k}^{\varepsilon}\right)^{2}}{1-a_{k}^{2}}\nonumber \\
 & = & \frac{T-a_{k}\left(2-2a_{k}^{T}+a_{k}T\right)}{\left(1-a_{k}\right)^{2}}\sigma_{k}^{2}\label{eq:2.10}
\end{eqnarray}

So we have,

\begin{equation}
\widetilde{\sigma}_{k}\left(T\right)=\frac{\sqrt{T-a_{k}\left(2-2a_{k}^{T}+a_{k}T\right)}}{\left(1-a_{k}\right)}\sigma_{k}\label{eq:2.11}
\end{equation}

Equation \ref{eq:2.7} is obtained from (\ref{eq:2.6}). Equation
\ref{eq:2.9} holds because $\varepsilon_{k}\left(v\right)$ are independent
random variables. Equations \ref{eq:2.9} and \ref{eq:2.10} follows
from the definition of $\sigma_{i}^{\varepsilon}$ and $\sigma_{i}$
as given by (2) and the paragraph before it. Finally, (\ref{eq:2.2})
and (\ref{eq:2.11}) are same as (3) and (4) respectively. This completes
the proof.

\section{Revenue Function for Heterogeneous Market\label{sec:RevFuncDerivation}}

In this section, we will derive an expression for the revenue function
$\mathcal{R}_{k}\left(\mathcal{S},T\right)$ for a heterogeneous market.
Let's redefine few notations from the main paper for the sake of continuity.
Recall the following from the main paper. $Y_{k}\left(c,T\right)$
is the net revenue of the $k^{th}$ operator in $c^{th}$ epoch if
lease duration is $T$. $\widehat{Y}_{k}\left(c,T\right)$ is the
bid of the $k^{th}$ operator in $c^{th}$ epoch if lease duration
is $T$. For notational simplicity, lets denote $Y_{k}\left(c,T\right)$
by $Y_{k}$ and $\widehat{Y}_{k}\left(c,T\right)$ by $\widehat{Y}_{k}$.
According to equation 5, the joint probability distribution of $\widehat{Y}_{k}$
and $Y_{k}$ is\vspace{-1.0em}

\begin{equation}
\begin{bmatrix}Y_{k}\\
\widehat{Y}_{k}
\end{bmatrix}\sim\mathcal{N}\left(\begin{bmatrix}\widetilde{\mu}_{k}\left(T\right)\\
\widetilde{\mu}_{k}\left(T\right)
\end{bmatrix},\begin{bmatrix}\widetilde{\sigma}_{k}^{2}\left(T\right) & \rho_{k}\widetilde{\sigma}_{k}^{2}\left(T\right)\\
\rho_{k}\widetilde{\sigma}_{k}^{2}\left(T\right) & \widetilde{\sigma}_{k}^{2}\left(T\right)
\end{bmatrix}\right)\,;\,\forall c\label{eq:2.1.5.a}
\end{equation}

According to equation 6, the marginal distribution as $\widehat{Y}_{k}$
is

\noindent 
\begin{equation}
\widehat{Y}_{k}\sim\mathcal{N}\left(\widetilde{\mu}_{k}\left(T\right),\widetilde{\sigma}_{k}^{2}\left(T\right)\right)\,;\,\forall c\label{eq:2.1.6.a}
\end{equation}

\noindent \textbf{Proposition 5.} \textit{Say that in every epoch,
one channel is allocated to each of the $\widetilde{M}=\min\left(M,s\right)$
operators having the $\widetilde{M}$ highest sum of revenue in that
epoch. Let $\mathcal{S}_{k}=\mathcal{S}-\left\{ k\right\} $ and $\mathcal{C}\left(A,a\right)$
denote all the possible combinations of size $a-1$ from set $A$
. Let the function $f_{k}\left(Y,\widehat{Y},T\right)$ denote the
probability density function (pdf) corresponding to the joint distribution
of bid $\widehat{Y}$ and true revenue $Y$ of the $k^{th}$ operator
as given by (\ref{eq:2.1.5.a}). Similarly, let the function $F_{k}\left(\widehat{Y},T\right)$
denote the cumulative distribution function (cdf) corresponding to
the marginal distribution of bid $\widehat{Y}$ of the $k^{th}$ operator
as given by (\ref{eq:2.1.6.a}). Then the revenue function of the
$k^{th}$ operator is}\vspace{-1.2em}

\begin{equation}
\mathcal{R}_{k}\left(\mathcal{S},T\right)=\stackrel[-\infty]{\infty}{\int}\:\stackrel[-\infty]{\infty}{\int}Y\cdot\mathcal{G}_{k}\left(\widehat{Y},\mathcal{S},T\right)\cdot f_{k}\left(Y,\widehat{Y},T\right)\,dY\,d\widehat{Y}\label{eq:2.2.3}
\end{equation}

\noindent \vspace{-1.0em}

\noindent \textit{where,}\vspace{-1.5em}

\begin{equation}
\mathcal{G}_{k}\left(\widehat{Y},\mathcal{S},T\right)=\stackrel[m=1]{\widetilde{M}}{\sum}\:\underset{W\in\mathcal{C}\left(\mathcal{S}_{k},m\right)}{\sum}\mathcal{H}_{k}\left(\widehat{Y},T,W\right)\label{eq:2.2.3.1}
\end{equation}

\noindent \vspace{-1.5em}

\begin{equation}
\mathcal{H}_{k}\left(\widehat{Y},T,W\right)=\underset{j\in W}{\prod}\left(1-F_{j}\left(\widehat{Y},T\right)\right)\underset{j\in\mathcal{S}_{k}-W}{\prod}F_{j}\left(\widehat{Y},T\right)\label{eq:2.2.3.2}
\end{equation}

\textit{Proof.} Consider the term $E\left[Y_{k}\left(1,T\right)|w_{1}^{m}=k\right]P\left[w_{1}^{m}=k\right]$
of (8). We have,\vspace{-1.0em}

\begin{eqnarray}
 &  & E\left[Y_{k}\left(1,T\right)|w_{1}^{m}=k\right]P\left[w_{1}^{m}=k\right]\nonumber \\
 & = & \underset{Y}{\sum}YP\left[Y_{k}=Y|w_{1}^{m}=k\right]P\left[w_{1}^{m}=k\right]\nonumber \\
 & = & \underset{\widehat{Y},Y}{\sum}YP\left[Y_{k}=Y,\widehat{Y}_{k}=\widehat{Y}|w_{1}^{m}=k\right]P\left[w_{1}^{m}=k\right]\label{eq:6.1.1}\\
 & = & \underset{\widehat{Y},Y}{\sum}YP\left[{\textstyle w_{1}^{m}=k\,|\,Y_{k}=Y,\widehat{Y}_{k}=\widehat{Y}}\right]P\left[Y_{k}=Y,\right.\nonumber \\
 &  & \qquad\qquad\qquad\qquad\qquad\qquad\qquad\qquad\left.\widehat{Y}_{k}=\widehat{Y}\right]\label{eq:6.1.2}\\
 & = & \underset{\widehat{Y},Y}{\sum}YP\left[w_{1}^{m}=k|\widehat{Y}_{k}=\widehat{Y}\right]P\left[Y_{k}=Y,\widehat{Y}_{k}=\widehat{Y}\right]\label{eq:6.1.3}
\end{eqnarray}

Equation \ref{eq:6.1.1} is obtained by marginalizing over bid $\widehat{Y}_{k}$,
(\ref{eq:6.1.2}) is obtained using \textit{Bayes' Theorem} and (\ref{eq:6.1.3})
is true because given $\widehat{Y}_{k}$, $w_{1}^{m}$ is independent
of $Y_{k}$ (spectrum allocation depends on operators' bid in an epoch
and not on its net revenue in an epoch). $\widetilde{M}=\min\left(M\,,\,s\right)$
channels are allocated in every epoch. As mentioned in the main paper
(first paragraph of section II-B), $w_{1}^{m}$ is the operator who
has the $m^{th}$ highest value of $\widehat{Y}_{k}$ where $m$ is
the channel index. Let $\bigcap$ denote the logical AND operator.
Then the event $w_{1}^{m}=k$ is equivalent to
\[
\underset{j\in W}{\bigcap}\,\widehat{Y}_{j}\geq\widehat{Y}_{k}\:,\:\underset{j\in\mathcal{S}_{k}-W}{\bigcap}\,\widehat{Y}_{j}\leq\widehat{Y}_{k}
\]
for some $W\in\mathcal{C}\left(\mathcal{S}_{k},m\right)$. Hence,
the term $P\left[w_{1}^{m}=k\,|\,Y_{k}=Y\right]$ of (\ref{eq:6.1.3})
can be written as

\vspace{-1.0em}

\[
P\left[w_{1}^{m}=k\,|\,\widehat{Y}_{k}=\widehat{Y}\right]\qquad\qquad\qquad\qquad\qquad\qquad
\]

\vspace{-1.5em}

\[
=\underset{W\in\mathcal{C}\left(\mathcal{S}_{k},m\right)}{\sum}P\left[\underset{j\in W}{\bigcap}\,\widehat{Y}_{j}\geq\widehat{Y}\:,\:\underset{j\in\mathcal{S}_{k}-W}{\bigcap}\,\widehat{Y}_{j}\leq\widehat{Y}\right]\qquad\qquad
\]
\vspace{-1.5em}

\begin{equation}
={\textstyle \underset{W\in\mathcal{C}\left(\mathcal{S}_{k},m\right)}{\sum}\left(\underset{j\in W}{\prod}P\left[\widehat{Y}_{j}\geq\widehat{Y}\right]\cdot\underset{j\in\mathcal{S}_{k}-W}{\prod}P\left[\widehat{Y}_{j}\leq\widehat{Y}\right]\right)}\quad\label{eq:6.2}
\end{equation}

Equation \ref{eq:6.2} holds because the bids of any two operators
are \textit{not correlated} and are hence \textit{independent}. Using
(8), (\ref{eq:6.1.3}) and (\ref{eq:6.2}) we get,

\vspace{-1.0em}

\begin{equation}
\mathcal{R}_{k}\left(\mathcal{S},T\right)=\underset{\widehat{Y},Y}{\sum}Y\cdot\mathcal{G}_{k}\left(\widehat{Y},\mathcal{S},T\right)\cdot P\left[Y_{k}=Y,\widehat{Y}_{k}=\widehat{Y}\right]\label{eq:6.3}
\end{equation}

\noindent where,\vspace{-1.25em}

\begin{equation}
\mathcal{G}_{k}\left(\widehat{Y},\mathcal{S},T\right)=\stackrel[m=1]{\widetilde{M}}{\sum}\:\underset{W\in\mathcal{C}\left(\mathcal{S}_{k},m\right)}{\sum}\mathcal{H}_{k}\left(\widehat{Y},T,W\right)\label{eq:6.3.1}
\end{equation}

\vspace{-1.25em}

\begin{equation}
\mathcal{H}_{k}\left(\widehat{Y},T,W\right)=\underset{j\in W}{\prod}P\left[\widehat{Y}_{j}\geq\widehat{Y}\right]\cdot\underset{j\in\mathcal{S}_{k}-W}{\prod}P\left[\widehat{Y}_{j}\leq\widehat{Y}\right]\label{eq:6.3.2}
\end{equation}

The joint probability distribution of $Y_{k}$ and $\widehat{Y}_{k}$
is governed by (\ref{eq:2.1.5.a}). Let $f_{k}\left(Y,\widehat{Y},T\right)$
denote the corresponding joint probability density function of $Y_{k}$
and $\widehat{Y}_{k}$. The marginal probability distribution of $\widehat{Y}_{k}$
is governed by (\ref{eq:2.1.6.a}). Let $F_{k}\left(\widehat{Y},T\right)$
denote the corresponding cumulative distribution function of $\widehat{Y}_{k}$.
Therefore,\vspace{-1.0em}

\begin{eqnarray*}
P\left[Y_{k}=Y,\widehat{Y}_{k}=\widehat{Y}\right] & = & f_{k}\left(Y,\widehat{Y},T\right)\,dY\,d\widehat{Y}\\
P\left[\widehat{Y}_{k}\geq\widehat{Y}\right] & = & \left(1-F_{k}\left(\widehat{Y},T\right)\right)\\
P\left[\widehat{Y}_{k}\leq\widehat{Y}\right] & = & F_{k}\left(\widehat{Y},T\right)
\end{eqnarray*}
which when substituted in (\ref{eq:6.3}) and (\ref{eq:6.3.2}) yields
(\ref{eq:2.2.3})-(\ref{eq:2.2.3.2}). This completes the proof.

\section{Revenue Function for Homogeneous Market}

We want to derive a simplified expression of revenue function for
a market that is homogeneous in $\mu_{k}$, $\sigma_{k}$, $a_{k}$
and $\rho_{k}$, i.e. $\mu_{k}=\mu$, $\sigma_{k}=\sigma$, $a_{k}=a$
and $\rho_{k}=\rho\,,\,\forall k$. In a homogeneous market, $\widetilde{\mu}_{k}\left(T\right)$
and $\widetilde{\sigma}_{k}\left(T\right)$ in (3) and (4) respectively,
are same for all the operators. We have,\vspace{-1.0em}

\begin{eqnarray}
\widetilde{\mu}\left(T\right) & = & \mu T\label{eq:4.0.1}\\
\widetilde{\sigma}\left(T\right) & = & \frac{\sqrt{T-a\left(2-2a^{T}+aT\right)}}{\left(1-a\right)}\sigma\label{eq:4.0.2}
\end{eqnarray}
In a homogeneous market, we can drop the subscript $k$ from $f_{k}\left(Y,\widehat{Y},T\right)$
and $\mathcal{G}_{k}\left(\widehat{Y},\mathcal{S},T\right)$ in (\ref{eq:2.2.3})
because $f_{k}\left(Y,\widehat{Y},T\right)$ and $\mathcal{G}_{k}\left(\widehat{Y},\mathcal{S},T\right)$
is same for all the operators. Let $s=\left|\mathcal{S}\right|$,
$\widetilde{M}=\min\left(M,s\right)$ and $\begin{pmatrix}s-1\\
m-1
\end{pmatrix}=\frac{\left(s-1\right)!}{\left(m-1\right)!\left(s-m\right)!}$. Then the simplified revenue function is\vspace{-1.0em}

\begin{equation}
\mathcal{R}\left(s,T\right)=\stackrel[-\infty]{\infty}{\int}\:\stackrel[-\infty]{\infty}{\int}Y\cdot\mathcal{G}\left(\widehat{Y},s,T\right)\cdot f\left(Y,\widehat{Y},T\right)\,dY\,d\widehat{Y}\;\text{{where}}\label{eq:4.1}
\end{equation}

\vspace{-1.0em}

\begin{equation}
\mathcal{G}\left(\widehat{Y},s,T\right)=\stackrel[m=1]{\widetilde{M}}{\sum}\begin{pmatrix}s-1\\
m-1
\end{pmatrix}\left(1-F\left(\widehat{Y},T\right)\right)^{m-1}F\left(\widehat{Y},T\right)^{s-m}\label{eq:4.2}
\end{equation}

Note that revenue function in a homogeneous market is dependent on
the number of interested operators $s$ and not on the set of interested
operators $\mathcal{S}$. $f\left(Y,\widehat{Y},T\right)$ and $F\left(\widehat{Y},T\right)$
are the pdf and cdf respectively of the normal distribution given
by (\ref{eq:2.1.5.a}). We have,\vspace{-1.0em}

\begin{equation}
f\left(Y,\widehat{Y},T\right)=\frac{\exp\left(-\frac{1}{2}\begin{bmatrix}\frac{Y-\widetilde{\mu}\left(T\right)}{\widetilde{\sigma}\left(T\right)}\\
\frac{\widehat{Y}-\widetilde{\mu}\left(T\right)}{\widetilde{\sigma}\left(T\right)}
\end{bmatrix}^{T}\begin{bmatrix}1 & \rho\\
\rho & 1
\end{bmatrix}^{-1}\begin{bmatrix}\frac{Y-\widetilde{\mu}\left(T\right)}{\widetilde{\sigma}\left(T\right)}\\
\frac{\widehat{Y}-\widetilde{\mu}\left(T\right)}{\widetilde{\sigma}\left(T\right)}
\end{bmatrix}\right)}{2\pi\sqrt{1-\rho^{2}}\widetilde{\sigma}^{2}\left(T\right)}\label{eq:4.3}
\end{equation}

\begin{equation}
F\left(\widehat{Y},T\right)=\frac{1}{2}\left(1+\text{erf}\left(\frac{\widehat{Y}-\widetilde{\mu}\left(T\right)}{\sqrt{2}\widetilde{\sigma}\left(T\right)}\right)\right)\label{eq:4.4}
\end{equation}

Substituting $Y=\widetilde{\sigma}\left(T\right)y+\widetilde{\mu}\left(T\right)$
and $\widehat{Y}=\widetilde{\sigma}\left(T\right)\widehat{y}+\widetilde{\mu}\left(T\right)$
in (\ref{eq:4.1}) we get,\vspace{-1.0em}

\begin{equation}
\mathcal{R}\left(s,T\right)=\alpha\left(\rho,s\right)\widetilde{\mu}\left(T\right)+\beta\left(\rho,s\right)\widetilde{\sigma}\left(T\right)\quad\text{{where}}\label{eq:4.4.0}
\end{equation}

\begin{eqnarray}
\alpha\left(\rho,s\right) & = & \stackrel[-\infty]{\infty}{\int}\:\stackrel[-\infty]{\infty}{\int}G\left(\widehat{y},s\right)h\left(y,\widehat{y},\rho\right)\,dy\,d\widehat{y}\label{eq:4.4.1}\\
\beta\left(\rho,s\right) & = & \stackrel[-\infty]{\infty}{\int}\:\stackrel[-\infty]{\infty}{\int}yG\left(\widehat{y},s\right)h\left(y,\widehat{y},\rho\right)\,dy\,d\widehat{y}\label{eq:4.4.2}\\
G\left(\widehat{y},s\right) & = & \stackrel[m=1]{\widetilde{M}}{\sum}\begin{pmatrix}s-1\\
m-1
\end{pmatrix}\left(1-H\left(\widehat{y}\right)\right)^{m-1}H\left(\widehat{y}\right)^{s-m}\label{eq:4.4.3}\\
h\left(y,\widehat{y},\rho\right) & = & \frac{\exp\left(-\frac{1}{2}\begin{bmatrix}y\\
\widehat{y}
\end{bmatrix}^{T}\begin{bmatrix}1 & \rho\\
\rho & 1
\end{bmatrix}^{-1}\begin{bmatrix}y\\
\widehat{y}
\end{bmatrix}\right)}{2\pi\sqrt{1-\rho^{2}}}\label{eq:4.4.4}\\
\overline{H}\left(\widehat{y}\right) & = & \frac{1}{2}\left(1+\text{erf}\left(\frac{\widehat{y}}{\sqrt{2}}\right)\right)\label{eq:4.4.5}
\end{eqnarray}

We want to simplify $\alpha\left(\rho,s\right)$ in (\ref{eq:4.4.1})
further. We have,\vspace{-1.0em}

\begin{eqnarray}
\alpha\left(\rho,s\right) & = & \stackrel[-\infty]{\infty}{\int}G\left(\widehat{y},s\right)\left(\frac{\exp\left(-\frac{1}{2}\widehat{y}^{2}\right)}{\sqrt{2\pi}}\right)\,d\widehat{y}\label{eq:4.5.1}\\
 & = & \stackrel[-\infty]{\infty}{\int}G\left(\widehat{y},s\right)\,d\left(\overline{H}\left(\widehat{y}\right)\right)\label{eq:4.5.2}
\end{eqnarray}

Equation \ref{eq:4.5.1} is obtained by re-writting (\ref{eq:4.4.1})
as $\mbox{{\ensuremath{\alpha\left(\rho,s\right)}=\ensuremath{\stackrel[-\infty]{\infty}{\int}}G\ensuremath{\left(\widehat{y},s\right)\left(\stackrel[-\infty]{\infty}{\int}h\left(y,\widehat{y},\rho\right)\,dy\right)\,}d\ensuremath{\widehat{y}}}}$
and then observing that the inner integral is equal to $\frac{1}{\sqrt{2\pi}}\exp\left(-\frac{1}{2}\widehat{y}^{2}\right)$.
Equation \ref{eq:4.5.2} follows from (\ref{eq:4.5.1}) because $\mbox{{\ensuremath{\overline{H}\left(\widehat{y}\right)}=\ensuremath{\stackrel[-\infty]{\infty}{\int}\frac{1}{\sqrt{2\pi}}\exp\left(-\frac{1}{2}\widehat{y}^{2}\right)\,}d\ensuremath{\widehat{y}}}}$.
So we have,
\begin{equation}
\alpha\left(\rho,s\right)=\stackrel[-\infty]{\infty}{\int}\stackrel[m=1]{\widetilde{M}}{\sum}\begin{pmatrix}s-1\\
m-1
\end{pmatrix}\left(1-\overline{H}\right)^{m-1}\overline{H}{}^{s-m}\,d\overline{H}\label{eq:4.6}
\end{equation}

By using Binomial Expansion of $\left(1-\overline{H}\right)^{m-1}$
followed by some algebraic manipulation, we can show that the RHS
of (\ref{eq:4.6}) is equal to $\frac{\widetilde{M}}{s}$. So the
final simplified form of the revenue function for homogeneous market
is
\begin{equation}
\mathcal{R}\left(s,T\right)=\frac{\widetilde{M}}{s}\widetilde{\mu}\left(T\right)+\beta\left(\rho,s\right)\widetilde{\sigma}\left(T\right)\label{eq:4.7}
\end{equation}

\noindent where $\widetilde{\mu}\left(T\right)$, $\widetilde{\sigma}\left(T\right)$
and $\beta\left(\rho,s\right)$ are given by (\ref{eq:4.0.1}), (\ref{eq:4.0.2})
and (\ref{eq:4.4.2}) respectively.

\section{Proof of the Properties of the Revenue and the Objective Function}

\subsection{Proof that $\mathcal{R}\left(s,T\right)$ is monotonic increasing
in $T$}

We want to prove that revenue function in a homogeneous market is
monotonic increasing in $T$. Revenue function in a homogeneous market
is given by (\ref{eq:4.7}). So proving that $\mathcal{R}\left(s,T\right)$
is monotonic increasing in $T$ is same as proving that $\widetilde{\mu}\left(T\right)$
and $\widetilde{\sigma}\left(T\right)$ are monotonic increasing in
$T$. $\widetilde{\mu}\left(T\right)$ and $\widetilde{\sigma}\left(T\right)$
are given by (\ref{eq:4.0.1}) and (\ref{eq:4.0.2}) respectively.
It is obvious that $\widetilde{\mu}\left(T\right)=\mu T$ is monotonic
increasing in $T$. To prove that $\widetilde{\sigma}\left(T\right)$
is monotonic increasing in $T$, it is enough to show that\vspace{-1.0em}

\begin{equation}
\left(T+1\right)-a\left(2-2a^{T+1}+a\left(T+1\right)\right)\geq T-a\left(2-2a^{T}+aT\right)\label{eq:5.1.1}
\end{equation}

Inequality \ref{eq:5.1.1} simplifies to $1+a\geq2a^{T+1}$. Since
$a\in\left[0,1\right)$ and $T\geq1$, $a^{T+1}\leq a^{2}$. Therefore,
proving $1+a\geq2a^{T+1}$ is same as proving $1+a\geq2a^{2}$ or
equivalently $1\geq a\left(2a-1\right)$. For $a\in\left[0,1\right)$,
$\left(2a-1\right)\leq1$ and hence $1\geq a\left(2a-1\right)$. This
completes the proof.

\subsection{Proof that $U\left(s,T\right)$ is monotonic decreasing in $T$}

We want to prove that objective function in a homogeneous market is
monotonic decreasing in $T$. Objective function in a homogeneous
market is given by (14). We have,\vspace{-1.0em}

\begin{eqnarray}
 &  & U\left(s,T\right)\nonumber \\
 & = & \frac{s}{T}\mathcal{R}\left(s,T\right)\nonumber \\
 & = & \frac{s}{T}\left(\frac{\widetilde{M}}{s}\widetilde{\mu}\left(T\right)+\beta\left(\rho,s\right)\widetilde{\sigma}\left(T\right)\right)\label{eq:5.2.1}\\
 & = & \mu\widetilde{M}+\frac{s\beta\left(\rho,s\right)\sigma}{\left(1-a\right)}\frac{\sqrt{T-a\left(2-2a^{T}+aT\right)}}{T}\label{eq:5.2.2}
\end{eqnarray}

Equation \ref{eq:5.2.1} is obtained by substituting $\mathcal{R}\left(s,T\right)$
from (\ref{eq:4.7}). Equation \ref{eq:5.2.2} is obtained by substituting
$\widetilde{\mu}\left(T\right)$ and $\widetilde{\sigma}\left(T\right)$
from (\ref{eq:4.0.1}) and (\ref{eq:4.0.2}) respectively. In (\ref{eq:5.2.2}),
$\frac{\sqrt{T-a\left(2-2a^{T}+aT\right)}}{T}$ is monotonic decreasing
in $T$. Hence, $U\left(s,T\right)$ is monotonic decreasing in $T$.
This completes the proof.

\subsection{Proof that $U\left(s,T\right)$ is monotonic increasing in $s$}

The proof is divided into two steps. In the first step, we show that
$U\left(s,T\right)$ in (\ref{eq:5.2.2}) can be equivalently expressed
as
\begin{equation}
U\left(s,T\right)=\mu\widetilde{M}+\frac{s\beta\left(1,s\right)\left(\rho\sigma\right)}{\left(1-a\right)}\frac{\sqrt{T-a\left(2-2a^{T}+aT\right)}}{T}\label{eq:5.3.1}
\end{equation}

Qualitatively, (\ref{eq:5.3.1}) shows that a homogeneous market with
standard deviation $\sigma$ and bid correlation coefficient $\rho$
have the same objective function as a homogeneous market with standard
deviation $\rho\sigma$ and bid correlation coefficient $1$. If bid
correlation coefficient is $1$, it represents a degenerate case where
the net revenue of the $k^{th}$ operator in epoch $c$ for lease
duration $T$, $Y_{k}\left(c,T\right)$, is equal to the bid of the
$k^{th}$ operator in epoch $c$ for lease duration $T$, $\widehat{Y}_{k}\left(c,T\right)$.
Mathematically, $\widehat{Y}_{k}\left(c,T\right)=Y_{k}\left(c,T\right)\,;\,\forall k,c,T$.
In our second step, we prove that if $\widehat{Y}_{k}\left(c,T\right)=Y_{k}\left(c,T\right)\,;\,\forall k,c,T$,
then $U\left(s,T\right)$ is monotonic increasing in $s$.

The first step is equivalent to showing $\beta\left(\rho,s\right)=\rho\beta\left(1,s\right)$.
$\beta\left(\rho,s\right)$ is given by (\ref{eq:4.4.2}). Equation
\ref{eq:4.4.2} can be equivalently written as
\begin{equation}
\beta\left(\rho,s\right)=\stackrel[-\infty]{\infty}{\int}G\left(\widehat{y},s\right)\left(\stackrel[-\infty]{\infty}{\int}yh\left(y,\widehat{y},\rho\right)\,dy\right)\,d\widehat{y}\label{eq:5.3.2}
\end{equation}

\noindent where $G\left(\widehat{y},s\right)$ and $h\left(y,\widehat{y}\right)$
is given by (\ref{eq:4.4.3}) and (\ref{eq:4.4.4}) respectively.
We will now evaluate the integral, $\stackrel[-\infty]{\infty}{\int}yh\left(y,\widehat{y}\right)\,dy$,
of (\ref{eq:5.3.2}). Lets consider the term $\begin{bmatrix}y\\
\widehat{y}
\end{bmatrix}^{T}\begin{bmatrix}1 & \rho\\
\rho & 1
\end{bmatrix}^{-1}\begin{bmatrix}y\\
\widehat{y}
\end{bmatrix}$ in (\ref{eq:4.4.4}). Using simple algebra , we can show that
\begin{equation}
\begin{bmatrix}y\\
\widehat{y}
\end{bmatrix}^{T}\begin{bmatrix}1 & \rho\\
\rho & 1
\end{bmatrix}^{-1}\begin{bmatrix}y\\
\widehat{y}
\end{bmatrix}=\frac{\left(y-\rho\widehat{y}\right)^{2}}{1-\rho^{2}}+\widehat{y}^{2}\label{eq:5.3.4}
\end{equation}

Substituting (\ref{eq:5.3.4}) in (\ref{eq:4.4.4}) and then substituting
the resulting $h\left(y,\widehat{y}\right)$ in the integral $\stackrel[-\infty]{\infty}{\int}yh\left(y,\widehat{y}\right)\,dy$
we have,\vspace{-1.0em}

\[
\stackrel[-\infty]{\infty}{\int}yh\left(y,\widehat{y},\rho\right)\,dy\quad\quad\quad\quad\quad\quad\quad\quad\quad\quad\quad\quad\quad\quad\quad
\]

\noindent \vspace{-1.5em}

\begin{equation}
=\overline{h}\left(\widehat{y}\right)\stackrel[-\infty]{\infty}{\int}y\frac{1}{\sqrt{2\pi}\sqrt{1-\rho^{2}}}\exp\left(-\frac{1}{2}\left(\frac{y-\rho\widehat{y}}{\sqrt{1-\rho^{2}}}\right)^{2}\right)\,dy\label{eq:5.3.5}
\end{equation}

\noindent where $\overline{h}\left(\widehat{y}\right)=\frac{1}{\sqrt{2\pi}}\exp\left(-\frac{\widehat{y}^{2}}{2}\right)$.
The integral in (\ref{eq:5.3.5}) is the expectation of a gaussian
random variable $Y$ whose mean and standard deviation is $\rho\widehat{y}$
and $\sqrt{1-\rho^{2}}$ respectively. Hence, integral in (\ref{eq:5.3.5})
is simply equal to $\rho\widehat{y}$. To this end we have,\vspace{-1.0em}

\begin{equation}
\stackrel[-\infty]{\infty}{\int}yh\left(y,\widehat{y},\rho\right)\,dy=\overline{h}\left(\widehat{y}\right)\cdot\left(\rho\widehat{y}\right)\label{eq:5.3.6}
\end{equation}

Substituting (\ref{eq:5.3.6}) in (\ref{eq:5.3.2}) we get,
\begin{eqnarray*}
\beta\left(\rho,s\right) & = & \rho\stackrel[-\infty]{\infty}{\int}\widehat{y}G\left(\widehat{y},s\right)\overline{h}\left(\widehat{y}\right)\,d\widehat{y}\\
 & = & \rho\left(1\cdot\stackrel[-\infty]{\infty}{\int}\widehat{y}G\left(\widehat{y},s\right)\overline{h}\left(\widehat{y}\right)\,d\widehat{y}\right)\\
 & = & \rho\beta\left(1,s\right)
\end{eqnarray*}

This concludes the proof of the first step. In our second step, we
have to prove that if $\widehat{Y}_{k}\left(c,T\right)=Y_{k}\left(c,T\right)\,;\,\forall k,c,T$,
then $U\left(s,T\right)$ is monotonic increasing in $s$. Rather
than proving this for a homogeneous market where $Y_{k}\left(c,T\right)$
is governed by Proposition 1, we take a more general approach. We
prove the following: In a heterogeneous market, $U\left(\mathcal{S},T\right)\leq U\left(\mathcal{S}\bigcup\left\{ a\right\} ,T\right)$
where $a\notin\mathcal{S}$, for any random variable $Y_{k}\left(c,T\right)$
if $\widehat{Y}_{k}\left(c,T\right)=Y_{k}\left(c,T\right)\,;\,\forall k,c,T$.

Refering to (9) and (10), we can say that the objective function $U\left(\mathcal{S},T\right)$
is the expectation of 
\begin{equation}
X\left(\mathcal{S},T\right)=\frac{1}{\mathcal{T}}\stackrel[c=1]{C}{\sum}\,\stackrel[m=1]{\widetilde{M}}{\sum}Y_{w_{c}^{m}}\left(c,T\right)\label{eq:5.3.7}
\end{equation}

Recall that $w_{c}^{m}$ is the index of the operator who is allocated
the $m^{th}$ channel in $c^{th}$ epoch. $w_{c}^{m}$ is decided
based on operators bids $\widehat{Y}_{k}\left(c,T\right)$ which in
our case is equal to $Y_{k}\left(c,T\right)$. Since $w_{c}^{m}$
is among the set of interested operators $\mathcal{S}$, $X\left(\mathcal{S},T\right)$
is a function of $\mathcal{S}$. If we can prove that $X\left(\mathcal{S},T\right)\leq X\left(\mathcal{S}\bigcup\left\{ a\right\} ,T\right)$
for all revenue process $x_{k}\left(t\right)$, it directly implies
that $U\left(\mathcal{S},T\right)\leq U\left(\mathcal{S}\bigcup\left\{ a\right\} ,T\right)$.
This is because if there are two random variables $Z_{1}$ and $Z_{2}$
such that $Z_{1}\leq Z_{2}\:;\:\forall Z_{1},Z_{2}$, then $E\left[Z_{1}\right]\leq E\left[Z_{2}\right]$.

Consider a list $\mathcal{Y}_{\mathcal{S}}$ consisting of $Y_{k}\left(c,T\right)$
for all the operators in set $\mathcal{S}$. Since the channels are
allocated to the operators having the $\widetilde{M}$ highest $Y_{k}\left(c,T\right)$,
the term $\stackrel[m=1]{\widetilde{M}}{\sum}Y_{w_{c}^{m}}\left(c,T\right)$
of (\ref{eq:5.3.7}) is equal to the sum of the $\widetilde{M}$ highest
values in list $\mathcal{Y}_{\mathcal{S}}$. Similarly, consider a
list $\mathcal{Y}_{\mathcal{S}\bigcup\left\{ a\right\} }$ which is
same as $\mathcal{Y}_{\mathcal{S}}$ but with an additional value
$Y_{a}\left(c,T\right)$. Definitely, the sum of the $\widetilde{M}$
highest values is greater for list $\mathcal{Y}_{\mathcal{S}\bigcup\left\{ a\right\} }$
than list $\mathcal{Y}_{\mathcal{S}}$. This is because the values
in list $\mathcal{Y}_{\mathcal{S}}$ is a subset of the values in
list $\mathcal{Y}_{\mathcal{S}\bigcup\left\{ a\right\} }$. Therefore,
the term $\stackrel[m=1]{\widetilde{M}}{\sum}Y_{w_{c}^{m}}\left(c,T\right)$
of (\ref{eq:5.3.7}) is greater for $\mathcal{S}\bigcup\left\{ a\right\} $
than $\mathcal{S}$. This is true for all $c$'s and for any revenue
process $x_{k}\left(t\right)$. This proves that $X\left(\mathcal{S},T\right)\leq X\left(\mathcal{S}\bigcup\left\{ a\right\} ,T\right)$
for any revenue process $x_{k}\left(t\right)$. This concludes the
proof.

\section{Proof of Proposition 4\label{sec:Proof-of-Proposition4}}

For a homogeneous market with complete information, the perceived
objective function, $\widetilde{U}\left(T\right)$, of $OP1$ is

\begin{equation}
\widetilde{U}\left(T\right)=\begin{cases}
\frac{N}{T}\mathcal{R}\left(N,T\right) & ;\left\lceil \theta\right\rceil \leq T\leq\Lambda\\
0 & ;o.w.
\end{cases}\label{eq:3.1.0}
\end{equation}

\noindent where $\theta$ is the solution to $\mathcal{R}\left(N,\theta\right)=\lambda$.
We will now explain (\ref{eq:3.1.0}). The largest set of interested
operators according to the regulator, $\widetilde{\mathcal{S}}^{L}\left(T\right)$,
in a homogeneous market, is equal to $\left\{ 1,2,\ldots,N\right\} $
if the lease duration $T\leq\Lambda$ and $\mu T\geq\lambda$ (refer
to (19)). Otherwise, $\widetilde{\mathcal{S}}^{L}\left(T\right)=\emptyset$.
If $\widetilde{\mathcal{S}}^{L}\left(T\right)=\emptyset$, then $\widetilde{\mathcal{S}}\left(T\right)=\emptyset$
. If $\widetilde{\mathcal{S}}^{L}\left(T\right)=\left\{ 1,2,\ldots,N\right\} $,
the minimum revenue an operator earns in an epoch is $\mathcal{R}\left(N,T\right)$.
Hence, according to (23), the perceived set of interested operators
is\vspace{-1.5em}

\begin{equation}
\widetilde{S}\left(T\right)=\begin{cases}
\left\{ 1,2,\ldots,N\right\}  & ;T\leq\Lambda,\mathcal{R}\left(N,T\right)\geq\lambda\\
\emptyset & ;o.w.
\end{cases}\label{eq:3.1.1}
\end{equation}

As discussed in Section~II-C, in a homogeneous market, $\mathcal{R}\left(N,T\right)$
is monotonic increasing in $T$. Therefore, the least $T$ satisfying
the constraint $\mathcal{R}\left(N,T\right)\geq\lambda$ is $\left\lceil \,\theta\right\rceil $
where $\theta$ is the solution to $\mathcal{R}\left(N,\theta\right)=\lambda$.
The ceiling function $\left\lceil \cdot\right\rceil $ is needed because
we consider a time slotted model. For any $T\geq\left\lceil \theta\right\rceil $,
we have $\mathcal{R}\left(N,T\right)\geq\lambda$. Therefore, the
condition $T\leq\Lambda,\mathcal{R}\left(N,T\right)\geq\lambda$ is
equivalent to $\left\lceil \theta\right\rceil \leq T\leq\Lambda$
and hence (\ref{eq:3.1.1}) is same as

\begin{equation}
\widetilde{S}\left(T\right)=\begin{cases}
\left\{ 1,2,\ldots,N\right\}  & ;\left\lceil \theta\right\rceil \leq T\leq\Lambda\\
\emptyset & ;o.w.
\end{cases}\label{eq:3.1.2}
\end{equation}

According to (17), the perceived objective function for homogeneous
market is

\begin{equation}
\widetilde{U}\left(T\right)=\frac{\widetilde{s}\left(T\right)}{T}\mathcal{R}\left(\widetilde{s}\left(T\right),T\right)\label{eq:3.1.3}
\end{equation}
where $\widetilde{s}\left(T\right)=\left|\widetilde{S}\left(T\right)\right|$.
Finally, (\ref{eq:3.1.0}) can be obtained directly by using (\ref{eq:3.1.2})
and (\ref{eq:3.1.3}). This completes the explanation of (\ref{eq:3.1.0}).
According to (\ref{eq:3.1.0}), if $\left\lceil \theta\right\rceil >\Lambda$,
then there exists no $T$ such that $\left\lceil \theta\right\rceil \leq T\leq\Lambda$.
Hence, $\widetilde{U}\left(T\right)=0\,;\,\forall T$. Therefore,
optimal lease duration $T^{*}$ can be set to any value and the optimal
value of objective function $U^{*}=0$. If $\left\lceil \theta\right\rceil \leq\Lambda$,
then $\widetilde{U}\left(T\right)=\frac{N}{T}\mathcal{R}\left(N,T\right)$
in the range $\left\lceil \theta\right\rceil \leq T\leq\Lambda$.
As discussed in Section~II-C, the objective function is monotonic
decreasing in a homogeneous market. Hence, $T^{*}=\left\lceil \theta\right\rceil $
maximizes $\widetilde{U}\left(T\right)=\frac{N}{T}\mathcal{R}\left(N,T\right)$
in the range $\left\lceil \theta\right\rceil \leq T\leq\Lambda$ and
accordingly $U^{*}=\frac{N}{\left\lceil \theta\right\rceil }\mathcal{R}\left(N,\left\lceil \theta\right\rceil \right)$.

\section{Proof of Proposition 5\label{sec:Proof-of-Proposition5}}

$\gamma_{k}$ and $\Gamma_{k}$ in line 12 of Algorithm~1 is computated
$\mathcal{O}\left(N^{2}\right)$ times. This simply follows from the
observation that the for loop in line 6 executes $\left|Q^{L}\right|=2N$
times and the for loop in line 12 executes $\left|\mathcal{X}_{i}^{L}\right|<N$
times. As discussed before, each computation of $\gamma_{k}$ and
$\Gamma_{k}$ involves $\mathcal{O}\left(\log_{2}\left(\Theta_{i}^{L}-\theta_{i}^{L}\right)\right)$
computations of $\widetilde{\mathcal{R}}_{k}\left(\mathcal{S},T\right)$.
Since the maximum value of lease duration in the list $Q^{L}$ is
$\widehat{\Lambda}^{L}$, we have $\Theta_{i}^{L}\leq\widehat{\Lambda}^{L}\,;\,\forall i$.
This suggests that $\widetilde{\mathcal{R}}_{k}\left(\mathcal{S},T\right)$
is computed $\mathcal{O}\left(N^{2}\log_{2}\left(\widehat{\Lambda}^{L}\right)\right)$
times in line 12. $\widetilde{U}\left(\mathcal{X}_{j}\,,\,\theta_{j}\right)$
and $\widetilde{U}\left(\mathcal{X}_{j}\,,\,\Theta_{j}\right)$ in
lines 22 and 24 respectively are also computed $\mathcal{O}\left(N^{2}\right)$
times which can be explained as follows. Corresponding to every operator
in $\mathcal{X}_{i}^{L}$ (line 11), there are at most $2$ ordered
pairs. Hence, $\left|Q\right|\leq2\left|\mathcal{X}_{i}^{L}\right|\leq2N$.
Since the for loop in line 6 executes $\left|Q^{L}\right|=2N$ times
and the for loop in line 18 executes $\left|Q\right|\leq2N$ times,
$\widetilde{U}\left(\mathcal{X}_{j}\,,\,\theta_{j}\right)$ and $\widetilde{U}\left(\mathcal{X}_{j}\,,\,\Theta_{j}\right)$
in lines 22 and 24 respectively are computed $\mathcal{O}\left(N^{2}\right)$
times. Each computation of $\widetilde{U}\left(\mathcal{X}_{j}\,,\,\theta_{j}\right)$
and $\widetilde{U}\left(\mathcal{X}_{j}\,,\,\Theta_{j}\right)$ involves
$\mathcal{O}\left(N\right)$ computations of $\widetilde{\mathcal{R}}_{k}\left(\mathcal{S},T\right)$
(refer to (16)). This suggests that $\widetilde{\mathcal{R}}_{k}\left(\mathcal{S},T\right)$
is computed $\mathcal{O}\left(N^{3}\right)$ times in lines 22 and
24. Finally, time complexity of Algorithm~1 is $\mathcal{O}\left(N^{2}\log_{2}\left(\widehat{\Lambda}^{L}\right)+N^{3}\right)$.


\vspace{-3.0em}
\begin{IEEEbiography}[{\includegraphics[clip,width=1in,height=1.15in]{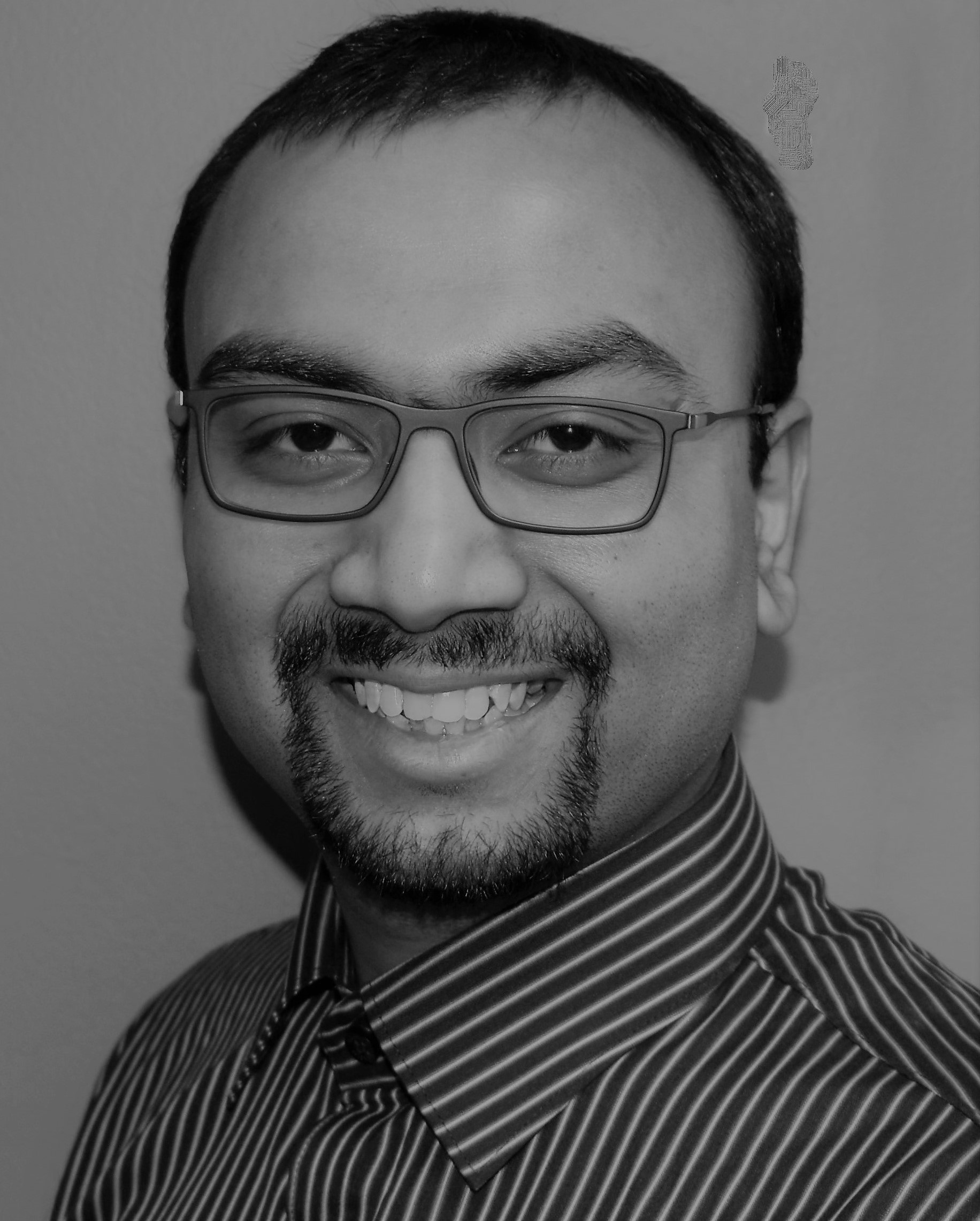}}]{Gourav Saha}
 received a B.E. degree from Anna University, Chennai, India, in 2012, M.S.from Indian Institute of Technology  Madras, India, in 2015, and Ph.D. from Rensselaer Polytechnic Institute, Troy, New York, in 2020, all in electrical engineering. He is currently a postdoctoral scholar in the Department of Electrical and Computer Engineering. His research experience includes control systems, online algorithms, game theory, and economics of wireless spectrum sharing market. His current research involves various stochastic control problems related to millimeter wave communication and cyber-physical systems.\end{IEEEbiography}

\vspace{-3.0em}
\begin{IEEEbiography}[{\includegraphics[clip,width=1in,height=1.25in]{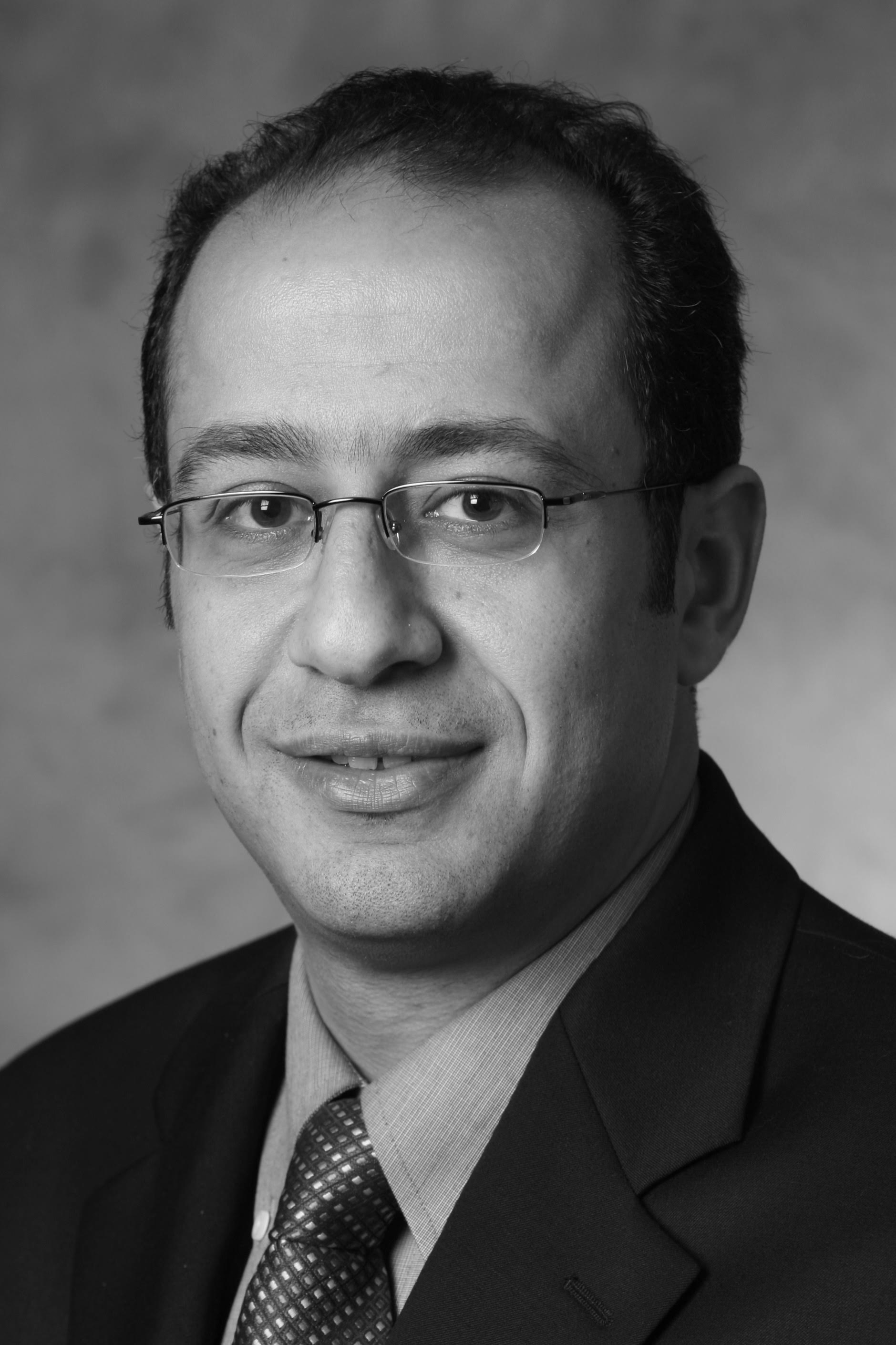}}]{Alhussein A. Abouzeid}
 received the B.S. degree with honors from Cairo University, Cairo, Egypt, in 1993, and the M.S. and Ph.D. degrees from University of Washington, Seattle, in 1999 and 2001, respectively, all in electrical engineering. He held appointments with Alcatel Telecom (1994-1997), AlliedSignal (1999), and Hughes Research Labs (2000). Since 2001 he has been with the Electrical, Computer and Systems Engineering Department at Rensselaer Polytechnic Institute where he is now a Professor. From 2008 to 2010 he served as Program Director in the Computer and Network Systems Division of the U.S. National Science Foundation (NSF). He was the founding director of WiFiUS, an international NSF-funded US-Finland virtual institute on wireless systems research. He received the Faculty Early Career Development Award (CAREER) from NSF in 2006, and the Finnish Distinguished Professor Fellow award from Tekes (now Business Finland) in 2014-2019. He has served as an Associate Editor for several IEEE and Elsevier journals, and on the organizing and technical committees of several IEEE/ACM conferences.\end{IEEEbiography}

\vspace{-3.0em}
\begin{IEEEbiography}[{\includegraphics[clip,width=1in,height=1.25in]{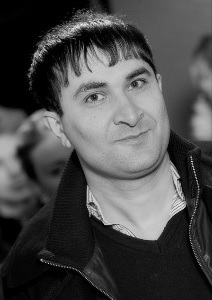}}]{Zaheer Khan}
 received the M.Sc. degree in electrical engineering from the University of Borås, Sweden, in 2007, and the PhD degree in electrical engineering from the University of Oulu, Finland, in 2011. He was a Research Fellow/Principal Investigator with the University of Oulu from 2011 to 2016. He held a tenure-track Assistant Professor position with the University of Liverpool, U.K., from 2016 to 2017. He is currently working as an Adjunct Professor at the University of Oulu. His research interests include design and  implementation of advanced signal processing, and data analytics algorithms for wireless networks  on  reconfigurable System-on-Chip (SoC) platforms, application of game theory to model distributed wireless networks, use of machine learning for industrial asset monitoring and proactive network resource allocation solutions, and wireless signal design.\end{IEEEbiography}

\vspace{-3.0em}
\begin{IEEEbiography}[{\includegraphics[clip,width=1in,height=1.25in]{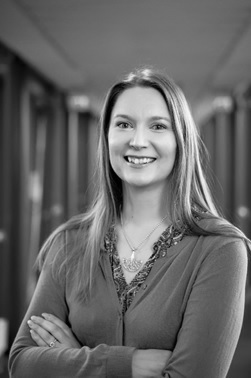}}]{Marja Matinmikko-Blue}
 is 6G Flagship Research Coordinator and Senior Research Fellow at CWC, University of Oulu where she holds an Adjunct Professor position on spectrum management. She conducts multi-disciplinary research on technical, business and regulatory aspects of mobile communication systems in close collaboration between industry, academia and regulators. She holds a Dr.Sc. degree in telecommunications engineering and a Ph.D. degree in management sciences from University of Oulu from 2012 and 2018. She has coordinated four national project consortia that have successfully demonstrated the world’s first licensed shared access spectrum sharing trials and introduced a new local 5G operator concept that has become a reality. She has published 150+ scientific publications and prepared 100+ contributions to regulatory bodies on spectrum management in national, European and international levels.\end{IEEEbiography}

\end{document}